\documentclass{ws-ijmpb}
\usepackage{epsfig,amssymb,amsmath,latexsym}
\usepackage{color}
\usepackage[colorlinks=true,linktoc=page,linkcolor=blue,citecolor=magenta]{hyperref}
\renewcommand\theequation{{\color{blue}\arabic{equation}}}
\newcommand{\beq}{\begin{equation}}
\newcommand{\eeq}{\end{equation}}
\newcommand{\bea}{\begin{eqnarray}}
\newcommand{\eea}{\end{eqnarray}}

\newcommand\non{\nonumber}

\newcommand\up{\uparrow}
\newcommand\dn{\downarrow}

\newcommand\smat{$\mathbb S$}

\newcommand\tstub{\mathcal T}

\newcommand\hmi{{\mathcal H}_{\textsf{int}}}

\newcommand\psiou{\Psi_{O\,\up}^{}}
\newcommand\psiiu{\Psi_{I\,\up}^{}}

\newcommand\psioudg{\Psi_{O\,\up}^\dagger}
\newcommand\psiiudg{\Psi_{I\,\up}^\dagger}

\newcommand\psiod{\Psi_{O\,\dn}^{}}
\newcommand\psiid{\Psi_{I\,\dn}^{}}

\newcommand\psioddg{\Psi_{O\,\dn}^\dagger}
\newcommand\psiiddg{\Psi_{I\,\dn}^\dagger}

\newcommand\epp{e^{i\,(k\,+\,k_F)\,x}}
\newcommand\epm{e^{i\,(-k\,+\,k_F)\,x}}

\newcommand\emp{e^{-i\,(k\,+\,k_F)\,x}}
\newcommand\emm{e^{-i\,(-k\,+\,k_F)\,x}}

\newcommand\ie{{\textsl{i.e.}}}

\newcommand{\etal}{{\it et al.}}

\newcommand\td{2{\textendash}D~}

\newcommand\odd{1{\textendash}D}

\newcommand\od{1{\textendash}D~}

\newcommand\afp{{\textsf{AFP~}}}

\newcommand\afpd{{\textsf{AFP}}}

\newcommand\cafp{{\textsf{CAFP~}}}

\newcommand\cafpd{{\textsf{CAFP}}}

\newcommand\rfp{{\textsf{RFP~}}}

\newcommand\rfpd{{\textsf{RFP}}}

\newcommand\tfp{{\textsf{TFP~}}}

\newcommand\tfpd{{\textsf{TFP}}}

\newcommand\sfp{{\textsf{SFP~}}}

\newcommand\sfpd{{\textsf{SFP}}}

\newcommand\agfpd{{\textsf{AGFP}}}

\newcommand\agfp{{\textsf{AGFP~}}}

\newcommand\cfp{{\textsf{CFP~}}}
\newcommand\cfpd{{\textsf{CFP}}}

\newcommand\gfpd{{\textsf{GFP}}}
\newcommand\gfp{{\textsf{GFP~}}}

\newcommand\lld{{\textsf{LL}}}
\newcommand\lln{{\textsf{LL~}}}

\newcommand\fld{{\textsf{FL}}}

\newcommand\ctd{{\textsf{CT}}}

\newcommand\ct{{\textsf{CT~}}}

\newcommand\wbsd{{\textsf{WBS}}}
\newcommand\sbsd{{\textsf{SBS}}}

\newcommand\wbs{{\textsf{WBS~}}}
\newcommand\sbs{{\textsf{SBS~}}}

\newcommand\wirgd{{\textsf{WIRG}}}
\newcommand\nsd{{\textsf{NS}}}
\newcommand\rgd{{\textsf{RG}}}
\newcommand\qwd{{\textsf{QW}}}

\newcommand\nsnd{{\textsf{NSN}}}
\newcommand\fsnd{{\textsf{FSN}}}
\newcommand\fsfd{{\textsf{FSF}}}
\newcommand\fnfd{{\textsf{FNF}}}
\newcommand\fsd{{\textsf{FS}}}
\newcommand\hfd{{\textsf{HF}}}

\newcommand\scd{{\textsf{SC}}}
\newcommand\tmrd{{\textsf{TMR}}}
\newcommand\card{{\textsf{CAR}}}
\newcommand\ard{{\textsf{AR}}}
\newcommand\ndcd{{\textsf{NDC}}}

\newcommand\ed{{\textsf{e}}}
\newcommand\hd{{\textsf{h}}}

\newcommand\wirg{{\textsf{WIRG~}}}
\newcommand\ns{{\textsf{NS~}}}
\newcommand\rg{{\textsf{RG~}}}
\newcommand\qw{{\textsf{QW~}}}

\newcommand\nsn{{\textsf{NSN~}}}
\newcommand\fsn{{\textsf{FSN~}}}
\newcommand\fsf{{\textsf{FSF~}}}

\newcommand\hf{{\textsf{HF~}}}

\renewcommand\sc{{\textsf{SC~}}}
\newcommand\scurrentd{{\textsf{SC}}} 
\newcommand\tmr{{\textsf{TMR~}}}
\newcommand\car{{\textsf{CAR~}}}
\newcommand\ar{{\textsf{AR~}}}
\newcommand\ndc{{\textsf{NDC~}}}

\newcommand\lbd{{\textsf{LB}}}
\newcommand\lb{{\textsf{LB~}}}

\newcommand\e{{\textsf{e~}}}

\newcommand\qwsd{{\textsf{QWs}}}
\newcommand\qws{{\textsf{QWs~}}}
\newcommand\ceod{{\textsf{CEO}}}

\def\ket#1{| ~#1~ \rangle}

\def\scxone#1{\langle ~#1~ \rangle}

\def\me#1#2#3{\langle ~#2\, |~#1~|\, #3~\rangle}

\def\And{{\rm and\ }}

\def\dfrac#1#2{{\displaystyle\frac{#1}{#2}}}

\newif\ifboo \boofalse

\begin{document}

\markboth{Arijit Saha}
{Electron-electron interaction effects on transport
through mesoscopic superconducting hubrid junctions}

%
\catchline{}{}{}{}{}
%

\title{Electron-electron interaction effects on transport\\ 
through mesoscopic superconducting hybrid junctions}

\author{ARIJIT SAHA
\footnote{arijit.saha@unibas.ch}}
\address{Department of Physics, University of Basel\\
Klingelbergstrasse 82, CH-4056 Basel, Switzerland}



\maketitle

\begin{history}
\received{Day Month Year}
\revised{Day Month Year}
\end{history}

\begin{abstract}
Effects due to the proximity of a superconductor has motivated a lot of research work in the last several decades 
both from theoretical and experimental point of view. In this review we are going to describe the 
physics of systems containing normal metal~-~superconductor interface. Mainly we discuss transport properties 
through such hybrid structures. In particular, we describe the effects of electron electron interaction on 
transport through such superconducting junction of multiple one-dimensional quantum wires. The latter can be described 
in terms of a non-Fermi liquid theory called Luttinger liquid. In this review, from the application point of view, we also 
demonstrate the possible scenarios for production of pure spin current and large tunnelling magnetoresistance in such hybrid 
junctions and analyze the influence of electron-electron interaction on the stability of the production of pure spin current.

\end{abstract}



\section{Introduction \label{sec:intro}}
In the past few decades, there has been an enormous amount of effort which has gone 
into designing one dimensional (\odd) quantum wires (\qwsd) in $GaAs-AlGaAs$ hetero-structure. 
Other than the observation of quantization of conductance, the expectation is to observe 
signatures of non Fermi liquid behavior in the transport and other optical properties has 
also put a lot of motivation in designing \od \qws experimentally. To engineer \od \qws a 
new crystal growth technique has been developed, which tightly confines the electron on three 
sides by smooth semiconductor hetero-junctions. Such quantum wires are called cleaved edge 
overgrowth (\ceod) quantum wire (\qwd)~\cite{1,2,3,4,5}. The electron mean free path in these 
\qws can be as long as $10 {\mu} m$. Also, experimentally measurable quantities in such \qwsd, 
are significantly affected by coulomb interactions between electrons inside them~\cite{1}. 
As long as one is interested in equilibrium phenomenon, the excitations contributing to any 
physical quantity are those excitations which are energetically close to the Fermi energy ($E_{F}$).
Particularly in \od, excitations are around the Fermi points, where the dispersion can be linearized 
and hence, a large variety of models at low energies in \od belong to a special class of non Fermi 
liquid called Luttinger liquids (\lld). The metallic state of this class of models is very different 
from that of usual Fermi liquid (\fld) theory. There is no fermionic quasi-particle, and
their elementary excitations are bosonic collective charge and spin fluctuations dispersing with different 
velocities. An incoming electron decays into such charge and spin excitations which then spatially
separate with time (charge-spin separation). The correlations between these excitations are anomalous and 
show up as interaction-dependent non-universal power-law behaviour in many physical quantities whereas 
those of ordinary metals are characterized by universal (interaction independent) powers. 
The above mentioned salient features of \lln have already been observed in cleaved edge overgrowth
\qwsd~\cite{1,2,3,4,5} and in carbon nanotubes~\cite{31}.

With the recent advancement in fabrication technology of semiconductor heterostructure 
it is now possible to study electronic transport in a variety of geometries of \qws 
among which the multi-point junctions is of special interest. Junction of several \qws 
has already been realized in Y-branched or multiple branched carbon nanotube~\cite{6}.
Recent studies of ballistic transport through a \qw have brought out the important role 
played by both scattering centers and the interactions between the electrons inside the 
\qwd. Theoretical studies using a real space renormalization group (\rgd) analysis show that 
repulsive interactions between electrons inside the \qw tend to increase the effective strength 
of the impurity as one goes to longer and longer length scales~\cite{7}; experimentally, this 
leads to a decrease in the conductance as the temperature is reduced or the wire
length is increased~\cite{1,8,9,10,11}. Considerable effort has also gone into understanding the 
effects of Fermi liquid leads~\cite{12,13,14}, multiple impurities~\cite{15,16,17}
and also contacts~\cite{18} in two terminal measurements. Then, significant efforts have also 
gone into understanding the next logical step which is the effect of inter electrons interaction 
on the conductance of more complicated geometrical structures such as three or more \qws meeting
at a junction. For the free electron case, the transport across the junction of multiple wires 
can be well understood under the framework of multi-terminal Landauer-Buttiker (\lbd) scattering 
theory. Although it is not straight forward to analyse transport through such junctions if
electron-electron ($\ed$-$\ed$) interactions are also present inside the \qw as \lb formalism does not
hold in this situation. This problem has been studied to some extent before in Ref.~\refcite{19} and
more recently in Ref.~\refcite{20}. To study this problem they use the technique of bosonization~\cite{21,22,23,24,25}
which can be used only for the weak and strong backscattering limits. On the other hand the problem 
of transport through a junction (characterized by an arbitrary scattering matrix \smat) of \od wires 
can be solved for any backscattering and in the weak $\ed$-$\ed$ interaction limit, by using a \rg 
technique introduced in Ref.~\refcite{26}. The main advantage of using this technique is the fact 
that one can access the intermediate fixed point which corresponds to scattering amplitudes with 
intermediate values~\cite{27,28}. Very recently, this \rg method has also been generalized for \qws 
with arbitary $\ed$-$\ed$ interactions~\cite{29,30}.

\begin{figure}
\begin{center}
\includegraphics[width=13.0cm,height=6.0cm]{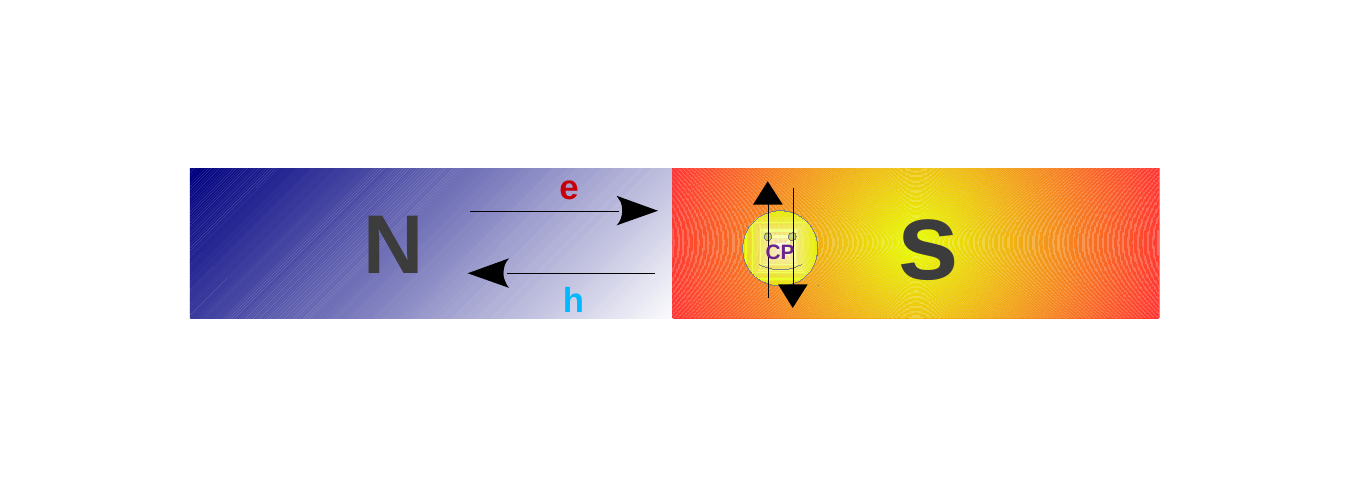}
\caption{(Color online) Cartoon of \ar process in which an incident electron is reflected back from the 
\ns interface as a hole and a Cooper pair transmits into the superconductor.}
\label{figone}
\end{center}
\end{figure}

On the other hand, effects due to the proximity of a superconductor has motivated a lot
of research work~\cite{32,33,34,35} in the last several decades. A direct manifestation of 
proximity effect is the phenomenon of Andreev reflection (\ard) which was predicted by Andreev 
in 1964~\cite{32}. In the case of a sufficientely clean normal$-$superconductor (\nsd) interface 
and a large superconducting gap the main contribution to transport comes from \ar in which an incident 
electron, below the superconducting gap $\Delta$ is reflected back as a hole from the interface and a 
Cooper pair (charge of 2e) jump into the superconductor (see also Fig.~\ref{figone}). In the \ar processes 
the pairing amplitude of the superconductor is induced in the normal-metal side, while the attractive 
interaction potential between electrons is identical to zero in the normal conductor. An even more intriguing 
example where the proximity effect manifests itself is the phenomenon of crossed Andreev reflection (\card) 
which can only take place in a normal metal$-$superconductor$-$normal metal (\nsnd) junction provided 
the distance between the two normal metals is less than or equal to the phase coherence length of the 
superconductor. This is a nonlocal process where an incident electron from one of the normal metal leads 
pairs up with another electron from the other lead to form a Cooper pair~\cite{40,41} and jumps into the 
superconductor. The signature of this nonlocal process also has been verified by a number of recent 
experiments involving \nsn junction~\cite{36,37,38,93} and also carbon nanotubes~\cite{39}. Also the 
relevance of \car regarding production of entangled electron pairs in nano devices for quantum computation
has attracted a lot of attention in recent times~\cite{42,43,44,45,46,47,48,49,50,51,106,107}. Non-local entangled
electron pairs involving \nsn junction can be realized in a Cooper pair beam splitter geometry~\cite{53,54,55,56,94}
in which such pairs are produced via \car processes when the superconductor is biased with respect to the normal
leads comprising the junction. Very recently, conclusive evidence of extremely high efficiency Cooper pair 
splitting via observing positive two-particle cross correlations of the shot noise of the split electrons via
\car process in the beam splitter geometry, has been put forwarded both theoretically~\cite{52} and 
experimentally~\cite{57,114}. 

In recent times, a lot of attention has also gone into understanding the effects of $\ed$-$\ed$ interactions
on \ar processes in case of \ns junctions in the context of \od quantum wires~\cite{58,59,60,61,64}, 
quantum dots~\cite{95,96,97,98}, carbon nanotubes~\cite{62} and proximity effects in \lld~\cite{63}. 
The power law dependence of the Andreev conductance for the \ns junction case was first obtained using weak 
interaction renormalization group (\wirgd) approach by Takane and Koyama in Ref.~\refcite{58}. This was in agreement 
with earlier results from bosonisation~\cite{59}, which, however, could only handle perturbative analysis around the 
strong back-scattering (\sbsd) and weak back-scattering (\wbsd) limits. The \wirg approach, on the other hand,  
can study the full cross-over from \wbs limit to \sbs limit. Hence the \wirg approach is very well-suited 
for studying problems where the aim is to look for non trivial fixed points with intermediate transmissions 
and reflections. The latter would be difficult using a bosonisation approach. For the \ns junction case, it was 
also shown that the power law exponent for the temperature dependence of conductance was twice as large as the 
exponent for a single barrier in a \qwd. This happens due to the existence of \ar process in which both electron 
and hole channels take part in transport. 

The organization of the rest of the review is as follows. In Sec.~\ref{sec:2} we give a generic description
of the model for junction of \od \qws which are described by the \lln theory. In particular in Sec.~\ref{sec:2.2} 
we describe the \wirg method applied to model the $\ed$-$\ed$ interactions in \od \qws and a comparative analysis
between \wirg method and Bosonization is presented in Sec.~\ref{sec:2.3}. In Sec.~\ref{sec:3.1} we analyse the
superconducting junction of many ($N \ge 2$) \od \qws by the \wirg procedure and in (Sec.~(\ref{3.1.1}-\ref{3.1.4}))
we present the \rg fixed point analysis of some special geometry. In Sec.~\ref{sec:3.2} we discuss the results of 
\rg flows for the conductance in different geometries comprising of superconducting junction. Our study reveals
the striking fact that due to the inter-play of the proximity and the interaction effects, one gets a novel 
non-monotonic behavior (non \lln behavior) of conductance for the case of \nsn junction as a function of the temperature. 
In Sec.~\ref{sec:4} we study the stability analysis of the \rg flow for the \nsn junction of \lln wires. In particular,
we compute the power laws associated with the \rg flow around the various fixed points of this system and also obtain 
the power law dependence of linear response conductance on voltage bias or temperature around them. In Sec.~\ref{sec:5}
we discuss the possible applications of different superconducting junctions of \od \qws and in particular we show
that one can have pure spin current (\scd) and large tunneling magnetoresistance (\tmrd) in such geometries. 
We also analyze the influence of $\ed$-$\ed$ interaction and see how it stabilizes or de-stabilizes the production of 
pure \scd. Finally in Sec.~\ref{sec:6}, we present our summary and possible outlooks of the topic.

\section{General model for junction \label{sec:2}}
\begin{figure}
\begin{center}
\includegraphics[width=13.0cm,height=6.0cm]{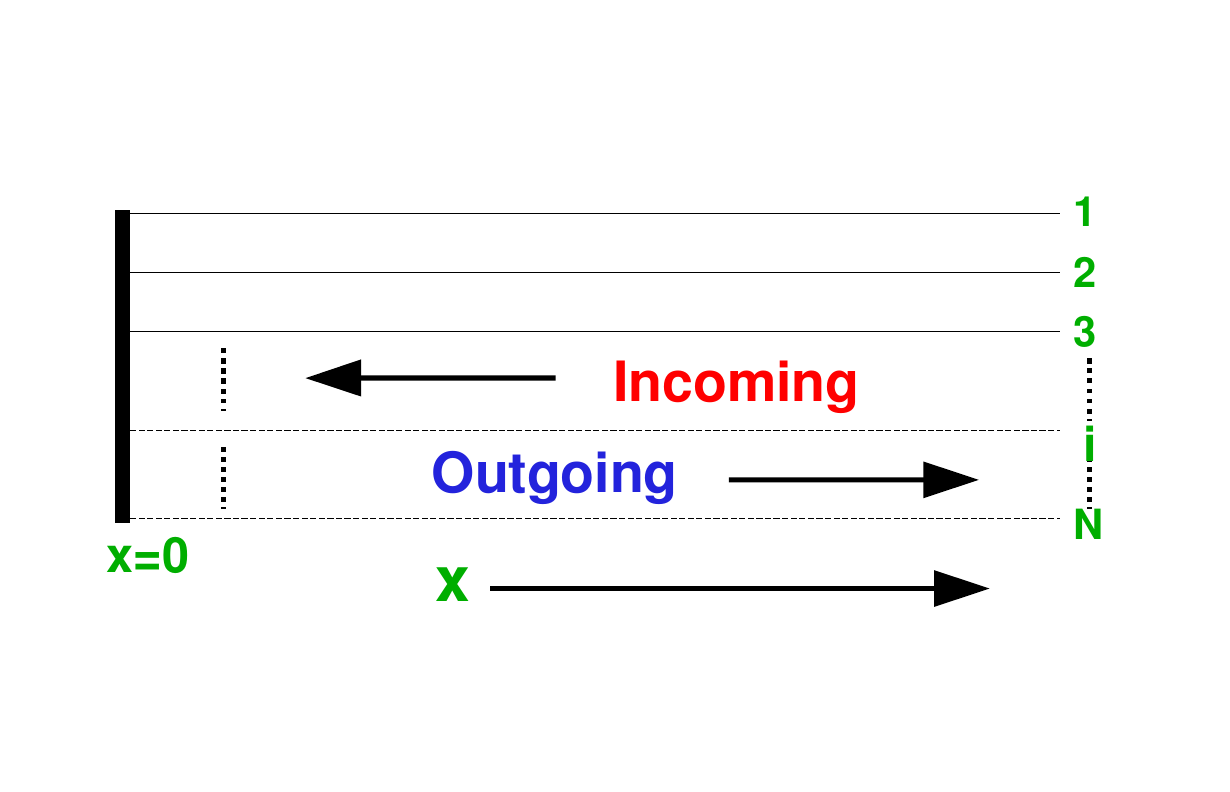}
\caption{(Color online) Cartoon of $N$ \qws meeting at a junction at $x=0$. Here the figure also 
illustrates the incoming and outgoing wave directions to and from the junction. The junction is shown 
as an extended line for better visibility but actually it's a point.}
\label{figtwo}
\end{center}
\end{figure}

The model for the junction essentially comprises of $N$ semi-infinite \qws meeting at a point. 
The \qws are parameterized by the coordinates $x_i$, $i=1,2,...,N$. The junction is the point 
where the $x_i$ are simultaneously equal to $0$. The convention is, each $x_i$ increases from 
$0$ as one goes outwards from the junction along wire $i$. The incoming and outgoing single particle 
wave functions on wire $i$ are denoted by $\psi_{Ii} (x_i)$ and $\psi_{Oi} (x_i)$ respectively
(ignoring the spin label $\sigma$ for the time being); see Fig.~\ref{figtwo}. For a given wave number 
$k>0$, these wave functions are proportional to the plane waves $\exp (-ikx_i)$ and $\exp (ikx_i)$. 

The coefficients of the plane waves are related to each other by a $N \times N$ scattering matrix 
denoted by \smat. Denoting the incoming and outgoing wave functions at the junction by the columns 
$\psi_I (0)$ and $\psi_O (0)$ respectively, we have the relation
\bea
&&\psi_O (0) ~=~\mathbb S~\psi_I (0) ~,
\eea
\smat~must be unitary for current conservation and in the absence of a magnetic field, the \smat~matrix 
must be symmetric also so that it respects time reversal symmetry. The diagonal entries of \smat~are the
reflection amplitudes $r_{ii}$, while the off-diagonal entries are the transmission amplitudes $t_{ij}$ 
to go from wire $j$ to wire $i$.

\subsection{\label{sec:2.1} Model for electron~-~electron interaction}
We start with the description of a single \qw for the moment, so that the label $i$ can be dropped.
At low temperatures or at low-energy, excitations are dominated by modes near the Fermi points $\pm k_F$ 
in \od \qwd. Around the $\pm k_F$ the second-quantized Fermi field $\Psi (x)$ corresponding to these
modes can be expanded in terms of the scattering basis as given below 
\bea
\Psi (x) ~=~ \Psi_I (x) ~e^{-ik_F x} ~+~ \Psi_O (x) ~e^{ik_F x} ~,
\label{psiio}
\eea
It is important to note that, the fields $\Psi_I$ and $\Psi_O$ defined in Eq.~\ref{psiio} vary slowly 
on the scale of the inverse Fermi momentum $k_F^{-1}$, since we have separated out the rapidly varying 
functions $\exp (\pm i k_F x)$. Henceforth we use the notation $\Psi_I$ and $\Psi_O$ for these slowly 
varying second-quantized fields, rather than the incoming and outgoing fields defined earlier. For such 
fields, we will only be interested in Fourier components with momenta $k$ which satisfy $|k| << k_F$. 
Hence we can linearize the dispersion relations such that the spectrum takes the form, $E= \pm \hbar v_F k$ 
for the fields $\Psi_O$ and $\Psi_I$ respectively, where $v_F$ is the Fermi velocity. Also it is practical 
to assume that the elements of \smat-matrix are energy independent \ie~independent of $k$ for small $k$.

The $\ed$-$\ed$ interaction part of the hamiltonian can be written as 
\bea
H_{\rm int} ~=~ \frac{1}{2} ~\int \int dx dy ~\rho (x) ~V (x-y) ~\rho (y),
\label{11}
\eea
where $V(x)$ is a real quantity and function of $x$. \\
The density $\rho$ is given in terms of the fermion fields as $\rho (x) =
\Psi^\dagger (x) \Psi (x)$.
Using Eq.~\ref{psiio}, we have 
\bea
\rho (x) ~&=&~ \Psi_I^\dagger \Psi_I ~+~ \Psi_O^\dagger \Psi_O \non \\
& & ~+~ \Psi_I^\dagger \Psi_O ~e^{i2k_F x} ~+~ \Psi_O^\dagger \Psi_I ~
e^{-i2k_F x} ~.
\label{dens}
\eea
Again let us assume a highly screened short-ranged interaction for $V(x-y)$ in Eq.~\ref{11}
such that the arguments $x$ and $y$ of the two density fields can be set equal to each other 
wherever possible. In doing so, we neglect terms with scaling dimension greater than 2, and 
are therefore irrelevant under \rg flow. Using the anti-commutation relations between
different fermion fields, we obtain
\bea
H_{\rm int} ~=~ g_2 ~\int dx ~\Psi_I^\dagger \Psi_I \Psi_O^\dagger \Psi_O~,
\label{hint2}
\eea
where $g_2$ is related to the Fourier transform of $V(x)$ as $g_2 = {\tilde V}(0)- {\tilde V} (2k_F)$ 
for the spinless electrons. It is interesting to note that $g_2$ is zero if $V(x)$ is a $\delta$-function 
potential as in that case ${\tilde V}(0)={\tilde V} (2k_F)$ for the spinless case. This argument reveals the fact 
that an ultra-short range interaction like the $\delta$-function potential is not useful. Thus the interaction 
may be short-ranged, but must have some finite range. Different \qws may have different values of the 
interaction parameter $g_2$ represented by $g_{2i}$. For later use, we define the dimensionless constants as
\bea
\alpha_i ~=~ \frac{g_{2i}}{2\pi \hbar v_F} ~,
\label{55}
\eea
where we assume that the Fermi velocity $v_F$ is the same on all wires. For \od interacting electrons 
(\lln theory) the most efficient and popular approach to analyze transport behaviour is to implement 
bosonization formalism~\cite{21,22,23,24,25}. For spinless fermions, the bosonic theory is characterized 
by two parameters, $v$, the renormalized Fermi velocity and a dimensionless parameter $K$ which
parameterizes the strength of inter electron interactions inside the \qwsd. Typically, $K$ governs 
the exponents which appear in the power-law fall-offs of various correlation functions in the theory. 
For a model defined on the entire real line with the interaction parameter $g_2$ or $\alpha$ defined above, 
we find that~\cite{25}
\bea
K ~&=&~ \bigl( \frac{1-\alpha}{1+\alpha} \bigr)^{1/2} ~.
\eea 
Thus $K=1$ corresponds to the noninteracting fermions (Fermi liquid theory), while $K<1$ and
$K>1$ to short-ranged repulsive and attractive interactions respectively. For weak interactions, 
we see that $v=v_F$ while $K = 1 -\alpha$ to first order in $\alpha$. In this review, we will be 
interested in the case in which the $\ed$-$\ed$ interaction is weak and repulsive, \ie, the parameters
$\alpha_i$ are all positive and small.
\subsection{\label{sec:2.2} Weak interaction renormalization group approach}
In this sub section we introduce the weak interaction renormalization group (\wirgd) method
which was first developed by Yue \etal~\cite{26} for the case of junction of two \qws and 
then extended by Lal \etal~\cite{27} to the case of junction of multiple \qwsd. In comparison 
to Bosonization, this method is instructive and  physically transparent. Using this method,
\rg equations for all the \smat-matrix elements can be evaluated to leading order in $\alpha$.
The basic idea behind this method is the following

In the presence of a non-zero reflection amplitude $r_{ii}$, the density of noninteracting fermions 
in wire $i$ gives rise to Friedel oscillations with wavenumber $2k_F$. In presence of weak $\ed$-$\ed$
interaction inside the \qwd, an electron scatters not only from the junction but also from
these Friedel oscillations by an amount proportional to the parameter $\alpha_i$. Yue \etal~use 
this idea to derive the \rg equations for an arbitrary \smat-matrix located at the junction of 
two semi-infinite \qwsd. In the limits of both weak back scattering ($r_{11} \rightarrow 0$) and 
strong back scattering ($|r_{11}| \rightarrow 1$), their results reduce to those known from
bosonization~\cite{7,21,22}.

Here we first derive the form of the density oscillations in one particular \qw given that there is a 
reflection coefficient $r$ for incoming electron waves along that wire. If the momentum of the incident 
electron is in the vicinity of Fermi momentum $k_F$, we can write the wave function in the form
\bea
\psi_k (x) ~=~ e^{-i(k+k_F)x} ~+~ r ~e^{i(k+k_F)x} ~,
\eea
where $|k| << k_F$. In the ground state of the noninteracting system, the density is given by
\bea
< \rho (x) > ~=~ \int_{-\infty}^0 ~\frac{dk}{2\pi} ~\psi_k^\star (x) \psi_k (x) ~,
\label{do}
\eea
where we use the fact that only states with energy less than $E_F$ (\ie, momenta less than $k_F$) 
are occupied, and we extend the lower limit of Eq.~\ref{do} to $-\infty$ for convenience. 
(Alternatively, we can impose a cut-off at the lower limit of the form $\exp (\epsilon k)$, 
and take the limit $\epsilon \rightarrow 0$ at the end of the calculation). We then find that 
$\rho$ has a constant piece $\rho_0$ (which can be eliminated by normal ordering the density operator), 
and a $x$-dependent piece given by
\bea
< \rho (x) > ~-~ \rho_0 ~=~ ~\frac{i}{4\pi x} ~(~ r^\star ~e^{-i2k_F x} -
r~e^{i2k_F x} ~) ~.
\label{fried}
\eea
Using the expression in Eq.~\ref{dens}, we find that the expectation value 
$<\Psi_I^\dagger \Psi_I + \Psi_O^\dagger \Psi_O>$ is a constant, while
\bea
<\Psi_O^\dagger \Psi_I> ~&=&~ ~\frac{ir^\star}{4\pi x} ~, \non \\
<\Psi_I^\dagger \Psi_O> ~&=&~ - ~\frac{ir}{4\pi x} ~.
\label{expec}
\eea
Note that there is also a contribution to $\rho (x)$ from the waves transmitted from 
the other wires; however those are independent of $x$ and can be absorbed in the constant 
piece $\rho_0$. Hence, the Friedel oscillations given by Eq.~\ref{fried} in a given \qw 
arise only from reflections within that wire. Also here the energy dependence of the
\smat-matrix elements has been neglected as only excitations $|E| << E_F$ has been taken
into account. Later \wirg with the explicit energy dependence of the \smat-matrix elements 
in case of resonant tunneling in \lln wires has been analysed in Ref.~[\refcite{28,88}].   

Next we derive the amplitude of scattering of the fermions from the Friedel oscillations, 
using a Hartree-Fock decomposition of the interaction in Eq.~\ref{hint2}. The reflection 
is caused by the following terms in the decomposition
\bea
H_{\rm int} ~&=&~ - g_2 ~\int_0^\infty ~dx ~(~ <\Psi_O^\dagger \Psi_I> 
\Psi_I^\dagger \Psi_O 
~+~ <\Psi_I^\dagger \Psi_O> \Psi_O^\dagger \Psi_I
~)~,
\non \\
&=&~- \frac{ig_2}{4\pi} ~\int_0^\infty ~\frac{dx}{x} ~( r^\star ~
\Psi_I^\dagger \Psi_O ~-~ r ~\Psi_O^\dagger \Psi_I )~,
\label{hf1}
\eea
where we have used Eq.~\ref{expec} to write the second line in Eq.~\ref{hf1}. 
Now we derive the amplitude to go from a given incoming wave with momentum $k$
to an outgoing wave (or vice versa) under the action of $\exp (-iH_{\rm int} t)$. 
The amplitude can be written as
\bea
& & - i ~\int \frac{dk^\prime}{2\pi} ~2\pi \delta (E_k - E_{k^\prime}) ~
|{\rm outgoing}, k^\prime > \non \\
& & ~~~~~~~~~~~~~~~~~~~~~~~~ \times <{\rm outgoing}, k^\prime | ~H_{\rm 
int} ~| {\rm incoming}, k> \non \\
& &~ =~ |{\rm outgoing}, k> ~\frac{ig_2 r}{4\pi \hbar v_F} ~\int_0^\infty ~
\frac{dx}{x}~ e^{-i2k x} ~, 
\label{trans}
\eea
where we have used Eq.~\ref{hf1}, the dispersion relation $E_k = \hbar v_F k$ 
(such that $\delta (E_k - E_{k^\prime}) = (1/\hbar v_F ) \delta (k - k^\prime)$), 
and the wave functions $\exp (\pm ikx)$ of the outgoing and incoming waves respectively 
to arrive at Eq.~\ref{trans}. The integral over $x$ in Eq.~\ref{trans} is divergent at 
the lower end; we therefore introduce a short-distance cut-off $d$ which is the inter 
particle spacing here. The amplitude in Eq.~\ref{trans} then reduces to 
\bea
- \frac{\alpha r}{2} ~{\rm ln} (kd) \ .
\label{inout}
\eea
plus contributions which remain finite as $kd \rightarrow 0$; we have used Eq.~\ref{55} here. 
Similarly, the amplitude to go from an outgoing wave to an incoming wave due to the scattering 
from Friedel oscillations is given by
\bea
\frac{\alpha r^\star}{2} ~{\rm ln} (kd) ~.
\label{outin}
\eea
\begin{figure}
\begin{center}
\includegraphics[width=10.0cm,height=6.0cm]{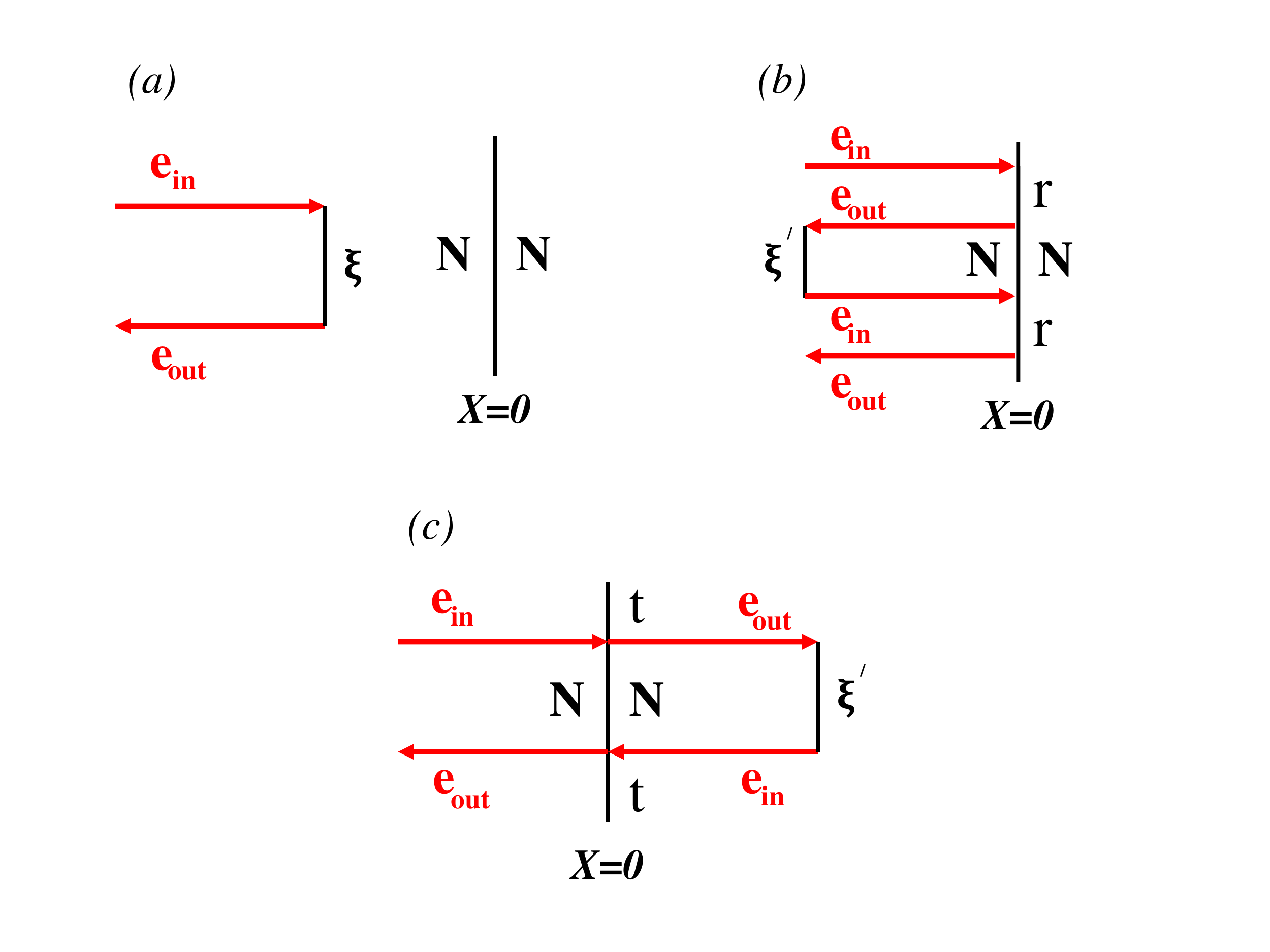}
\caption{(Color online) The processes that contribute to the amplitude for an incoming electron to
transform to an outgoing electron. Note that all the processes shown here are to first order
in the interaction parameters since they only involve a single scattering from a Friedel oscillation.
Process (a) involves scattering from the Friedel oscillation before the electron reaches the junction.
Process (b) involves two reflections from the junction along with a scattering from the Friedel oscillation 
in the same wire. On the other hand process (c) involves two transmissions through the junction and a 
scattering from the Friedel oscillation in the other wire. In the diagrams, $\xi=-\frac{\alpha r}{2} ~{\rm ln} (kd)$
and $\xi^{\prime}=\frac{\alpha r^\star}{2} ~{\rm ln} (kd)$.}
\label{figthree}
\end{center}
\end{figure}
These reflections from the Friedel oscillations can then be combined along 
with the bare \smat-matrix at the junction to calculate the corrections to it. 
For instance, let us consider $r_{ii}$ at first. To first order in the 
interaction parameters $\alpha_i$, this amplitude gets corrections from the 
following processes. An incoming electron wave on wire $i$ can either 
(i) turn into an outgoing electron on the same wire with the
amplitude given in Eq.~\ref{inout} (with $r$ replaced by $r_{ii}$ in that
expression), or (ii) get reflected from the junction with amplitude
$r_{ii}$ thereby turning into an outgoing wave, turn back into an incoming
wave after scattered back from the Friedel oscillations according to
Eq.~\ref{outin}, then get reflected again from the junction, or (iii)
transmit through the junction into wire $j$ (with $j \ne i$) with amplitude
$t_{ji}$, turn from an outgoing wave to an incoming wave on wire $j$
according to Eq.~\ref{outin} (with $r$ replaced by $r_{jj}$), then transmit
back through the junction to wire $i$ with amplitude $t_{ij}$. The above 
mentioned three processes (see Fig.~\ref{figthree}) effectively give correction 
to the bare $r_{ii}$ and the renormalized $r_{ii}$ therefore becomes
\bea
dr_{ii} ~&=&~ - ~A_{ii} {\rm ln} (kd) ~, \non \\
{\rm where} \quad A_{ii} ~&=&~ -~ \frac{1}{2} ~[~ - ~\alpha_i r_{ii} ~
+~ \alpha_i |r_{ii}|^2 r_{ii} \non \\
& & ~~~~~~~~~~~ +~ \sum_{j \ne i} ~\alpha_j t_{ij} r_{jj}^\star t_{ji} ~]~.
\label{aii}
\eea
Similarly, we calculate the correction to the transmission $t_{ji}$
from wire $i$ to wire $j$ and it is given by
\bea
dt_{ji} ~&=&~ - ~A_{ji} {\rm ln} (kd) ~, \non \\
{\rm where} \quad A_{ji} ~&=&~ - ~\frac{1}{2} ~[~ \alpha_i t_{ji}
|r_{ii}|^2 ~
+~ \alpha_j |r_{jj}|^2 t_{ji} \non \\
& & ~~~~~~~~~~ +~ \sum_{k \ne i,j} ~\alpha_k t_{jk} r_{kk}^\star t_{ki}
~]~.
\label{aji}
\eea

Hence, one can derive the \rg equations for the \smat-matrix which is considered to
be a function of a length scale $L$; by replacing $-{\rm ln} (kd)$ in Eqs.~(\ref{aii}-\ref{aji}) 
by $dl$, where $l = {\rm ln} (L/d)$. The \rg equations therefore take the form
\bea
\frac{dr_{ii}}{dl} ~&=&~ A_{ii} ~, \non \\
\frac{dt_{ij}}{dl} ~&=&~ A_{ij} ~,
\label{rg1}
\eea
where $A_{ii}$ and $A_{ij}$ are given by Eq.~\ref{aii} and Eq.~\ref{aji} respectively. 
We can also write Eqs.~\ref{rg1} in a simpler way. Given the matrix \smat~and the 
parameters $\alpha_i$ (which do not flow under \rg for the spinless case), we define 
a diagonal matrix $F$ whose entries are 
\bea
F_{ii} ~=~ - ~\frac{1}{2} ~\alpha_i r_{ii} ~.
\label{fmat}
\eea
Then the \rg equations for the \smat-matrix can be written in the matrix form as
\bea
\frac{d\mathbb S}{dl} ~=~ {\mathbb S} F^\dagger {\mathbb S} ~-~ F ~.
\label{77}
\eea
Note that the derivation of the \rg equation for the \smat-matrix presented above is 
correct only when the $\ed$-$\ed$ interaction strength inside the \qw is 
perturbative~\ie~first order in $\alpha$. A non-perturbatibe (arbitary strength of $\alpha$) 
\rg approach has been developed recently by~\cite{29,30} to study transport of interacting 
electrons through a potential barrier. In their work they calculate the linear response conductance 
of electrons in a \lln with arbitary $\ed$-$\ed$ interaction strength. Their result also agrees 
well with the known limiting cases of \wirgd~\cite{26,27}.

\subsection{\label{sec:2.3} \wirg vis-a-vis Bosonisation for junction}
In this subsection we present a comparative discussion between Bosonization and \wirg
for analyzing transport through a quantum scatterer (for instance, a simple
static barrier or a dynamical magnetic impurity like Kondo spin) in a \od \qwd.
The latter is qualitatively different from its higher dimensional counterpart~\cite{7}
as in \odd, due to $\ed$-$\ed$ interactions, the Fermi-liquid ground state is destroyed 
and the electrons form a non-Fermi liquid ground state known as {\textsl{Luttinger Liquid}}~\cite{66}.
The low energy dynamics of such \od system is governed mainly by coherent particle-hole excitations 
around the left and the right Fermi points. Hence, it is natural to use bosonic fields to describe 
these low lying excitations. The later can be implemented by re-expressing the original fermions 
using boson coherent state representation~\cite{21,22,24,25} which is referred to as {\textsl{bosonisation}}.
Although the bosonization approach only allows for a perturbative analysis for transport around the 
limiting cases of \sbs and \wbs for the quantum impurity problem. On the other hand, if we start
with a weakly interacting electron gas, it is possible to do a perturbative analysis in the $\ed$-$\ed$  
interaction strength around the free fermion Hamiltonian, but treating the strength of the quantum 
impurity {\textsl{exactly}}. This approach allows us to study transport through the impurity for 
any scattering strength. The advantage of this approach lies in the fact that even in presence of 
$\ed$-$\ed$ interaction, one can use single particle pictures such as the transmission and reflection
amplitudes (\smat-matrix) in order to characterize the impurity. Then the idea is to calculate 
correction to the transmission and reflection amplitudes perturbatively in the $\ed$-$\ed$ interaction strength.

Now in the next step, since we are working in \odd, the perturbative correction turns out to be logarithmically 
divergent. To obtain a finite result, one has to sum up all such divergent contributions to the transmission and 
reflection amplitudes to all relevant orders at a given energy scale. This was first done by Matveev {\textsl{\etal~}} 
in Refs.~\refcite{67} and \refcite{26} in the context of a single (scalar) scatterer for both spin-less and spin-full 
electrons using the ``poor man's scaling'' approach~\cite{69}. For the spin-less case, it was shown that the 
logarithmic correction to the bare transmission probability (to first order in interaction parameter parameterized 
by $\alpha$) was ${{\delta T = 2\,\alpha \,T_0\, \left(1\,-\,T_0\right) \,
{\ln}(kd) }} $ and the explicit \rg equation for transmission probability
was $dT/dl=-2 \alpha T (1-T)$
where $k$ is the momentum of the fermion measured from $k_F$, $d$ is a short distance cut-off and 
$\alpha$ is the $\ed$-$\ed$ interaction parameter given by
%
$\alpha =\alpha_1 - \alpha_2$ with $\alpha_1={{V(0)}/{2\pi \hbar
v_F}}$ and $\alpha_2={V(2k_F)} /{2\pi \hbar v_F} $.
%

The \rg equation upon integration gives rise to final answear for the transmission probability which can be 
written as,
\bea T(L) = \dfrac{T_0  e^{-2 \alpha l}}{\left[{1 - T_0 + T_0
e^{-2\alpha l}}\right]}
   = \dfrac{T_0  \left(\frac{d}{L}\right)^{2\alpha}}
   {\left[1 - T_0 + T_0 \left(\frac{d}{L}\right)^{2\alpha}\right]},
\label{eq:one} 
\eea
where, $l = -\ln(kd) = \ln({L/d})$ and $L$ is the length of the \qwd. Also $l$ can also be measured 
as a function of the temperature by introducing the thermal length, $L_T = ({\hbar v_F})/({k_B T})$.
In Eq.~\ref{eq:one} $T_0$ is the bare transmission probability at the short distance cut-off, $d$.
It is easy to see from Eq.~\ref{eq:one} that for very small values of $T_0$, the latter can be neglected 
in the denominator of the expression for $T(L)$ leading to a pure power law scaling behavior 
consistent with the power law known from bosonisation in the \wbs limit. Similarly for the spin-full 
electrons, it can be shown that the parameter $\alpha$ in the power law gets replaced by a new parameter, 
$\beta$ given by $\beta = {(g_2-2g_1)}/{\pi \hbar v_F}$ where $g_2 = g_2(k)$ and $g_1 = g_1(k)$ are 
momentum dependent functions or `running coupling constants' which is in sharp contrast to the spin less case.
At high momentum, (or equivalently, at the short distance cut-off scale),
$g_1(d) = V(2k_F)$ and $g_2(d) = V(0)$. Due to the presence of the extra logarithmic dependence originating 
from scaling of the interaction parameter itself (see Eq.~\ref{intrg1} and Eq.~\ref{intrg2} below), the expression 
for transmission probability~\cite{26,67}, no longer follows a pure power law scaling even for small values of $T_0$. 
Instead $T(L)$ is now given by
\bea T(L) = \dfrac{\left[T_0\left[1+\alpha_1 \ln
\left(\frac{L}{d}\right)\right]^\frac{3}{2}
\left(\frac{d}{L}\right)^{(2\alpha_2-\alpha_1)}\right]}{\left[1 -
T_0 + T_0 \left[1 + 2\alpha_1 \ln
\left(\frac{L}{d}\right)\right]^\frac{3}{2}
\left(\frac{d}{L}\right)^{(2\alpha_2-\alpha_1)} \right]},
\label{spcondlog}
\eea
using the length scale dependence of $g_1(L)$ and $g_2(L)$ given by~\cite{68} 
\bea g_1(L) &=& \dfrac{V(2k_F)}{\left[1 +
\frac{V(2k_F)}{{\pi v_{F}}} \ln (\frac{L}{d})\right]},
\label{intrg1}
\\
g_2(L) &=& V(0) - \frac{1}{2}\, V(2k_F) +
\frac{1}{2}\,\dfrac{V(2k_F)}{\left[1\,+\,\frac{{V(2k_F)}}{{\pi
v_{F}}}\ln(\frac{L}{d})\right]}. 
\label{intrg2} 
\eea

Note that in the absence of $\ed$-$\ed$ interaction induced back-scattering (\ie,~when $V(2k_F)=0$), 
there is no correction to the power law behavior. Hence, bosonisation, which ignores $\ed$-$\ed$
back-scattering always results in power law behaviour. Although, when $\ed$-$\ed$ interaction induced
back-scattering is included, the sign of $g_2-2g_1$ can change under \rg flow, and hence, there can be 
a qualitative change in the behavior of the conductance. The latter actually develops a non-monotonic
dependence on the temperature; it first grows and then drops to zero as one approaches very low temperature.
Although, except for this non-monotonic behavior of conductance for the spin-full case, there is no new 
physics which is achieved by studying the full crossover from \wbs to \sbsd. In conclusion, both bosonisation 
and \wirg methods predict that for the single impurity (scalar) problem there are only two fixed points$-$ 
{\textsl{(a)}} the perfectly back-scattering (no transmission) case is the stable fixed point and {\textsl{(b)}} 
the absence of back-scattering (perfect transmission) case is the unstable fixed point. There are no other 
fixed points with intermediate reflection and transmission.

It is first shown by Lal {\textsl{\etal}} in Ref.~\refcite{27}, using the \wirg approach that even though there 
are only two fixed points for the two$-$wire$-$junction, surprisingly enough, the three$-$wire$-$junction has 
a host of fixed points, some of which are isolated fixed points while others are one parameter or multi
parameter families of fixed points. It is also shown to be true for more than three wires. From this point of view, 
the physics of a two$-$wire$-$junction is different from its three$-$wire$-$counterpart. Also in Ref.~\refcite{27},
it is shown that for a multiple$-$wire$-$junction, the \rg equations for the full \smat-matrix characterizing the
junction take a very convenient matrix form shown in Eq.~\ref{77}. The advantage of writing the \rg equation in 
this way is that it immediately facilitates the hunt for various fixed points. All one needs to do is to set the 
matrix on the LHS of Eq.~\ref{77} to zero. The latter scheme formally provides us with all the fixed points 
associated with a given \smat-matrix. The three$-$wire$-$junction was also studied using bosonisation and
conformal field theory methods in the context of fixed point structure and conductance around 
them~\cite{19,20,78,79,80,81,82,91}, tunneling density of states~\cite{89} and power dissipation~\cite{90}.
The latter approaches confirmed some of the fixed points found using \wirgd. It also gave some extra fixed 
points which were related to charge fractionalisation at the junction, and which could not be seen 
within the \wirg approach. Very recenly, tunneling density of states~\cite{83} and transport properties of 
three$-$wire$-$junction comprising of normal or chiral \lln wires have been studied using the \wirg method 
with arbitary repulsive $\ed$-$\ed$ interaction strength and different interaction strength in each 
\qwd~\cite{84,85,86,87}. In their work they predict a new M (mystery) fixed point at which the conductance
takes an intermediate value and also this M fixed point becomes stable for attractive $\ed$-$\ed$ interaction.
Their results also match with the bosonization treatment of three$-$wire geometry~\cite{20} and Ref.~\refcite{27}
in the weak interaction limit.

The \wirg method was further extended to more complicated geometries made out of junctions of 
\qw which can host resonances and anti-resonances in Ref.~\refcite{71}. The scaling of the resonances 
and anti-resonances were studied for various geometries which included the ring and the stub geometry.
This approach was further extended in Refs.~[\refcite{72,73}] to study the multiple$-$wire$-$junction with a 
dynamical scatterer, \ie~ a (Kondo) spin degree of freedom. The coupled \rg equations involving the Kondo 
couplings, $J_{ij}$ as well as the \smat-matrices were solved. For different starting scalar \smat-matrices, 
the \rg flows of the Kondo couplings was studied. The temperature dependence of the conductances was shown 
to have an interesting interplay of the Kondo power laws as well as the interaction dependent power laws. 

Almost a decade ago, the \wirg method was also extended to the case of \ns junction~\cite{58,59}. 
In the vicinity of the superconductor, it is well-known that the system is described by holes as well 
as electrons~\cite{34}. Hence the \smat-matrix characterizing the junction not only includes the
electron channel but also the hole channel. Hence, both electron and hole channels take part in transport.
Naively, one might expect that in the presence of particle-hole symmetry, the only effect of including 
the hole channel would be to multiply the conductance by a factor of two (in analogy with inclusion of spin
and imposing spin up-spin down symmetry). However, it was shown~\cite{58,59} that in the vicinity of a superconductor,
the proximity induced scattering potential that exists between electron and holes, also gets renormalized by 
$\ed$-$\ed$ interactions. When this scattering is also taken into account, the correction to the scattering 
amplitude to first order in the interaction parameter depends on $(2g_2-g_1)$ instead of $(g_2-2g_1)$ in the spinful
case. It is worth stressing that this particular linear combination of the interaction parameters ($g_i$'s) 
is independent of the scaling as the logarithmic factors ($l = \ln(kd)$) in Eqs.~\ref{intrg1} and \ref{intrg2} 
cancel each other. Hence, there is no non-monotonic behavior of the conductance in this scenario. The \wirg predicted 
only two fixed points, the Andreev fixed point (perfect \ard) which turns out to be an unstable one
and the perfectly reflecting fixed point which is the stable fixed point. The \ns junction has also been studied 
using bosonisation~\cite{74}. It is easy to check that the power laws resulting from bosonisation agree with 
those obtained from the \wirg, when the $\ed$-$\ed$ interaction induced back-scattering (which is neglected
in the bosonisation method) is ignored.

In the next section, we apply the \wirg method to the superconducting junction of multiple \od quantum wires~\cite{75}. 
We note that we now have two complications - {\textsl{(a)}} multiple wires are connected to the
junction and {\textsl{(b)}} we have both electron and hole channels associated with the junction. 
So in this case, even for the \ns junction, we have two spin channels as well as the electron
and hole channels, so the scattering matrix is four component. For $N$ wires, the scattering matrix is 
$4N \times 4N$-dimensional. Although, we expect our method to work well even in this case,
there is one caveat we must keep in mind. We have incorporated the effect of the superconductor as a 
boundary condition on the \qw and neglected any internal dynamics of the superconductor itself.
This should work reasonably well as long as we are studying transport at energies much below the superconducting gap. 
Our main result here is that the conductance across the junction depends on both $g_1$ and $g_2$ and not on a a 
special combination $2g_2-g_1$ (as in \ns case) which does not get renormalized under \rg flow.
Hence, the cancellation of the logarithmic terms in the effective interaction parameter is specific to the \ns case 
and is not true in general. For $N \ge 2$ wires attached to a superconductor, we expect a non-monotonic form of 
the conductance. We also expect to get a host of fixed points with intermediate transmission and reflection, 
knowledge of which can be of direct relevance for spintronics~\cite{76} and application to device fabrication 
of such geometries.

\section{\label{sec:3} Superconducting junction of multiple \od quantum wires}
\begin{figure}[htb]
\begin{center}
\includegraphics[width=7.0cm,height=6.0cm]{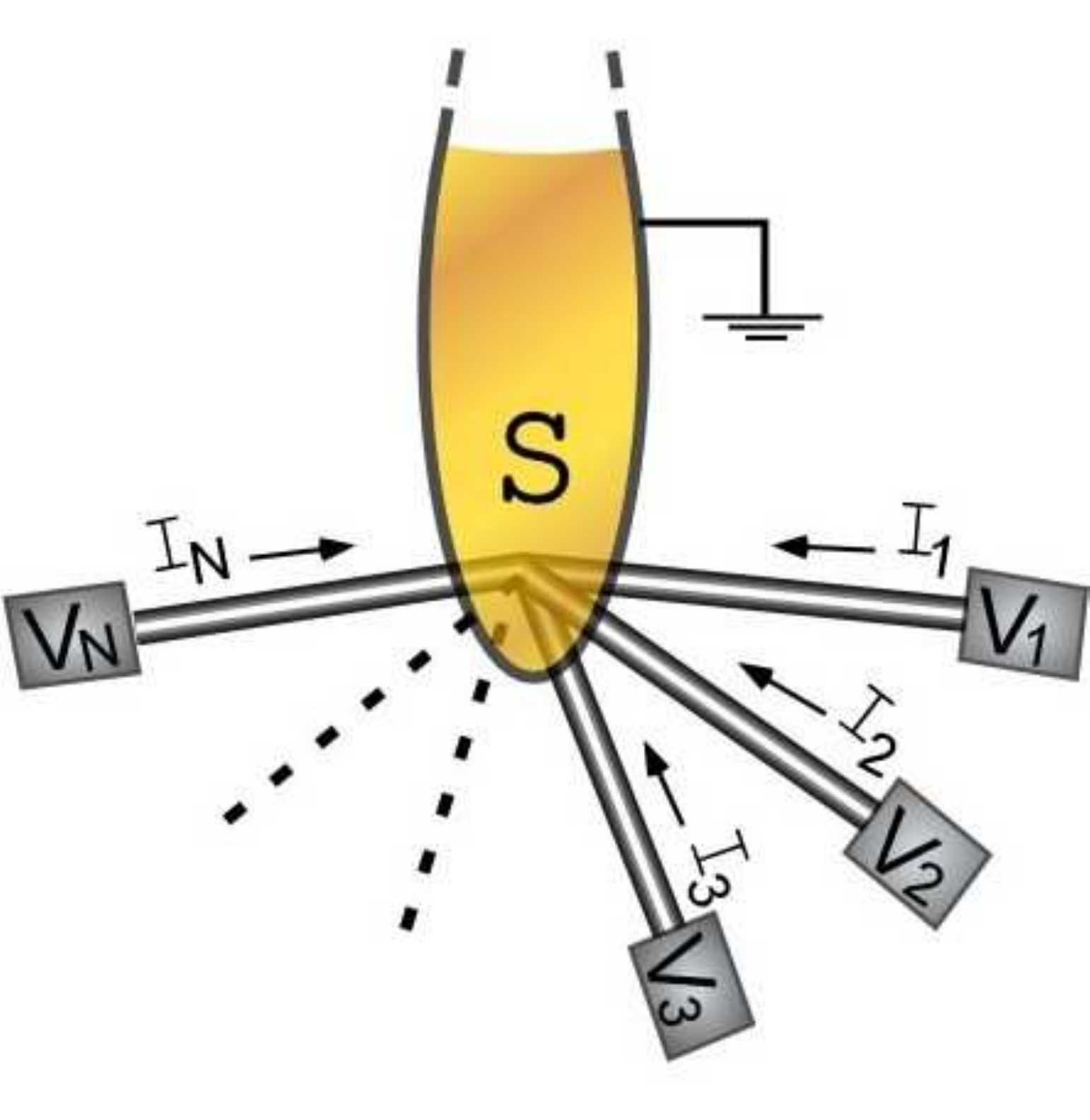}
\caption{(Color online) Multiple \qws connected to a superconducting junction. The dashed lines represents 
the fact that the model can be trivially extended to more than two wires. {\textquoteleft}$a${\textquoteright} 
is the effective length of the superconductor which is assumed to be of the order of phase coherence length of it.
Figure adapted from Ref.~\protect\refcite{75}.}
\label{figfour}
\end{center}
\end{figure}

In this section, let us consider multiple ($N$) quantum wires meeting at a junction on which
a superconducting material is deposited on top of it as depicted in Fig.~\ref{figfour}. The wires are parameterized by 
coordinates $x_i$, with the superconducting junction assumed to be at $x_i=0$. We consider a situation where 
the effective width `$a$' of the superconductor between two consecutive wires is of the order of
the phase coherence length of the superconductor (size of the Cooper pair). For our purpose, it is safe to ignore 
the finiteness of `$a$' and effectively treat the junction of \qw as a single point in space with an appropriate 
boundary condition called Andreev boundary condition. We parameterize the junction by the following quantum mechanical
amplitudes via a \smat-matrix. There are two kinds of reflection amplitudes: the normal reflection amplitude 
($r_{ii\,s_i\,s_i}$) and the \ar amplitude ($r_{A{ii}\,s_i\,s_i}$) on each \qwd. 
In addition, there are two kinds of transmission amplitudes between different wires whic are 
the \ct amplitude ($t_{ij \,s_i \,s_j}$) and the non-local \car amplitude ($t_{Aij \,s_i\,s_j}$). 
The indices $s_i,s_j$ refer to the spin of incoming and outgoing particles. As we consider a singlet 
superconductor at the junction, spin remains conserved in all the processes mentioned above. Thus, the boundary 
conditions can be parametrized by a $4N \times 4N$ scattering matrix for $N$ \qws connected to the 
superconducting junction.

Let us now consider the various symmetries that can be imposed to simplify the $4N \times 4N$ \smat-matrix
at the junction. We impose particle-hole symmetry, \ie, we assume that the reflection and transmissions are 
the same for particles (electrons) and holes. Further, in the absence of a magnetic field, spin symmetry is
conserved which implies that the various transmission and reflection amplitudes for spin up-down electrons 
and holes are equal. (This spin symmetry breaks down in the presence of magnetic fields, or in the case of 
ferromagnetic wires where the spin is fully polarized). Also, since we assume that all the wires, connected 
to the superconductor, are indistinguishable, we can impose a wire index symmetry. (This symmetry again can 
be broken if we take some ferromagnetic and some normal wires attached to the superconductor). On imposing 
these symmetries, the \smat-matrix for a two-wire system with the superconductor at the junction is given by
\bea {\mathbb S} \quad = \quad \begin{bmatrix} ~{\mathbb S}_{\up} & 0~ \\
~0 & {\mathbb S}_{\dn}~
\end{bmatrix}
\non \eea
\noindent with \bea
 {\mathbb S}_{\up} \quad =\quad   {\mathbb S}_{\dn} \quad
 = \quad \begin{bmatrix} ~r & t & r_A & t_A  ~\\
~t & r & t_A & r_{A} ~\\
~r_{A} & t_A & r & t ~\\
~t_A & r_{A} & t & r ~
\end{bmatrix}
,
\label{smat1} 
\eea
Here $r$ stands for normal reflection of electron or hole in each wire, and $r_{A}$ represents \ar from
electron to hole or vice-versa in each wire. On the other hand $t$ represents the elastic \ct amplitude 
($t = t_{12} = t_{21}$) while $t_{A}$ represents the \car amplitude ($t_{A} = t_{A12} = t_{A21}$) from
one wire to the other. For the spin symmetric case, there are two such matrices, one for spin up electrons 
and holes and the other one for spin down electrons and holes.
Note that this is the relevant \smat-matrix at energy scales (temperature and applied voltage on the wires) 
$k_B \,T, e V_i \ll$ $\Delta$, where $\Delta$ is the bulk superconducting gap energy.

The competition between \ct and \car has been analysed before~\cite{40,41,47} and also different ways of
separating the contributions experimentally have been considered~\cite{36,37,51,93}.
However, in this review article we analyse the intriguing effects of $\ed$-$\ed$ interactions within the 
\qws for the \nsn junction case. It is worth emphasizing here that if such \nsn junctions are made out of 
\od systems like $GaAs$ \qws or carbon nanotubes, then the effect of $\ed$-$\ed$ interactions can influence 
the transport properties and low energy dynamics of such systems significantly.

The Landauer$-$Buttiker conductance matrix for the \nsn case can be written, in the subgapped regime 
where $k_B \,T, eV_i \ll \Delta$, as~\cite{40}
\bea
\begin{bmatrix} ~I_1~ \\
~I_2~ \\
\end{bmatrix}
~=~
\begin{bmatrix} ~
G_{A}+G_{CA}+G_{CT}& G_{CA}-G_{CT}~ \\
~G_{CA}-G_{CT}& G_{A}+G_{CA}+G_{CT}~
\end{bmatrix}
\begin{bmatrix}
~V_1~ \\
~V_2~
\end{bmatrix}
,
\eea 
The conductances here are related to the elements of the \smat-matrix: 
$G_{A} \propto |r_{A}|^2$, $G_{CT} \propto |t|^2$ and $G_{CA} \propto |t_{A}|^2$. 
$G_{A}$ is the conductance due to the local \ar that occurs at a single \ns junction, 
whereas $G_{CT}$ and $G_{CA}$ are the conductance due to the elastic \ct and \car
processes respectively, both of which involve transmissions between two \qws and give 
contributions with opposite signs to the sub-gap conductance between the two wires, $G_{CA}-G_{CT}$.
The opposite signs in $G_{CA}-G_{CT}$ arise due to the electron and hole carriers with opposite charge.

\subsection{\label{sec:3.1} \wirg study of superconducting junctions}
We study the effects of inter-electron interactions on the \smat-matrix using the 
renormalization group (\rgd) method introduced in Sec.~\ref{sec:2} following
Ref.~\refcite{26}, and the generalizations to multiple \qws in Refs.~[\refcite{27,71}]. 
The basic idea of the method is as follows. The presence of back-scattering (reflection) induces 
Friedel oscillations in the density of non-interacting electrons. Within a mean field 
picture for the weakly interacting electron gas, the electron not only scatters off the
potential barrier but also scatters off these density oscillations with an amplitude 
proportional to the interaction strength. Hence by calculating the total reflection 
amplitude due to scattering from the scalar scatterer and also from the Friedel oscillations
created by the scatterer in the presence of the $\ed$-$\ed$ inside the \qwd, we can include 
the effect of $\ed$-$\ed$ interaction in calculating transport. The above mentioned idea
can now be generalized to the case where there is, besides non-zero reflection, also non-zero \ar
due to the presence of the superconducting boundary at the junction.

To derive the \rg equations in the presence of Andreev processes, we here follow a similar 
procedure as described in Sec.~\ref{sec:2}. The fermion fields on each wire 
can be written as,
\bea \psi_{is} (x) = \Psi_{I\,is}(x)\, e^{i\,k_F\,x} \,+\,
\Psi_{O\,is}(x)\, e^{-i\,k_F\,x} , \eea
where $i$ is the wire index, $s$ is the spin index which can be $\up,\dn$ 
and $I,O$ stands for outgoing or incoming fields. Note that $\Psi_{I}(x) (\Psi_O (x))$ 
are slowly varying fields on the scale of $k_F^{-1}$ and contain the annihilation operators 
as well as the slowly varying wave-functions. For a momentum in the vicinity of $k_F$, the incoming 
and outgoing fields (with the incoming field on the $i^{th}$ wire) can be Fourier expanded in terms 
of the scattering states and the electron field can be written as
\bea
\Psi_{is}(x) &=& \int ~dk~ \Big[\, b_{ks} \,\epp \,+\,
d_{ks}^\dagger \, \epm  \non\\
 &+& r \,b_{ks}\, \emp  \,+
 \, r^\star \,d_{ks}^\dagger \, \emm  \non\\
 &+&  r_A \,d_{ks}\, \emm  \,+\, r_A^\star \,b_{ks}^\dagger\,
 \emp \Big]
\non\\
\Psi_{(j\ne i)s}(x) &=& \int ~dk~ \Big[ tb_{ks} \,\epp \,+\,
t d_{ks}^\dagger \, \epm  \non\\
 &+&  t_A \,d_{ks}\, \emm  \,+\, t_A^\star \,b_{ks}^\dagger\,
 \emp \Big] ,
\eea
where $b_{ks}$ is the electron destruction operator and $d_{ks}$ is the hole destruction operator. 
Note that we choose to quantise the fermions in the basis of the space of solutions of the 
Dirac equation \ie~around the linear spectrum, in the presence of a potential which allows for 
normal as well as Andreev scattering. We have also allowed for both incident electrons and holes. 
We find that (dropping a constant background density),
\bea
\scxone{\rho_{is}(x)} = \scxone{\Psi_{is}^\dagger \Psi_{is}^{}} =
\frac{i}{4 \pi x}~ \Big[(r^\star e^{2ik_{F}x} - r e^{-2ik_{F}x})
\non \\ ~+~ (r e^{2ik_{F}x}-r^\star e^{-2ik_{F}x})\Big] \ ,
\label{density} 
\eea 
where the two terms corresponds to the density for electrons and holes respectively. 
Here we have also used the fact that due to the proximity of the superconductor, 
the amplitude to create (destroy) a spin $s$ electron and destroy
(create) a spin $s$ hole is non-zero {\textemdash} \ie, the
Boguliobov amplitudes $\scxone{d_{k-s}^\dagger b_{ks}^\dagger} =1=
\scxone{b_{ks}d_{k-s}}$, besides the normal amplitudes 
$\scxone{d_{ks}^\dagger d_{ks}}= \scxone{b_{ks}b_{ks}^\dagger}=1$. 
(This is of course true only close to the superconductor \ie~of the order of the
phase coherence length of the superconductor. We have checked that this gives the 
same result as solving the Boguliobov{\textendash}de Gennes equation as done in Ref.~[\refcite{58,59}]). 
Hence, besides the density, the expectation values for the pair amplitudes
$\scxone{\Psi_{is}^\dagger\Psi_{is}^\dagger}$ and its complex conjugate $\scxone{\Psi_{is}\Psi_{is}}$ 
are also non-zero and are given by (dropping the wire index)

\bea \scxone{\psi_{O\,\up}^\dagger \psi_{I\,\dn}^\dagger} ~=~ -
\scxone{\psi_{O \,\dn}^\dagger \psi_{I\,\up}^\dagger} ~=~ \frac
{-i\, r_{A}}{4 \pi x}
  \non \\\And \quad
\scxone{\psi_{O\,\up} \psi_{I \,\dn}} ~=~ - \scxone{\psi_{O\,\dn}
\psi_{I\,\up}} ~=~ \frac {-i \,r_{A}^\star} {4 \pi x}. 
\label{boguliobov}
\eea 
So, we see that the Boguliobov amplitudes also fall off as $1/x$ just like the 
normal density amplitudes which is $1/{x^{\alpha}}$ for strong $\ed$-$\ed$ interaction
inside the \qw and using bosonization~\cite{63}.

We now allow for short-range density-density interactions between the fermions
\bea
\hmi = \frac{1}{2} \, \int dx \,dy \, \left(\sum_{s\,=\,\up,\dn}
\rho_{s}\right) \,V(x-y)\,\left(\sum_{s\,=\,\up,\dn}
\rho_{s}\right) ,
\eea
to obtain the standard four-fermion interaction Hamiltonian for the spin-full fermions as
\bea
\hmi & = & \int dx \Big[g_1 \Big(\psiiudg  \psioudg \psiiu \psiou
\,+\, \psiiddg \psioddg \psiid \psiod
\nonumber\\
&+&   \psiiudg \psioddg \psiid \psiou \,+\, \psiiddg \psioudg \psiiu
\psiod\Big)
\nonumber \\
&+&  g_2  \Big(\psiiudg \psioudg \psiou \psiiu
+ \psiiddg \psioddg  \psiod \psiid
\nonumber \\
&+&  \psiiudg \psioddg \psiod \psiiu  \,+\,
\psiiddg \psioudg \psiou  \psiid\Big)\Big] ,
\label{hint}
\eea 
where $g_1$ and $g_2$ are the running coupling constants defined in Sec.~\ref{sec:2} (Eq.~\ref{intrg1} and
Eq.~\ref{intrg2}).

Now if we perform Hartree$-$Fock (\hfd) decomposition of the four-fermion interaction Hamiltonian (Eq.~\ref{hint}) 
allowing only particle number conserving amplitudes, then we get
\bea
\hmi^N &=& \int dx \Bigg[g_{1}\Big[\Big(\Psi_{I
\uparrow}^{\dagger}\Psi_{O\uparrow}^{}~+~\Psi_{I
\downarrow}^{\dagger}\Psi_{O
\downarrow}^{}\Big)+\Big(\scxone{\Psi_{O\uparrow}^{\dagger}\Psi_{
I\uparrow }^{ }}+\scxone {\Psi_{O\downarrow}^{\dagger} \Psi_{
I\downarrow}^{}}\Big) \non \\
&+&\Big(\Psi_{O\uparrow}^{\dagger}\Psi_{I\uparrow}^{}
+\Psi_ {O\downarrow}^{\dagger}\Psi_{I\downarrow}^{}\Big)\Big(\scxone
{\Psi_{I\uparrow}^{\dagger}\Psi_{O \uparrow}^{}}~+~\scxone {\Psi_{I
\downarrow}^{\dagger}\Psi_{O \downarrow}^{}}\Big)\Big] \non \\
&-& g_{2} \Big(\Psi_{I\uparrow}^{\dagger}\Psi_{O\uparrow}^{}\scxone {\Psi_{O
\uparrow}^{\dagger}\Psi_{I\uparrow}^{}}~+~\Psi_{O
\uparrow}^{\dagger}\Psi_{I\uparrow}^{}\scxone {\Psi_{I
\uparrow}^{\dagger}\Psi_{O\uparrow}^{}} \non \\
&+& \Psi_{I\downarrow}^{\dagger}\Psi_{O\downarrow}^{}\scxone {\Psi_{O
\downarrow}^{\dagger}\Psi_{I\downarrow}^{}}~+~\Psi_{O
\downarrow}^{\dagger}\Psi_{I\downarrow}^{}\scxone {\Psi_{I
\downarrow}^{\dagger}\Psi_{O\downarrow}^{}}\Big)\Bigg] ,
\label{hintn}
\eea

Using the expectation values for the fermion operators given in Eq.~\ref{density}, the effective 
interaction Hamiltonian (normal) can be derived in the following form on each half wire,
\bea
\hmi^N &=& \dfrac{-i(g_2-2g_1)}{4\pi}  \int_0^\infty \dfrac{dx}{x}
\Big[r^\star \left(\psiiudg \psiou + \psiiddg \psiod \right)
\nonumber \\
&-& r \left(\psioudg \psiiu + \psioddg \psiid \right)\Big].
\label{hintnmf}
\eea
(where we have assumed spin-symmetry \ie~ $r_{\up} = r_{\dn} = r$). Eq.~\ref{hintnmf} has been derived earlier in Ref.~\refcite{27}.

On the other hand, if we perform the \hf deconposition of the four fermion terms allowing for particle 
non-conserving terms (pairing amplitude), then we obtain
\bea
\hmi^A &=& \int dx\Bigg[g_{1}\Big(\scxone
{\Psi_{O\downarrow}^{\dagger}\Psi_{I\uparrow}^{\dagger}
}\Psi_{O\uparrow}^{}\Psi_{I\downarrow}^{}~+~\Psi_{O\downarrow}^{\dagger}
\Psi_{I\uparrow}^{\dagger}\scxone{\Psi_{O\uparrow}^{}\Psi_{I\downarrow}^{}} \non \\
&+&\scxone{\Psi_{O\uparrow}^{\dagger}\Psi_{I\downarrow}^{\dagger}}\Psi_{
O\downarrow}^{}\Psi_{I\uparrow}^{}~+~\Psi_{O\uparrow}^{\dagger}\Psi_{
I\downarrow}^{\dagger} \scxone{\Psi_{O\downarrow}^{}\Psi_
{I\uparrow}^{}}\Big) \non \\
&-&g_{2}\Big(\scxone{\Psi_{O
\downarrow}^{\dagger}\Psi_{I\uparrow}^{\dagger}}\Psi_{
O\downarrow}^{}\Psi_{I\uparrow}^{}~+~\Psi_{O\downarrow}^{\dagger}\Psi_{
I\uparrow}^{\dagger}\scxone{\Psi_{O\downarrow}^{}\Psi_{I\uparrow}^{}} \non \\
&+&\scxone{\Psi_{O\uparrow}^{\dagger}\Psi_{I\downarrow}^{\dagger}}\Psi_{
O\uparrow}^{}\Psi_{I\downarrow}^{}~+~\Psi_{O\uparrow}^{\dagger}\Psi_{
I\downarrow}^{\dagger}\scxone{\Psi_{O\uparrow}^{}\Psi_{I\downarrow}^{}}
\Big)\Bigg],
\label{hinta}
\eea

Using Eq.~\ref{boguliobov} in $\hmi^A$, we get the (Andreev) Hamiltonian which can be written as
\bea
\hmi^A &=& \dfrac{-i(g_1+g_2)}{4\pi}  \int_0^\infty  \dfrac
{dx}{x} \Big[-r_{A}^{\star}\big(\psiiudg \psioddg +
 \nonumber \\
&& \psioudg \psiiddg \big)
 + r_{A}\left( \psiod\psiiu + \psiid\psiou \right) \Big].
\label{hintamf}
\eea
Note that although Eq.~\ref{hintamf} appears to be charge non-conserving, 
charge conservation is taken care of by the $2e$ charge that flows into the 
superconductor every time there is an Andreev process taking place.

The amplitude to go from an incoming electron wave to an outgoing electron wave 
under $e^{-i{\hmi^N}t}$ (for electrons with spin) was derived in Ref.~\refcite{27} 
and is given by
\bea {-\alpha \, r_{s} \over 2} \, \ln (kd) \label{Friedeln} \ ,
\eea
where $\alpha = (g_2-2g_1) / 2\pi\hbar v_F $ and $d$ was a short distance cut-off. 
Analogously, the amplitude to go from an incoming electron ${\sf{e_{in}}}$ wave to an 
outgoing hole ${\sf{h_{out}}}$ wave under $e^{-i{\hmi^A} t}$ is given by 
\bea
e^{-i\,\hmi^A \, t}
\ket{{\mathsf{e_{in}}},s,k} ,
\hskip -2.5 cm
\non \\
&=& -i \int
\dfrac{dk^\prime}{2\pi}\Bigg[\ket{{\mathsf{h_{out}}},s^\prime,k^\prime}
 \me{\hmi^A}
{{\mathsf{h_{out}}},s^\prime,k^\prime}{{\mathsf{e_{in}}},s,k}\Bigg] ,
\hskip -0.5 cm
\non \\
&=& \dfrac{-i(g_1\,+\,g_2)\, r_A} {4 \, \pi \, \hbar \, v_F}
\int_{}^{} \dfrac{dx}{x}  e^{-2\,i\,k\,x} \,
\ket{{\mathsf{h_{out}}},s^\prime,k^\prime} \ .  
\eea
where $s \ne s^\prime$. Hence, the amplitude for an incoming electron to be scattered to an 
outgoing hole by the Andreev process is given by
\bea {\alpha^\prime \,r_{A} \over 2} \, {\ln (kd)} ,
\label{FriedelA} \eea
where $\alpha^\prime = (g_1+g_2)/ 2 \pi  \hbar  v_F $. Note also
that $\alpha$ and $\alpha'$ are themselves momentum dependent, 
since the $g_i$'s are momentum dependent \ie~running coupling constants. 
For the spinful case, the amplitude for an incoming electron to go to an outgoing 
electron on the same \qw is governed by the interaction parameter $\alpha =(g_2-2g_1) / 
2\pi\hbar v_F $ which has the possibility of chaging sign under \rg evolution, because 
of the relative sign between $g_1$ and $g_2$. On the other hand, $\alpha^\prime = (g_2+g_1) / 2\pi\hbar v_F $ 
can never change its sign as they are of the same sign.

\subsubsection{\ns Junction \label{3.1.1}}
In is shown before that the amplitudes in Eqs.~\ref{Friedeln} and \ref{FriedelA} 
are corrections to the reflections of electrons from Friedel oscillations (as an electron)
and from the pair potential (as a hole) respectively.
We can combine them with the bare \smat-matrix at the junction to find
the corrections to the amplitudes of the bare \smat-matrix. For an \ns
junction, there is only one wire coupled to the superconductor and
the \smat-matrix is just $2 \times 2$ for each value of the spin and is given by
\bea {\mathbb S}
\quad=\quad \begin{bmatrix} ~r &  r_{A}~ \\
~r_{A}& r~
\end{bmatrix} \ ,
\label{smat2} 
\eea
Here $r$ is the bare normal refelction amplitude and $r_A$ is the bare \ar amplitude from the \ns interface. 
So we only need to compute the corrections to $r$ and $r_A$ in this case.

\begin{figure}[htb]
\begin{center}
\includegraphics[width=10.0cm,height=6.0cm]{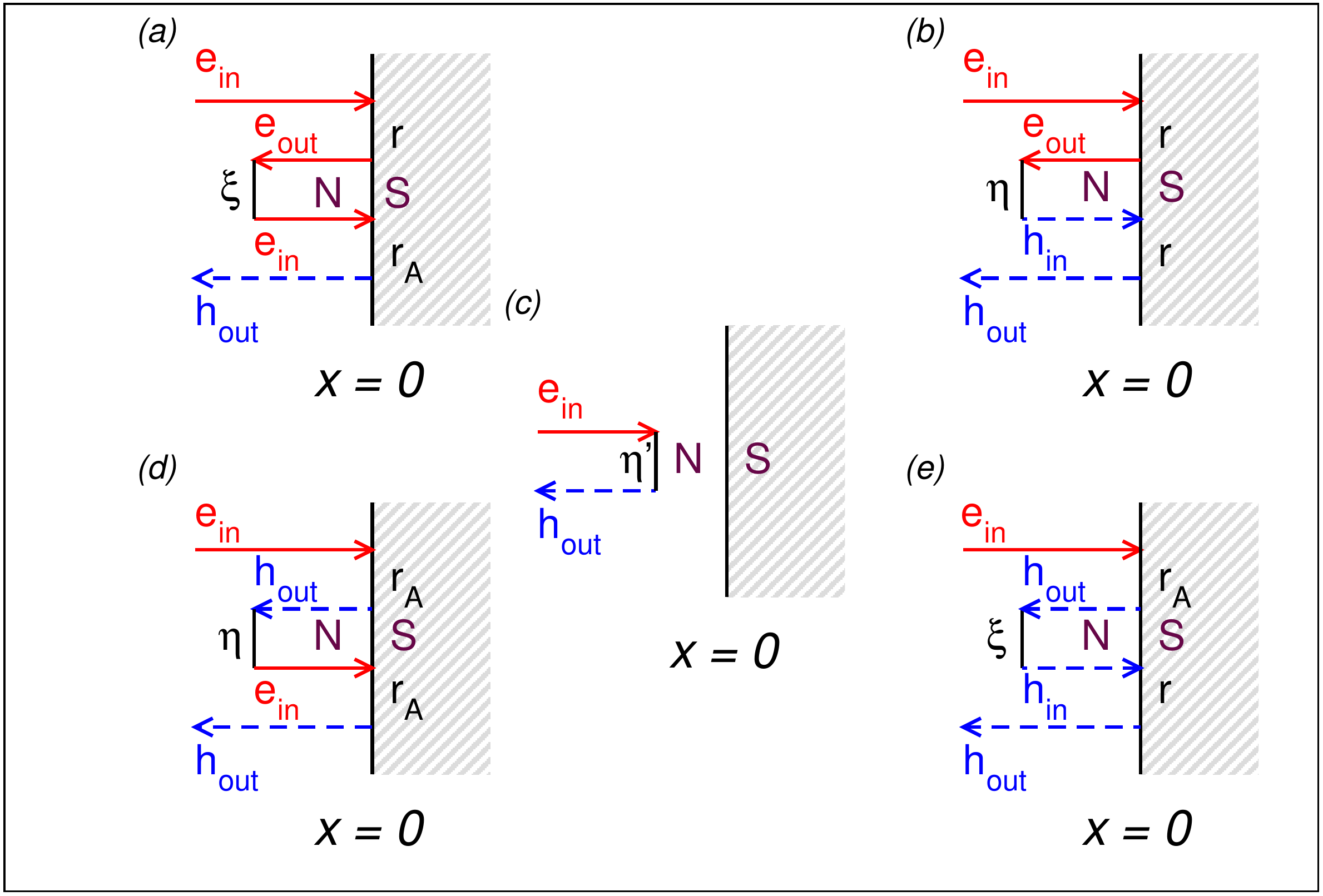}
\caption{(Color online) The processes that contribute to the amplitude 
for an incoming electron to transform to an outgoing hole. Note that all
the processes shown here are to first order in the interaction
parameters since they only involve a single scattering from a
Friedel oscillation or the pair potential. Process $(c)$ involves
scattering from a pair potential before the electron reaches the
junction. The remaining processes involve two reflections from the
junction and a scattering from the Friedel oscillation or the pair
potential. In the diagrams, $\xi = \frac{1}{2}\,\alpha
\,r^{\star}\, {\ln (kd)}$, $\eta = -\,\frac{1}{2}\,\alpha^\prime
\, r_{A}^{\star}\, \ln (kd)$ and $\eta^\prime=\frac{1}{2} \,
\alpha^\prime \, r_{A} \, \ln (kd)$. Figure adapted from Ref.~\protect\refcite{75}.}
\label{figfive}
\end{center}
\end{figure}

We find that there are five processes which contribute to the amplitude $r_A$ to first order in the 
repulsive $\ed$-$\ed$ interaction parameter. The following diagrams are illustrated in Fig.~\ref{figfive}.
Adding all the contributions, we obtain the renormalized \ar amplitude $r_{A}$ that takes an 
incoming electron to an outgoing hole and is given by
\bea 
\delta r_{A} &=&  \dfrac{\alpha^\prime}{2}  \big[\,r_{A}\,-\,
r_A^{\star} \left( r^{2} + r_{A}^{2} \right)\,\big]
 \, \ln(kd)  
~+~ \alpha \, |r|^2 \,  r_{A} \, \ln(kd) \ ,
\label{rans} 
\eea
%
which is in agreement with Ref.~[\refcite{58,59}]. On the other hand, for an incoming 
electron reflected back as an electron, we find the small correction in the
amplitude $\delta r$ given by~\cite{26,27}
\bea 
{{\delta r}} &=& {{-\alpha^\prime \, |r_A|^2 \, r \, \ln(kd)}}
{{+~ \frac{\alpha}{2} \, \left[\, r_A^2 \, r^{\star} \,-\,
r\left(1\,-\,|r|^2\right)\, \right] \ln(kd) }} \ ,
\label{ransnew0}
\eea
%
We replace $-\ln(kd)$ by $dl$ using the ``poor man's scaling" approach~\cite{69} to obtain the
\rg equation for $r_A$ as
\bea 
\frac {dr_{A}}{dl} &=&
 -\,\frac{\alpha^\prime}{2} \left[r_{A}
\,-\, r_A^{\star}\left(r^{2} \,+\, r_{A}^{2}  \right)
\right]
\,-\, \alpha \, |r|^2   r_{A} \ ,
\label{ransnew1} 
\eea
Using the unitarity of  the \smat-matrix ($\vert \, r_A \, \vert ^2 + \vert\, r^2 \, \vert = 1$ 
and $r_A^\star  r + r_A r^\star = 0$), we can simplify the RHS of the Eq.~\ref{ransnew1} to obtain

\bea 
\frac {dr_{A}}{dl} &=&
-\, \left(\,\alpha \, + \, \alpha^\prime \,\right) \, r_A
\left(\,1 - \vert r_A \vert ^2 \, \right)  \ ,
\label{ransnew2} 
\eea
Note that the combination $\alpha\,+\,\alpha^\prime = (2\,g_2 - g_1)/2\pi \hbar v_F$ 
which appears in the \rg equation (Eq.~\ref{ransnew2}) does not flow under \rgd. 
The latter can be seen from Eqs.~\ref{intrg1} and \ref{intrg2} which shows that
$ (2\,g_2 - g_1)/2\pi \hbar v_F = (2V(0) -V(2k_F))/2\pi\hbar v_F$.
The following observation implies that $r$ and $r_A$ either monotonically increase or
decrease as a power law depending on the sign of $\alpha\,+\,\alpha^\prime$. 
From Eq.~\ref{ransnew2}, we also observe that $\vert r_A \vert = 0$ and $ \vert r_A \vert = 1$
correspond to the insulating and the Andreev fixed points of the \ns junction respectively. 
One can easily see from the \rg equations that $\vert r_A \vert = 0$ is a stable fixed point 
and $\vert r_A \vert = 1$ is an unstable fixed point. Due to a small perturbation around the
unstable $\vert r_A \vert = 1$ fixed point, the system always flows towards the stable
$\vert r \vert = 0$ fixed point in which the \qw is completely disconneced from the
superconducting junction.

\subsubsection{\nsn Junction \label{3.1.2}}
\begin{figure}[b]
\hskip +2.2cm
\includegraphics[width=4.2cm,height=3.5cm]{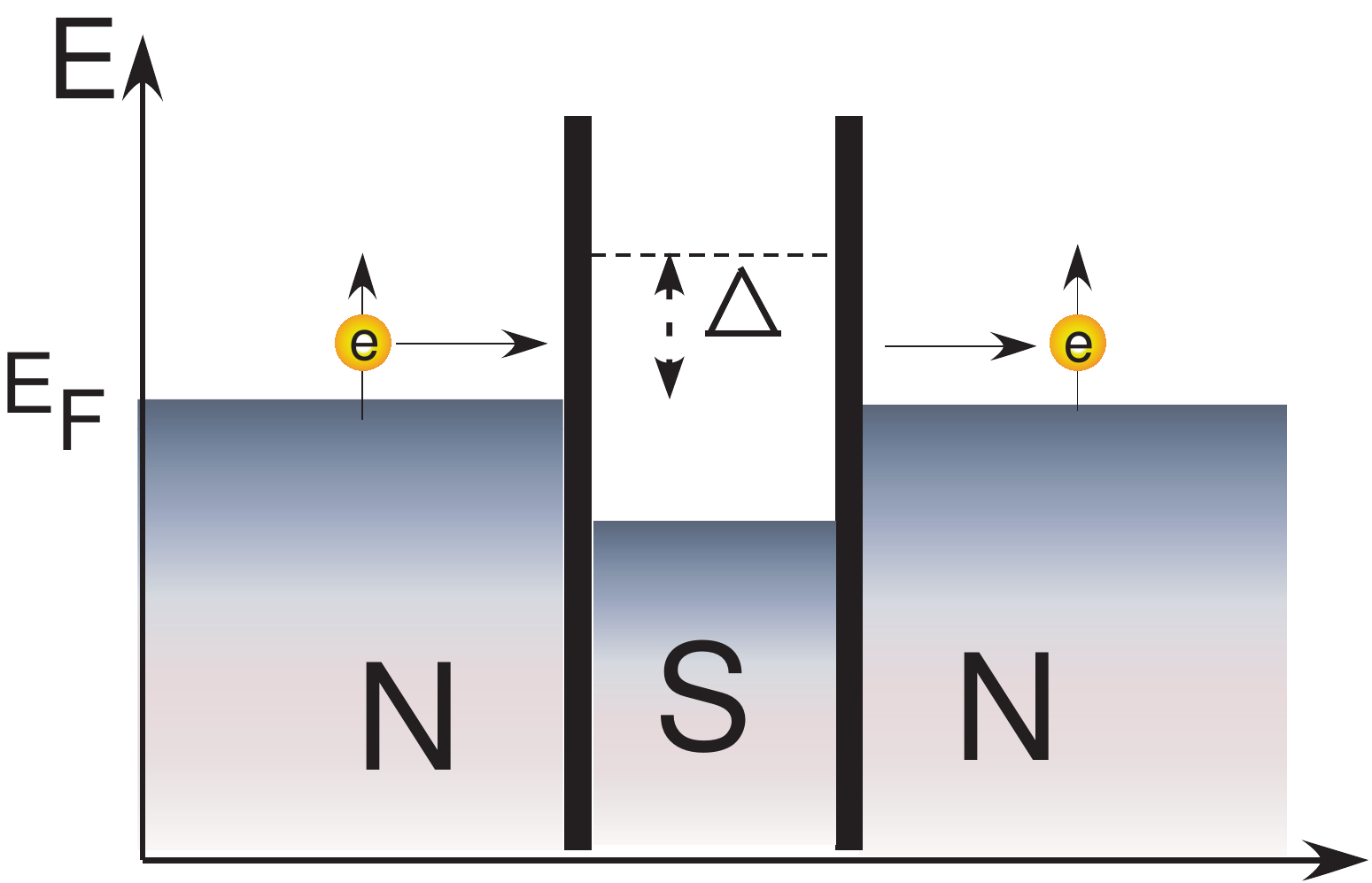}
\vskip -3.5cm \hskip +8cm
\includegraphics[width=4.2cm,height=3.5cm]{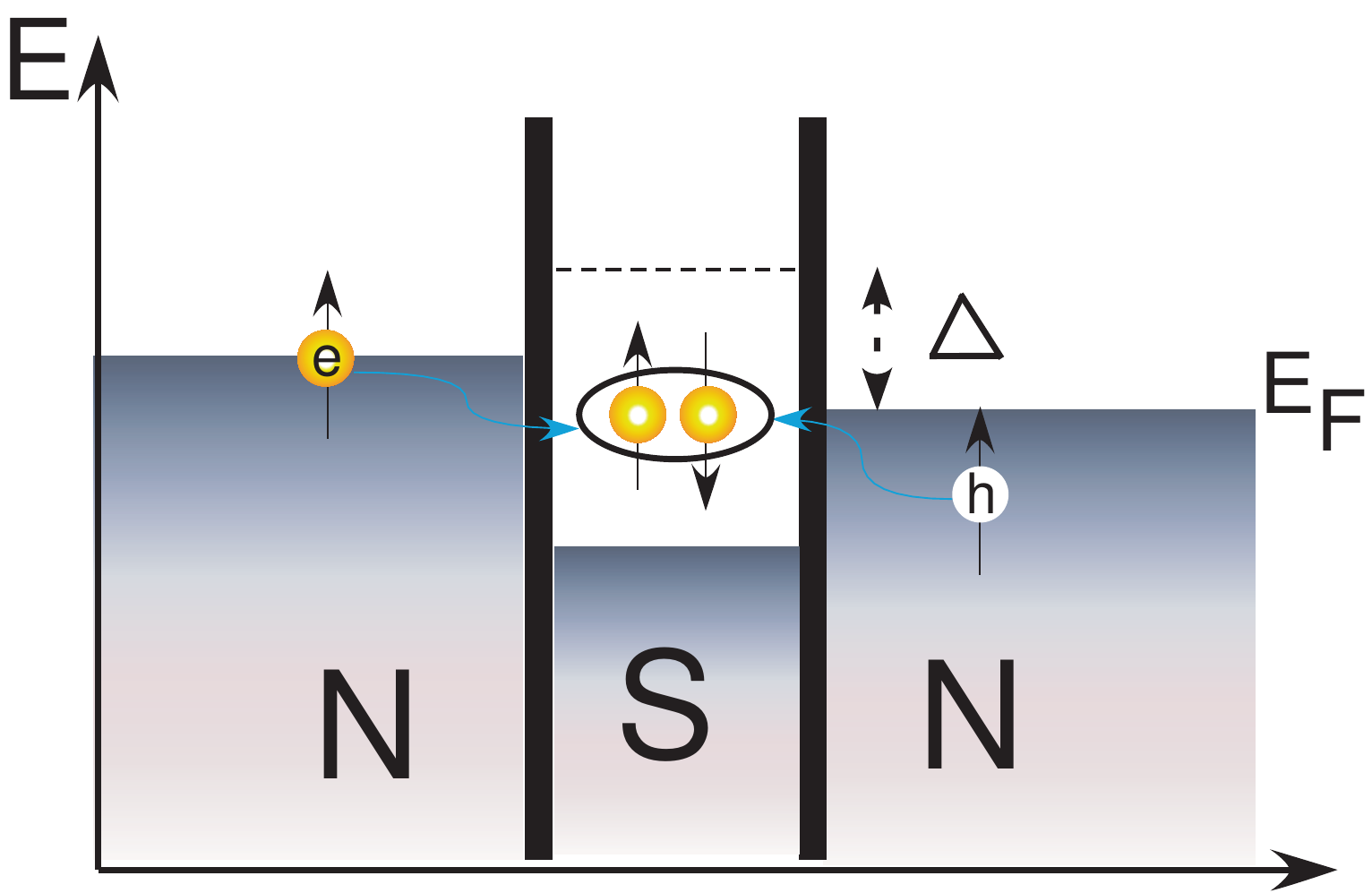}
\caption{(Color online) Electron \ct with bare amplitude $t$ is shown in the left figure and 
\car with bare amplitude $t_A$ is shown in the right diagram. Figure adapted from Ref.~\protect\refcite{75}.}
\label{figsix}
\end{figure}

\begin{figure}[htb]
\begin{center}
\vskip +1.2cm
\includegraphics[width=10.0cm,height=6.0cm]{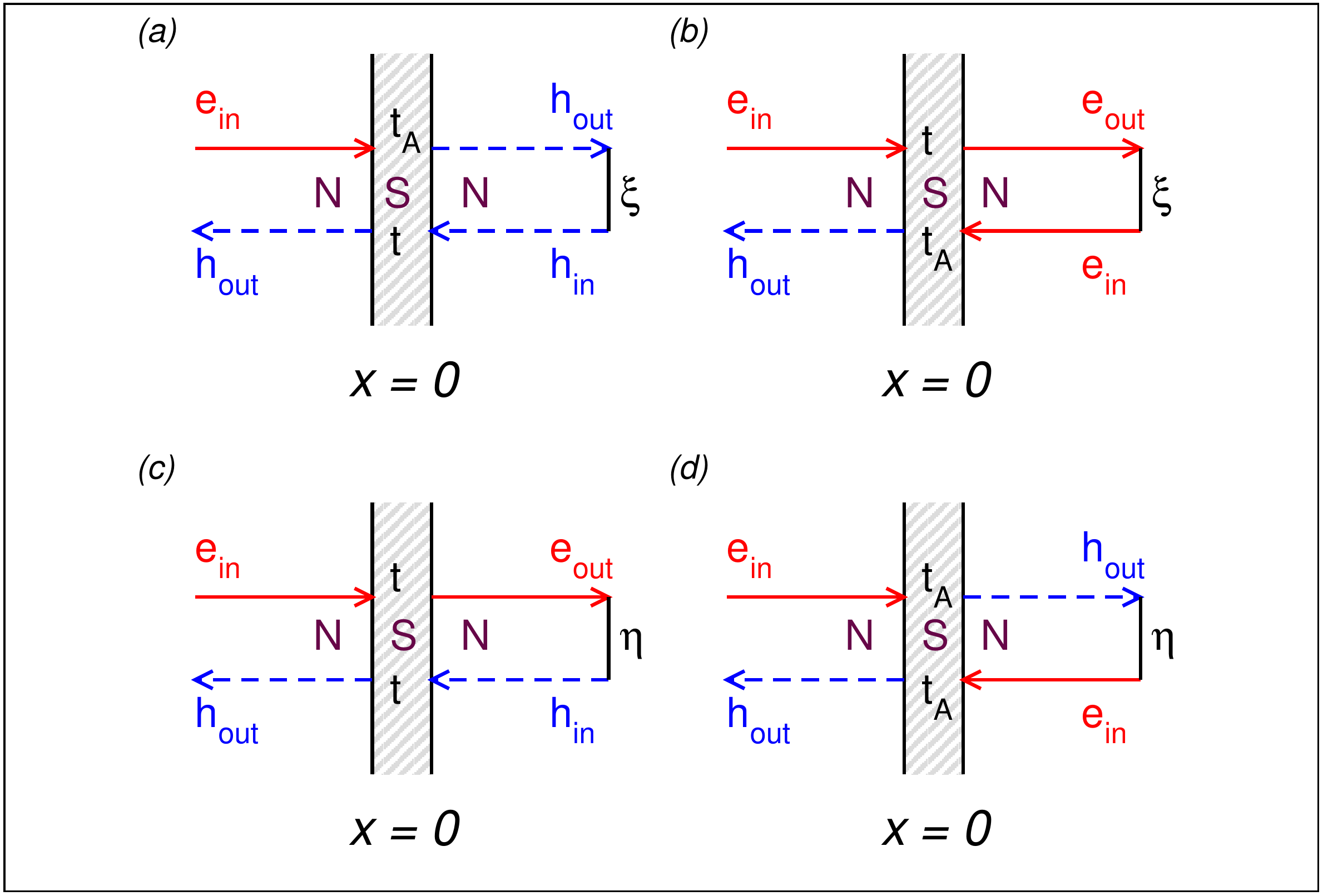}
\end{center}
\caption{The extra processes that contribute to the amplitude for an incoming electron to transform to an 
outgoing hole on the same wire, due to the second wire. In processes (a) and (b) the incident electron from 
the first wire are transmitted to the second wire and reflected by the Friedel oscillation whereas 
(c) and (d) are transmitted to the second wire and reflected by the pair potential. Figure adapted from Ref.~\protect\refcite{75}.} 
\label{figseven}
\end{figure}

In this subsection, we consider a \nsn junction. Here in
addition to the two reflection channels, we also have two channels
for transmission - the direct transmission of an electron to an electron
through \ct process and the transmission of an incoming electron to a outgoing 
hole via \card. These two processes are depicted in Fig.~\ref{figsix}. 
The \smat-matrix at the junction is $8\times 8$ in this case and is 
given in Eq.~\ref{smat1}.
The number of scattering processes from the Friedel oscillations and pair potentials
that contribute to the renormalization of the scattering amplitudes in this case is 
thirty four, since we also need to include terms that transmit electrons
or holes through the junction. For instance, for the renormalization of the \ar term, 
besides the terms corresponding to the \ns junction, we also have to include processes 
in which the incident electron from wire $1$, goes through the junction to wire $2$, 
Andreev reflects from the pair potential on wire $2$ and then comes back through the junction 
to wire $1$, as shown pictorially in Fig.~\ref{figseven}(c).

Collecting all the nine processes that contribute to first order in $\alpha$ and $\alpha'$ 
to the reflection amplitude, we find that 
\bea 
\dfrac{dr}{dl} &=& - \bigg[\dfrac{\alpha}{2} \, \left[
(t^2\,+\,r_{A}^2\,+\,t_{A}^2)\,r^\star\,-\,r(1-|r|^2)\right]
~-~\alpha'\, (r \,|r_{A}|^2 \,+\, r_{A}^\star \,t_{A}\,t) \bigg] \ ,
\label{rnsn} 
\eea 
Similarly, adding up the contributions from the nine processes that contribute to $r_A$, we find that
\bea 
\dfrac{dr_{A}}{dl} &=&
-\bigg[\alpha(|r|^2\,r_{A}\,+\,t\,t_{A}\,r^\star)
~+~
\dfrac{\alpha'}{2}(r_{A}\,-\,(r^2\,+\,r_{A}^2\,+\,t^2\,
+\,t_{A}^2)\,r_{A}^\star) 
\bigg] \ ,
\label{ransn} 
\eea
Moreover, for the \nsn case, besides the reflection parameters, we also need to
compute the renormalizations of the transmissions (\ct and \card) to first order
in $\alpha$ and $\alpha^\prime$. The \rg equations for $t$ and $t_A$ are also obtained 
by considering all possible processes that ultimately have one incoming electron and 
one outgoing electron (for $t$) and one incoming electron and one outgoing hole 
(for $t_A$) which are either reflected once from the Friedel potential or
the pair potential. Finally they are found to be
\bea 
\dfrac{dt}{dl} &=&
-\,\bigg[ \alpha \, ( |r|^2\,t \,+\, r^\star \,r_A \,t_A) 
~-~  \alpha^\prime\,(|r_A|^2 \,t \,+\, r \,r_A^\star\, t_A) \bigg]\ ,
\label{tnsn}\\
\dfrac{dt_A}{dl} &=& -\,\bigg[\alpha (r^\star\, r_A\,t \,+\,
|r|^2\, t_A) 
~-~ \alpha^\prime\,(r\,t\,r_A^\star \,+\,|r_A|^2\,t_A)\bigg]\ .
\label{tansn} 
\eea

It has been emphasized in Sec.~\ref{sec:2} that for the normal junction case the 
\rg equations can be expressed in a compact from (Eq.~\ref{77})~\cite{27}. Similarly
we can express the \rg equations for the superconducting junction case also
in a compact matrix form as given below,
\bea
\frac{d{\mathbb {S}}}{dl} = {\tilde F} - {\mathbb {S}} {\tilde F}^\dagger {\mathbb {S}} \ .
\eea
where the matrix \smat~is given in Eq.~\ref{smat1} and $\tilde F$ depends on the 
interaction parameters $\alpha= (g_2-2g_1)/2\pi \hbar v_F$ and $\alpha^\prime = (g_1+g_2)/2\pi
\hbar v_F$ in the \qwd. $\tilde F$ is {\textsl{non-diagonal}} matrix (unlike the case
in Ref.~\refcite{27}) and is given by
\bea {\tilde F} ~=~
\begin{bmatrix}
~\frac{\alpha r}{2}& 0 & \frac{-\alpha^\prime r_{A}} {2} &0~  \\
~0& \frac{\alpha r}{2}& 0& \frac{-\alpha^\prime r_{A}}{2} ~ \\
~\frac{-\alpha^\prime r_{A}}{2} &0 & \frac{\alpha r}{2}& 0 ~\\
~0& \frac{-\alpha^\prime r_{A}}{2} & 0 & \frac{\alpha r}{2}~ \\
 \end{bmatrix}~.
\label{fmat} 
\eea
It is now easy to check that all the \rg equations can be reproduced from the matrix equation. 
The matrix form also makes the generalization to $N$ wires case notationally simple and makes 
the search for various fixed point much easier. But note that these equations have to be augmented 
by Eqs.~\ref{intrg1} and \ref{intrg2} to get the full set of \rg equations.

Let us now look at some of the fixed points of the \smat-matrix. Clearly, the fixed points 
occur when ${\tilde F}-{\mathbb {S}}{\tilde F}^\dagger {\mathbb {S}}=0$ or when
${\tilde F}{\mathbb {S}}^\dagger$ is hermitian. There are several possibilities and we
list below some important of them.

\begin{enumerate}
\item[]{{\textsl{Case I:}} Any one of the four amplitudes is non-zero}\\
{\textsl{(a)}} $t=1$, $r=r_{A}=t_A=0$, fully transmitting fixed point (\tfpd)\\
{\textsl{(b)}} $r=1$, $r_{A}=t_A=t=0$ fully reflecting fixed point (\rfpd)\\
{\textsl{(c)}} $r_A=1$, $r=t=t_A=0$, fully Andreev reflecting fixed point (\afpd)\\
{\textsl{(d)}} $t_A=1$, $r=t=r_A=0$, fully crossed Andreev reflecting fixed point (\cafpd)
\item[]{{\textsl{Case II:}} Any two amplitudes are non-zero}\\
When both $r$ and $r_{A}$ are zero, the RHS of the \rg equations identically vanishes 
as both the Friedel oscillation amplitude as well as the pair potential amplitude in the wire 
become zero. Hence any value of $t$ and $t_A$ remains unrenormalized under \rgd.
\item[]{{\textsl{Case III:}} Any three of them are non-zero}\\
We did not find any fixed point of this kind.
\item[]{{\textsl{Case IV:}} All four of them are non-zero}\\
Here, we get a fixed point when $r_1=r_2=t=t_A=1/2$ and $r_{A1}=r_{A2}=-1/2$. 
The latter is the most symmetric \smat-matrix possible for the \nsn case. Since it is a 
symmetry-dictated fixed point with intermediate transmission and reflection, we shall
refer to it as symmetric fixed point (\sfpd).
\end{enumerate}
%

\subsubsection{\fsd, \fsf and \fsn Junctions \label{3.1.3}}
We can also consider junctions where one or more of the wires are spin-polarised, 
with Fermi distributions for the spin up and down electrons being different. 
As long as at least one of the wires is ferromagnetic, the spin up-spin down symmetry 
of the system is broken. This implies that we can no longer impose ${\mathbb {S}}_\up = 
{\mathbb {S}}_\dn$ on the \smat-matrix parametrising the scattering at the junction
as we had before in Eq.~\ref{smat1}. We now need to choose an \smat-matrix
with indices $\uparrow$ and $\downarrow$ denoting the spin. Experimental set-up
based on Ferromagnet$-$Superconductor (\fsd) hybrid structures has been investigated
in the recent past~\cite{37,113}. For the Ferromagnet$-$Superconductor$-$Normal (\fsnd) case 
(and the Ferromagnet$-$Superconductor$-$Ferromagnet (\fsfd) case where the ferromagnets 
on the two sides are not identically polarized) the wire index symmetry is also broken due
to the spin assymetry in the two wires. Hence, the \smat-matrix chosen must also break 
the wire-index symmetry. Note that for the ferromagnetic wire, the amplitude to destroy 
a spin $s$ electron and create a spin $s$ hole cannot be non-zero, even in the proximity 
of the superconductor. The Boguliobov amplitudes $\scxone{d_{ik-s}^\dagger b_{iks}^\dagger}$
and $\scxone{b_{iks}d_{ik-s}}$ decay exponentially fast (with a length scale set by the ferro$-$anti-ferro 
gap) in the ferromagnetic wire. So, in our \smat-matrix, $r_A$ is zero and there is no
pair potential due to the proximity effect in ferromagnetic wire. Also as mentioned earlier, 
we must keep in mind that the influence of the bulk ferromagnet on the spectrum of the superconductor 
and on the \qw have to be negligibly small. This will be true only if the superconductor is
large enough. Hence, for such junctions, the renormalization of the \smat-matrix sets in only 
scattering due to the Friedel oscillations. Also note that in these wires, since the bulk 
does not have both the spin species, $g_1$ and $g_2$ do not get renormalized \ie~they are not
running coupling constant anymore. All the cases mentioned above will therefore involve the 
full $4N \times 4N$ \smat-matrix since there is no reduction in number of independent elements 
of the \smat-matrix which can occur when symmetries are imposed.

\subsubsection{Three$-$Wire$-$Junction$-$A Beam Splitter \label{3.1.4}}
In this subsection, we consider the standard beam splitter geometry comprising of a 
superconductor at the junction of three quantum wires. In this case, we show that 
there is a fixed point that is analogous to the Andreev fixed point of the \ns junction. 
The \smat-matrix representing this fixed point is symmetric under all possible permutations 
of the three \qws and allows for the maximum Andreev transmission (in all channels 
simultaneously within unitarity constraints). The \smat-matrix is given by 
$r_A = -1/3$ and $t_{A}=t_{A}^\prime=2/3$ with $r=t=t'=0$. We refer to
this fixed point as the {\textsl{Andreev$-$Griffith's fixed point}} (\agfpd)
which is also an intermediate fixed point with non zero scattering amplitudes.
Very recently, similar Cooper pair beam splitter geometry has been realized
experimentally in the context of two-particle correlations of the shot noise of the 
split electrons via \car process~\cite{53,54,57,114}. In literature the Griffith's fixed point 
represents the most symmetric \smat-matrix for a normal three wire junction. 
It is given by $r=-1/3$ and $t=2/3$ where $r$ is the reflection within each wire and 
$t$ is the transmission from one wire to the other. The boundary condition for the three 
wire junction corresponding to the above mentioned \smat-matrix was obtained by 
Griffith~\cite{92} hence we refer to it as the Griffith's fixed point (\gfpd). 

For an analytic treatment of this case, we consider a simplified situation where 
there is a complete symmetry between two of the wires, say $1$ and $2$, and the 
\smat-matrix is real. In addition, the elements of the \smat-matrix corresponding to
normal transmission or reflection of an incident electron (hole) to a reflected or 
transmitted electron (hole) are set to zero so that only Andreev scattering processes 
participating in transport. Then the \smat-matrix is given by
%
\bea 
{\mathbb {S}} =
\begin{bmatrix} ~0 & 0 & 0 &
r_{A} & t_{A} & t_{A}^\prime~\\
~0 & 0 & 0 & t_{A} & r_{A} & t_{A}^\prime ~\\
~0 & 0 & 0 & t_{A}^\prime & t_{A}^\prime & r_{A}^\prime~ \\
~r_{A} & t_{A} & t_{A}^\prime & 0 & 0 & 0~\\
~t_{A} & r_{A} & t_{A}^\prime & 0 & 0 & 0~\\
~t_{A}^\prime & t_{A}^\prime & r_{A}^\prime & 0 & 0 & 0~\\
\end{bmatrix} \ ,
\label{smatbs} 
\eea
where, $r_A$ and $t_A$ and $t'_A$ are real parameters which satisfy~\cite{65}
\begin{eqnarray}
t_A&=&1+r_A~, \non\\ r'_A&=&-1-2r_A~,
\non\\
t'_A&=&\sqrt{(-2r_A)(1+r_A)}~, \non\\{\mathrm{and}} \quad  -1 &\le
& r_A \le 0 \label{unit}
\end{eqnarray}
by unitarity. Using Eq.~\ref{unit}, the simplified \rg equation for the single parameter 
$r_A$ is given by
\begin{equation}
\dfrac{dr_A}{dl} =  \alpha'~ \left[r_A  (1+r_A) (1+3r_A) \right]\ .
\end{equation}
\begin{figure}[htb]
\begin{center}
\includegraphics[width=10.0cm,height=6.0cm]{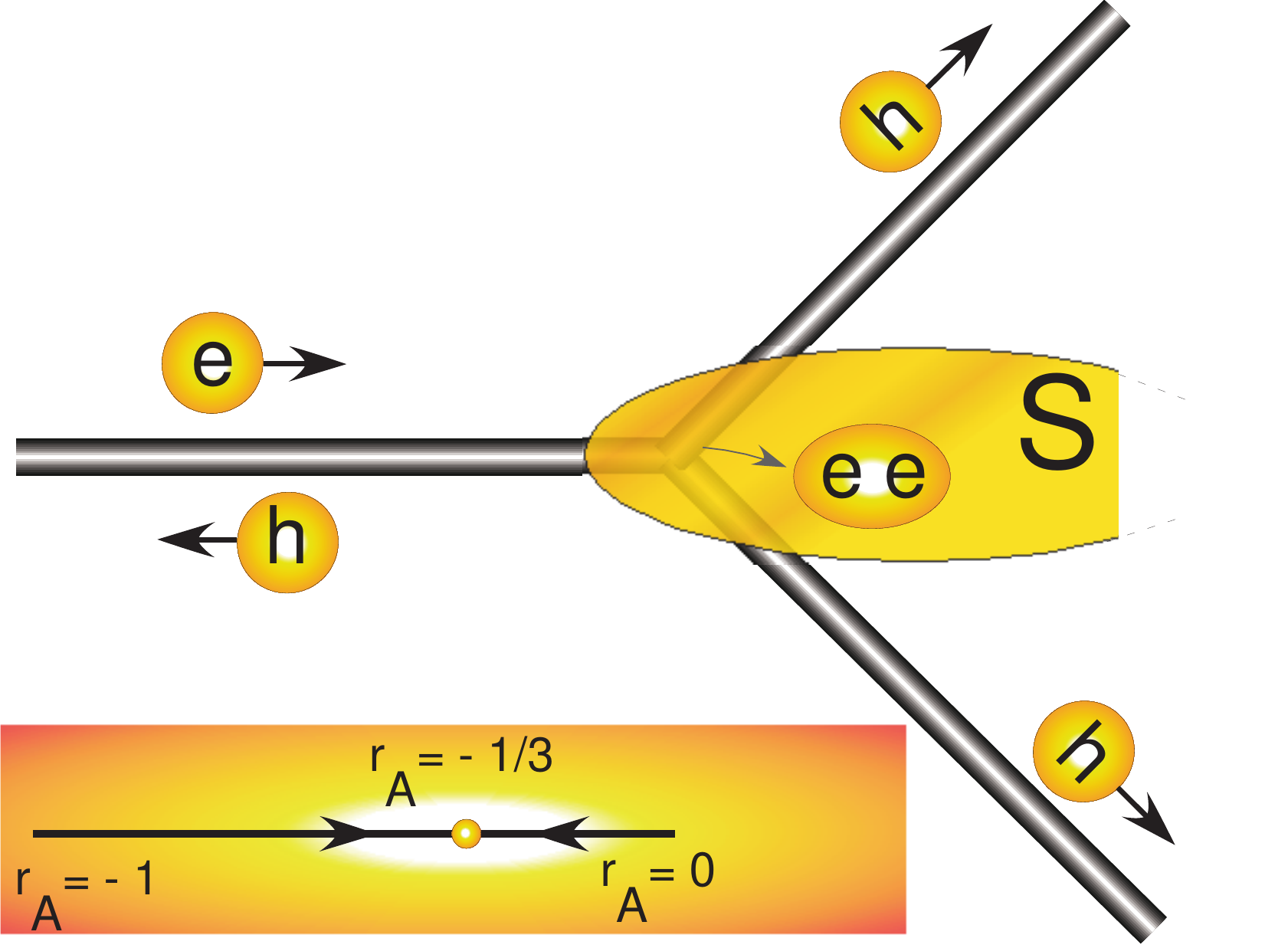}
\end{center}
\caption{(Color online) Schematic representation of the beam splitter geometry where a
three$-$wire$-$junction is hooked to the stable fixed point, \agfpd. 
An incident electron in one wire is either reflected back as a hole in the same wire (\ard)
or is transmitted as a hole (\card) in another wire along with the addition of the two 
electrons into the superconductor forming a Cooper pair. The direction of \rg flow
from two unstable fixed points to the stable fixed point (\agfpd) is also depicted on the 
bottom left side of the diagram. Figure adapted from Ref.~\protect\refcite{75}.}
\label{figeight}
\end{figure}
So, within the real parametrization we have two unstable fixed points, given by 
$r_A=0$ and $r_A=-1$ and a stable fixed point given by $r_A=-1/3$. The $r_A=0$ fixed point 
corresponds to a situation where there is perfect \car between wires $1$ and $2$
and wire $3$ gets cut off from the remaining two wires (labelled by $1$ and $2$)
and is in the perfect \ar limit with the superconductor. The $r_A=-1$ fixed point corresponds 
to a situation where all the three wires are disconnected from each
other and are in perfect \ar limit individually with the superconductor. 
The third fixed point given by $r_A=-1/3$ corresponds to a perfect Andreev limit of the 
three wire junction where an incident electron is either Andreev-reflected into the
same wire as a hole or is transmitted as a hole via \car into another wire. 
This is essentially the \agfpd. It is very interesting to note that the original 
Griffith's fixed point was a repulsive fixed point~\cite{27,65} whereas the \agfp is 
an attractive fixed point. This can be understood as follows.
In the current situation, there is no scattering from the Friedel oscillations as the
junction is assumed to be reflection-less ($r=0$), whereas there exists a
proximity induced pair potential, which induces an effective
attractive interaction between the electrons. Hence, the physics is
very similar to the well-known \lln physics, which says that for 
attractive interaction between the electrons, back-scattering is an irrelevant operator.
Hence the stable fixed point here will be the one which will have
maximal transmission between the wires. So, it is not surprising
that the \agfp turns out to be a stable fixed point. Thus, for a
reflection-less junction, we have found a stable fixed point with
intermediate transmission and reflection analogus to the \gfp in three wire junction~\cite{27}.

\subsection{\label{sec:3.2} Results of \rg flows for the conductance}
In this subsection, we consider various physical cases regarding the \rg
fixed points and discuss the outcome of \rg flows for the \lb conductances in each case.
\subsubsection{\ns Junction \label{sec:3.2.1}}
\begin{figure}[htb]
\begin{center}
\includegraphics[width=9.0cm,height=6.0cm]{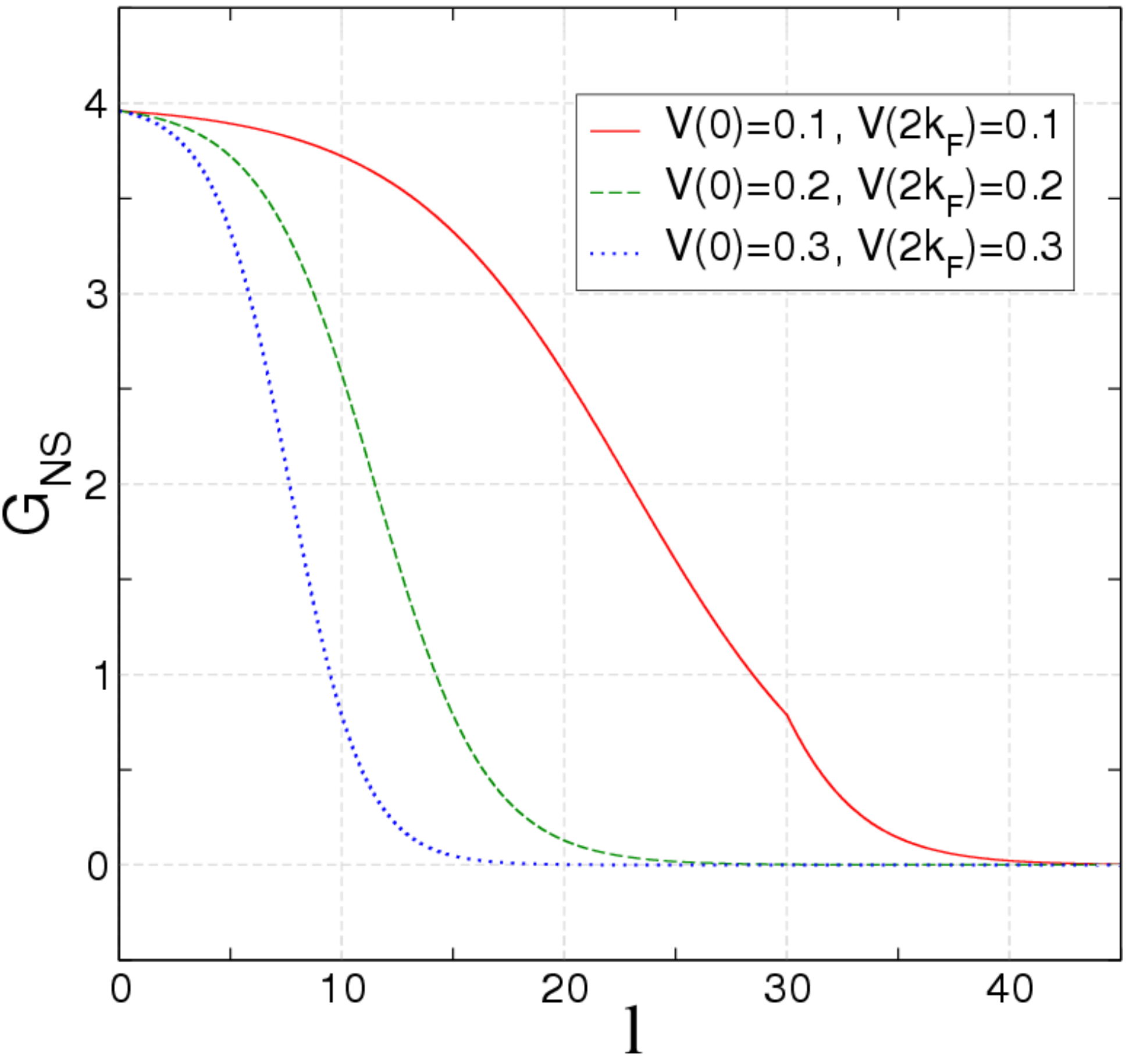}
\end{center}
\caption{(Color online) Conductance of the \ns junction is plotted in units of $e^2/h$ as a function of the 
dimensionless parameter $l$ where $l=ln(L/d)$ and $L$ is either $L_{T}=\hbar v_{F}/k_{B} T$ 
at zero bias or $L_{V}=\hbar v_{F}/eV$ at zero temperature and $d$ is the short distance 
cut-off for the \rg flow. The three curves correspond to three different values of $V(0)$ and $V(2k_{F})$.
Figure adapted from Ref.~\protect\refcite{75}.}
\label{fignine}
\end{figure}

We start with the results for the \ns junction, just to contrast with
the results of the \nsn junction in the next case. Here, we have only 
two parameters, $r$ and $r_A$. The conductance occurs only due to the
\ar amplitude, $r_A$ which obeys the \rg equation given by Eq.~\ref{rans}. 
As mentioned earlier, there is no flow of the particular linear combination 
of the interaction parameters $2g_2 - g_1$ that occurs in the equation and
hence the \rg flow of the conductance is therefore monotonic. The conductance 
as a function of the length scale for different interaction parameters 
$V(0)$ and $V(2k_{F})$ is plotted in Fig.~\ref{fignine}. Here $L$ simply denotes 
the length at which the \rg is cut-off. So if we take very long wires $L_W \gg L_T$,
then the cut-off is set by the temperature, and the plot shows the variation 
of the conductance as a function of $L_T$ starting from the high temperature limit, 
which here is the superconducting gap $\Delta$. We observe that as we lower the temperature, 
the Andreev subgapped conductance decreases monotonically with a power law set by the
\lln parameter and finally becomes zero when the \qw is disconnected from the 
superconducting junction at $r=1$ stable fixed point. Also it was established in
Ref.~\refcite{59} that the power law scaling of conductance ($~|r_{A}|^{2}$) calculated 
from \wirg and bosonization were found to be in  agreement with each other for the 
limiting cases of $|r_{A}|^{2} \cong 1$ and $~|r_{A}|^{2}\cong 0$ (which are the only 
limits where bosonisation results are valid) provided effects due to electron-electron 
induced back-scattering in the wires is neglected.
\vskip +1.0cm

\subsubsection{Ballistic \nsn Junction \label{sec:3.2.2}}
Here, we consider the case of a reflection-less ballistic junction between 
the superconductor and the two \qwsd, \ie~$r=0$. This implies that the renormalization 
of the \smat-matrix due to the Friedel oscillations is absent. The only
renormalization of the \smat-matrix elements occur due to reflections from the 
proximity effect induced pair potential inside the \qwd.
Let us now consider various interesting cases:\\
\begin{enumerate}
\item[] {\textsl{(a)}} $r = 0, r_A = 0, t \neq 0, t_A \neq 0$.\\
In this case, since we have both $r = 0$ and $r_A = 0$, there is
no \rg flow of the transmissions and the conductance is frozen at
the value that it had for the bare \smat-matrix. The most
interesting situation in this case arises when $t = t_A$. For this
case, the probability for an incident electron in one wire, to
transmit in the other wire as an electron due to $t$ or as hole
due to $t_A$ is equal, leading to perfect cancellation of charge
current.
\vskip +0.3cm
\item[] {\textsl{(b)}} $r = 0, t = 0, r_A \neq 0, t_A \neq 0$.\\
For this case, one can easily check from the \rg equations
(Eqs.~\ref{rnsn}-\ref{tansn}) that if we start our \rg flow with
the given parameters at high energies, then the value of $r,t$
remain stuck to the value zero under the \rg flow. Hence, in this
case the two parameter subspace $r_A \neq 0, t_A \neq 0$ remains
secluded under the \rg flow. The \rg equation for $t_A$ is given
by
\begin{eqnarray}
\dfrac{dt_A}{dl} &=& \alpha' \, t_A \, (1 - |t_A|^2)\ .
\label{33}
\end{eqnarray}
The above equation can be integrated to obtain an expression for
\car probability ($T_A = |t_A|^2$),

\bea 
T_{A}(L)=\frac{T_{A}^{0} \left[[1+2\alpha_{1}
\ln(\frac{L}{d})]^\frac{3}{2}(\frac{d}{L})^{-(2\alpha_{2}
-\alpha_{1})}\right]}{R_{A}^{0}+T_{A}^{0}\left[[1+2\alpha_{1}
\ln(\frac{L}{d})]^\frac{3}{2}(\frac{d}{L})^{-(2\alpha_{2}
-\alpha_{1})}\right]} \ .
\label{ta} 
\eea 
where $T_A^0$ and $R_A^0$ are the \car and \ar probabilities respectively 
at the short distance cut-off, $L=d$.
We notice that the \rg equation and its solution are very similar to
that for the single scatterer problem~\cite{67} apart from a sign difference 
on the RHS of Eq.~\ref{ta} and the dependance of the interaction parameter 
$\alpha'$ on $g_1$ and $g_2$. This implies that even if we start with a 
small crossed Andreev transmission across the junction, the \rg flow will 
take us towards the limit of perfect transmission. This is in sharp contrast 
to the normal transmission across a single scatterer. For the single barrier
problem, the equation for the \rg flow of $t$ was by
\begin{eqnarray}
\dfrac{dt}{dl} &=&- \alpha \, t \, (1 - |t|^2) \ .
\end{eqnarray}
Hence, $t=0$ is the stable fixed point in this case. But if the $\ed$-$\ed$
interactions had been attractive, then the sign on the RHS would have been positive 
and $t=1$ would have been the stable fixed point. Thus, the \rg flow of $t_A$ for 
the case when $r_A\ne 0,t=r=0$, and repulsive interactions, is very similar to the 
\rg flow for $t$ when $r\ne 0,t_A=r_A=0$ but with attractive interactions.  
In both cases transmission is relevant and $t=1$ and $t_{A}=1$ are the stable fixed
points. On the other hand the \rg flow of $t_{A}$ for the case of $r_{A} \neq 0, t=0, r=0$ 
and attractive $\ed$-$\ed$ interaction ($V(0), V(2k_{F})<0$) in the wire is very similar 
to the \rg flow for $t$ for the case $r \neq 0, r_{A}=0, t_{A}=0$ and repulsive $\ed$-$\ed$
interaction ($V(0), V(2k_{F})>0$). In both cases transmission is irrelevant and $t=0$ and 
$t_{A}=0$ are the stable fixed points. At an intuitive level, one can perhaps say that 
even if we start with repulsive inter-electron interactions inside the \qwd, the proximity-induced 
pair potential leads to a net attractive interaction between the electrons, which is responsible 
for the counter-intuitive \rg flow. 

Also note that while solving the above \rg equation for $t_A$, we have to take into account the 
\rg flow of the interaction parameter ($\alpha^\prime$) itself. This will lead to
non-power law (non Luttinger) behavior for the conductance close to $|t_A|\,\simeq\,0$ or $|t_A|\,\simeq\,1$. 
It is worth pointing out that the non-power law part appearing in Eq.~\ref{ta} is identical to Ref.~\refcite{67}, 
even though the interaction parameter for their case was proportional to $g_2-2g_1$ and for our case it is $g_2+g_1$.
But of course the latter will not lead to any non-monotonic behavior as $\alpha^\prime$ can not change sign under 
\rg flow. So the stable fixed point for this case is the \cafpd.
\vskip +0.3cm
\item[] {\textsl{(c)}} $r = 0, t_A = 0, r_A \neq 0, t \neq 0$.\\
This case is identical to the case (b) discussed above except for the fact that we have to 
replace $t_A$ in the previous case by $t$. In this case also the two parameter subspace $r_A \neq 0,
t\neq 0$ remains secluded under \rg flow. The \rg equation for $t$ is given by
\begin{eqnarray}
\dfrac{dt}{dl} &=& \alpha' t (1 - |t|^2)\ .
\end{eqnarray}
Here also, $t=1$ remains the stable fixed point and $t=0$ is the unstable fixed point.
\vskip +0.3cm
\item[] {\textsl{(d)}} $r = 0, t \neq 0, r_A \neq 0, t_A \neq 0$.\\
In this case if we start from a symmetric situation, \ie~$t=t_A$, we can see from the 
\rg equations in Eqs.~\ref{tnsn} and \ref{tansn} that both $t$ and $t_A$ have identical \rg flows. 
So, the sub-gap conductance $G = G_{CA} - G_{CT}$ vanishes identically and remains zero through 
out the \rg flow. Hence this \smat-matrix can facilitate production of pure \scd~\cite{76} if we
inject spin polarized electrons from one of the leads as the charge current gets completely filtered 
out at the junction. In Sec~\ref{sec:5} we shall discuss the \nsn and \fsn junction from the 
spintronics application point of view in greater details. 
\end{enumerate}

\subsubsection{Ballistic \fsf Junction \label{sec:3.2.3}}
Here, we consider the case where both the wires are spin polarized. 
In this case we can have two interesting possibilities, \ie~either 
both the wires have aligned spin polarization (ferromagnetic) or they 
have them anti-aligned (anti-ferromagnetic). In either case the \ar
amplitude is zero on each wire due to reasons explained earlier.\\

\begin{enumerate}
\item[]{\textsl{(a)}} When the two wires have their spins aligned,
$t \ne 0$, but $t_A=0$ because for \car to happen we need up (down)
spin polarization in one wire and down (up) spin polarization on
the other wire which is not possible in this case.
\vskip +0.3cm
\item[]{\textsl{(b)}} When the two wires have their spins anti-aligned, 
then $t=0$, but $t_A\ne 0$ because the up (down) electron from one wire 
cannot tunnel without flipping its spin into the other wire. As there is 
no mechanism for flipping the spin of the electron at the junction, 
such processes are not allowed.
\end{enumerate}

Hence these two cases can help in separating out and measuring amplitudes of the 
direct tunneling process (\ctd) and the \car process experimentally~\cite{37,93}. 
Both these are examples of case II, since they have both $r=0$ and $r_A=0$. 
In this case, neither $t$ nor $t_A$ change under \rg flow and hence conductance 
is not influenced by $\ed$-$\ed$ interaction at all.

\subsubsection{Non-ballistic \nsn Junction without \ar on individual wires \label{sec:3.2.4}}
Here we consider an \nsn junction with finite reflection in each wire and no \ar in 
the individual wires. So the renormalization of the \smat-matrix is purely due to the 
Friedel oscillations and there are no contributions originating from scattering due to the 
proximity induced pair potential. Below we discuss two cases : \\

\begin{figure}[htb]
\begin{center}
\includegraphics[width=9.0cm,height=6.0cm]{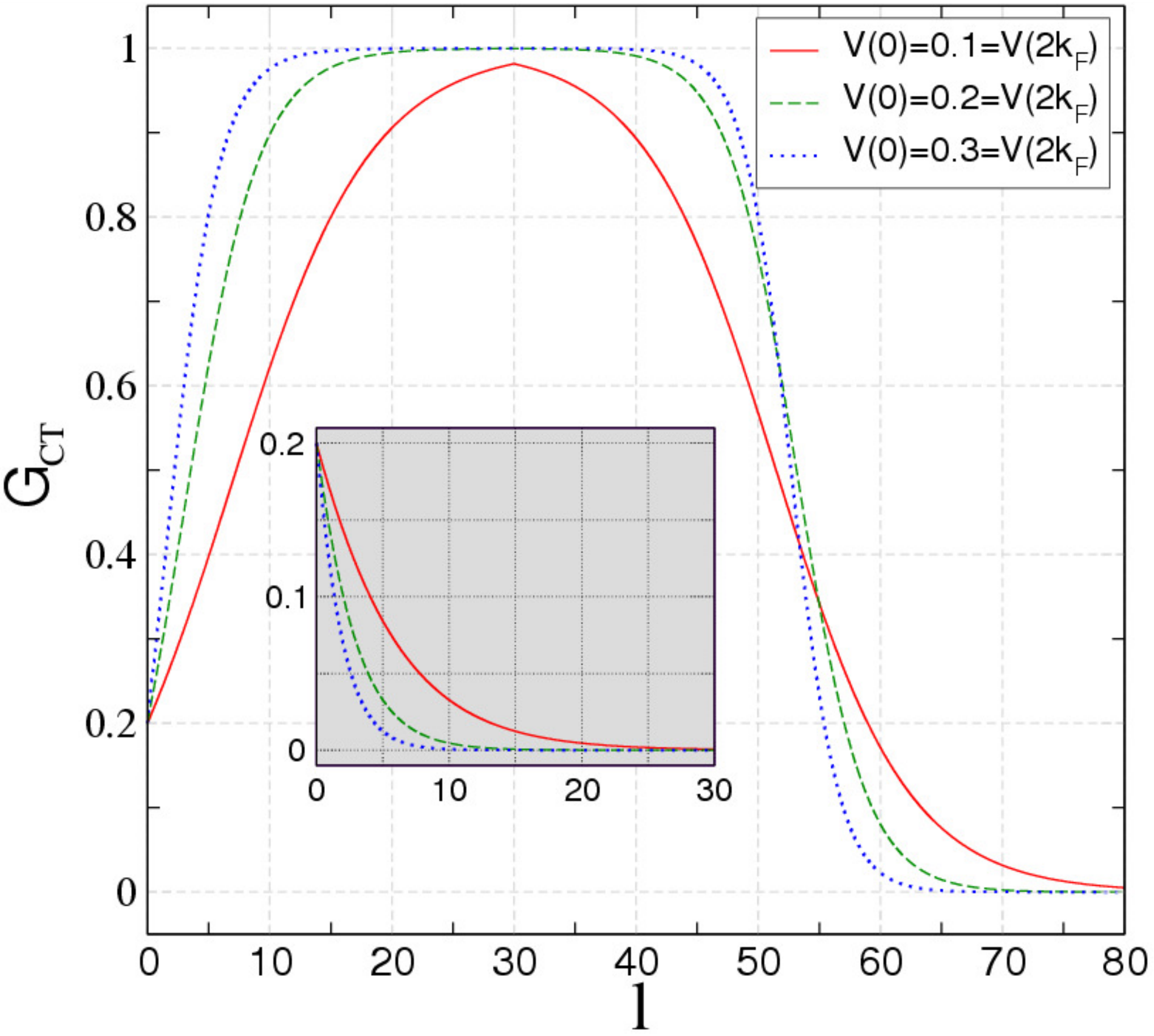}
\end{center}
\caption{(Color online) Conductance $G_{CT}$ of the \nsn junction (when the two leads have parallel spins) 
in units of $e^2/h$ as a function of the dimensionless parameter $l$ where $l=ln(L/d)$ and $L$ is either
$L_{T}=\hbar v_{F}/k_{B} T$ at zero bias or $L_{V}=\hbar v_{F}/eV$ at zero temperature and $d$ is the 
short distance cut-off for the \rg flow. The three curves correspond to three different values of $V(0)$ 
and $V(2k_{F})$. The inset shows the behavior of the same conductance for fixed values of $\alpha$ 
\ie~for the spinless case. Figure adapted from Ref.~\protect\refcite{75}.}
\label{figeleven}
\end{figure}

\begin{enumerate}

\item[]{\textsl{(a)}} $r_A = 0, t = 0, r \neq 0, t_A \neq 0$.\\
%
The \rg equations (Eqs.~\ref{rnsn}-\ref{tansn}) predict that $r_A,t$ will remain zero 
under the \rg flow and $r, t_A$ form a secluded sub-space. The \rg equation for this 
case is given by
\begin{eqnarray}
\dfrac{dt_A}{dl} &=& -\,\alpha\, t_A (1 - |t_A|^2)\ .
\label{targ}
\end{eqnarray}
Note the change in sign on the RHS with respect to the \rg equation for $t_A$ (Eq.~\ref{33}) 
for the ballistic case. This change in sign represents the fact that the ballistic case
effectively represents a situation corresponding to attractive $\ed$-$\ed$ interaction 
while the present case corresponds to a purely repulsive  $\ed$-$\ed$ interaction.
In Fig.~\ref{figten} we show the behavior of conductance ($G_{CA}$) for this case. 
The conductance in the main graph shows a non-monotonic behavior due to the running 
coupling constant $\alpha$ in Eq.~\ref{targ}.
\begin{figure}[t]
\begin{center}
\includegraphics[width=9.0cm,height=6.0cm]{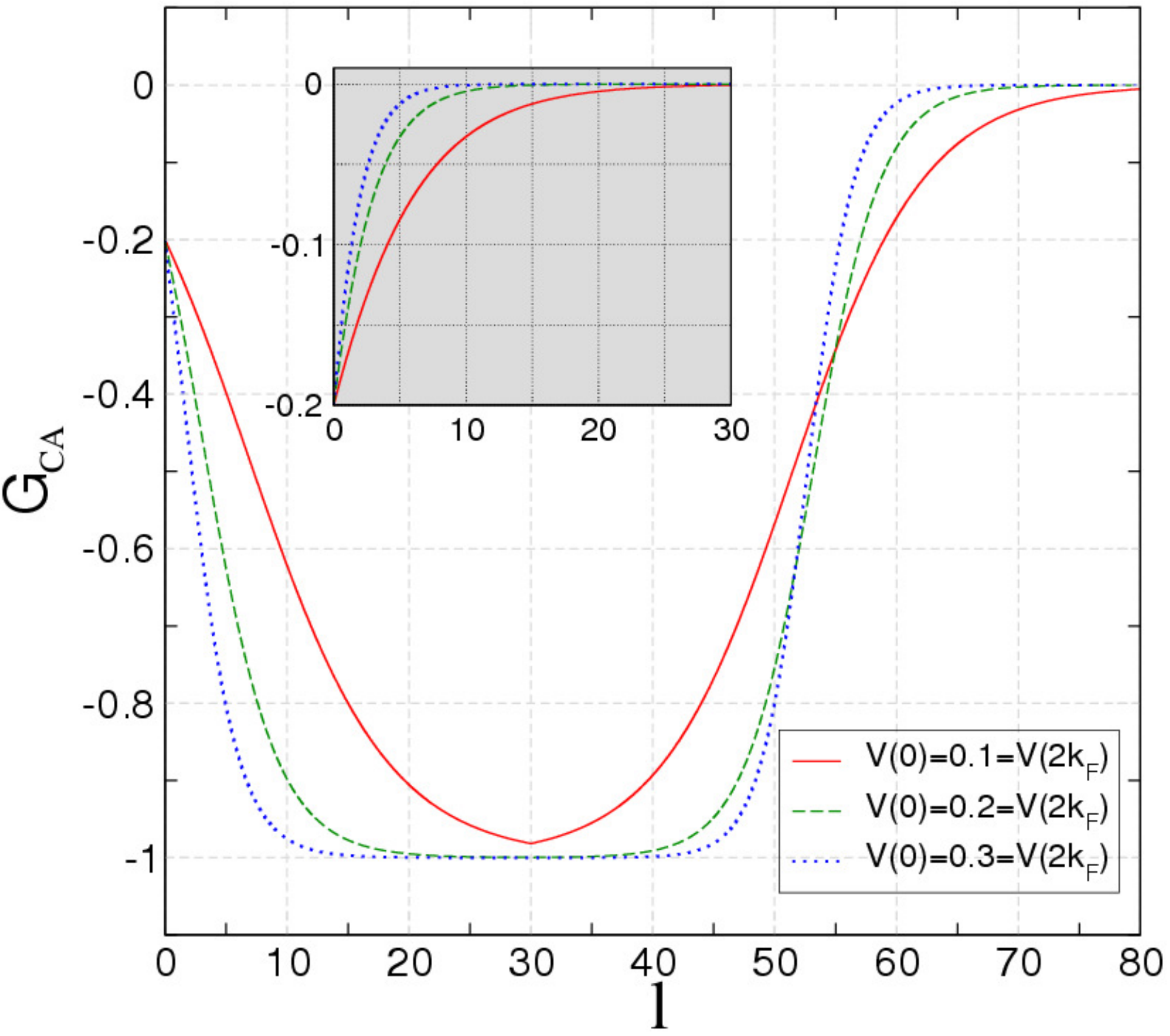}
\end{center}
\caption{(Color online) Conductance $G_{CA}$ of the \nsn junction is plotted (when the 
two leads have anti-parallel spins) in units of $e^2/h$ as a function of the dimensionless 
parameter $l$ where $l=ln(L/d)$ and $L$ is either $L_{T}=\hbar v_{F}/k_{B} T$ at zero bias or 
$L_{V}=\hbar v_{F}/eV$ at zero temperature and $d$ is the short distance cut-off for the \rg flow.
The three curves correspond to three different values of $V(0)$ and $V(2k_{F})$. The inset shows 
the behavior of the same conductance for fixed values of $\alpha$ \ie~for the spinless case.
Figure adapted from Ref.~\protect\refcite{75}.}
\label{figten}
\end{figure}
To contrast, we also show in the inset, the behavior of the conductance for the spinless case
when the renormalization of $\alpha$ in not taken into account. Thus, it is apparent from the plot 
that the non-monotonicity in the behavior of conductance manifests itself solely from the \rg evolution 
of the running coupling constant $\alpha$. The inset and the main graph, both start from the same value 
of $t_A$. Even though this case is theoretically interesting to explore, its experimental realization
may not be viable. This is because of the following reasons. Here we have $r_A = 0$ on both wires, 
which can only happen if the wires are ferromagnetic. However, we also know that if the wires are
ferromagnetic, there is no scaling of $\alpha$ parameter and hence there will be no interesting 
non-monotonic trend in the conductance. So it is hard to find a physical situation where $r_A = 0$ 
and at the same time, there is renormalization of the interaction parameter $\alpha$. 
Lastly note that the conductance $G_{CA}$ is negative. The process responsible for the 
conductance, (\ie~\card), converts an incoming electron to an outgoing hole or
vice-versa, resulting in the negative sign.

\vskip +0.3cm
\item[]{\textsl{(b)}} $r_A = 0, t_A = 0, r \neq 0, t \neq 0$.\\ 
This case is identical to the previous case with the replacement of $t_A$ by $t$. 
Fig.~\ref{figeleven} shows the the \ct conductance $G_{CT}$ as a function of the length scale. 
It shows a similar non-monotonic behavior with positive values for the conductance. The inset shows 
the behavior of $G_{CT}$ when the renormalization of $\alpha$ in not taken into account.
\end{enumerate}

\subsubsection{Non-ballistic \nsn Junction with \ar on individual wires \label{sec:3.2.5}}
\begin{figure}[htb]
\begin{center}
\includegraphics[width=9.0cm,height=6.0cm]{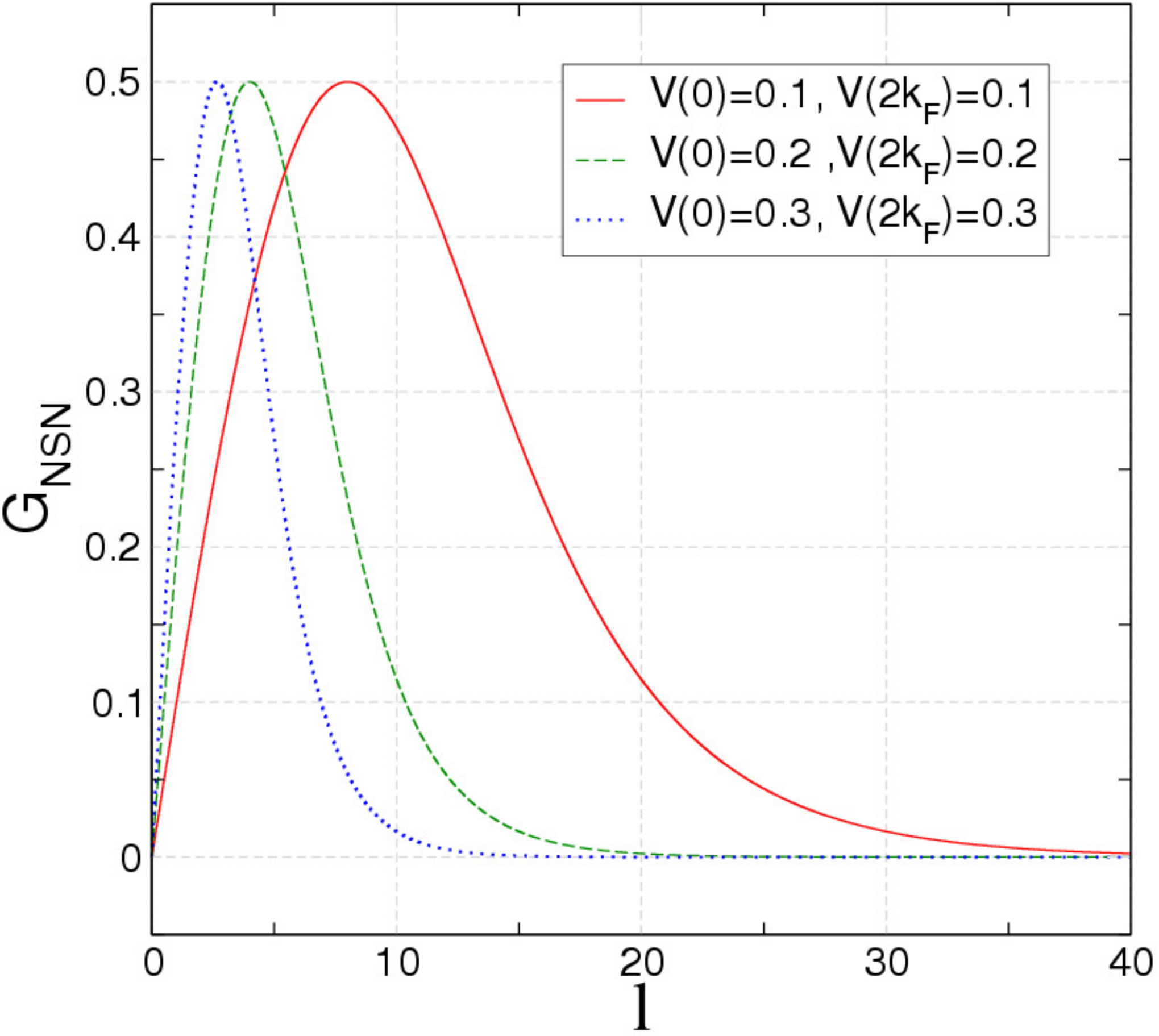}
\end{center}
\caption{(Color online) Conductance of the \nsn junction $G_{NSN}=|t_A|^2-|t|^2$ 
is plotted in units of $2e^2/h$ as a function of the dimensionless parameter
$l$ where $l=ln(L/d)$ and $L$ is either $L_{T}=\hbar v_{F}/k_{B} T$ at zero
bias or $L_{V}=\hbar v_{F}/eV$ at zero temperature and $d$ is the short
distance cut-off for the \rg flow. The three curves correspond to three
different values of $V(0)$ and $V(2k_{F})$. Figure adapted from Ref.~\protect\refcite{75}.}
\label{figtwelve}
\end{figure}

For this case $r_A \neq 0, t_A \neq 0, r \neq 0, t \neq 0$ \ie~all the scattering amplitudes are non zero. 
This is the most interesting case, where both $r$ and $r_A$ are non-zero, and we get an interplay of the 
effects due to scattering from Friedel oscillations and from
proximity induced pair potential inside the \qwd. Here, all the four parameters are non-zero and flow under \rgd, 
as do the interaction parameters $\alpha$ and $\alpha'$. An example where the system starts in the
vicinity of the unstable fixed point {\textsf{SFP}} (as mentioned in Case IV in the Subsection.~\ref{3.1.2}) 
is shown in Fig.~\ref{figtwelve}. The \nsn conductance here is defined as $G_{NSN} = G_{CA} - G_{CT}$. 
Here also we observe a strong non-monotonicity in the conductance which comes about due to the
interplay of the electron and the hole channels taking part in transport, which contribute
to the conductance with opposite signs, coupled with the effects from the \rg flow of the interaction parameters.

\subsubsection{Non-ballistic \fsn Junction \label{sec:3.2.6}}
In this case, for the ferromagnetic wire $r_A = 0$, as for an up spin 
polarized \qwd, the incident electron can't pair up with a spin
down electron in the same \qwd. On the other hand for the
normal wire $r_A$ has a finite value. As explained earlier, the
interaction parameters $\alpha$ and $\alpha'$ on the ferromagnetic
side do not renormalize, whereas they do on the normal side.
Hence, even if we start from a situation where the interaction
parameter $\alpha$ and $\alpha'$ are symmetric for the two wires,
\rg flow will always give rise to an asymmetry in the interaction
strength. Therefore, it becomes a very interesting case to study
theoretically. The \smat-matrix for this case has neither 
spin up-spin down symmetry, nor the wire index (left-right for two wires) symmetry. 
Only the particle-hole symmetry can be retained while parameterizing the \smat-matrix. 
The latter case gets very complicated to study theoretically because the
minimum number of independent complex-valued parameters that are
required to parameterize the \smat-matrix is nine as opposed to four 
in the \nsn case. These are given by
$r_{\up\up}^{11}, ~r_{\up\up}^{22}, ~r_{\dn\dn}^{22},
~t_{A\up\up}^{12}, ~t_{A\dn\dn}^{21}, ~r_{A\up\up}^{22},
~r_{A\dn\dn}^{22}, ~t_{\up\up}^{12}, ~{\mathrm{and}}~
t_{\up\up}^{21}$.
Here, $1$($2$) is the wire index for the ferromagnetic
(normal) wire while, $\up$ and $\dn$ are the respective
spin polarization indices for the electron.

\begin{figure*}[htb]
\begin{center}
\includegraphics[width=9.0cm,height=6.0cm]{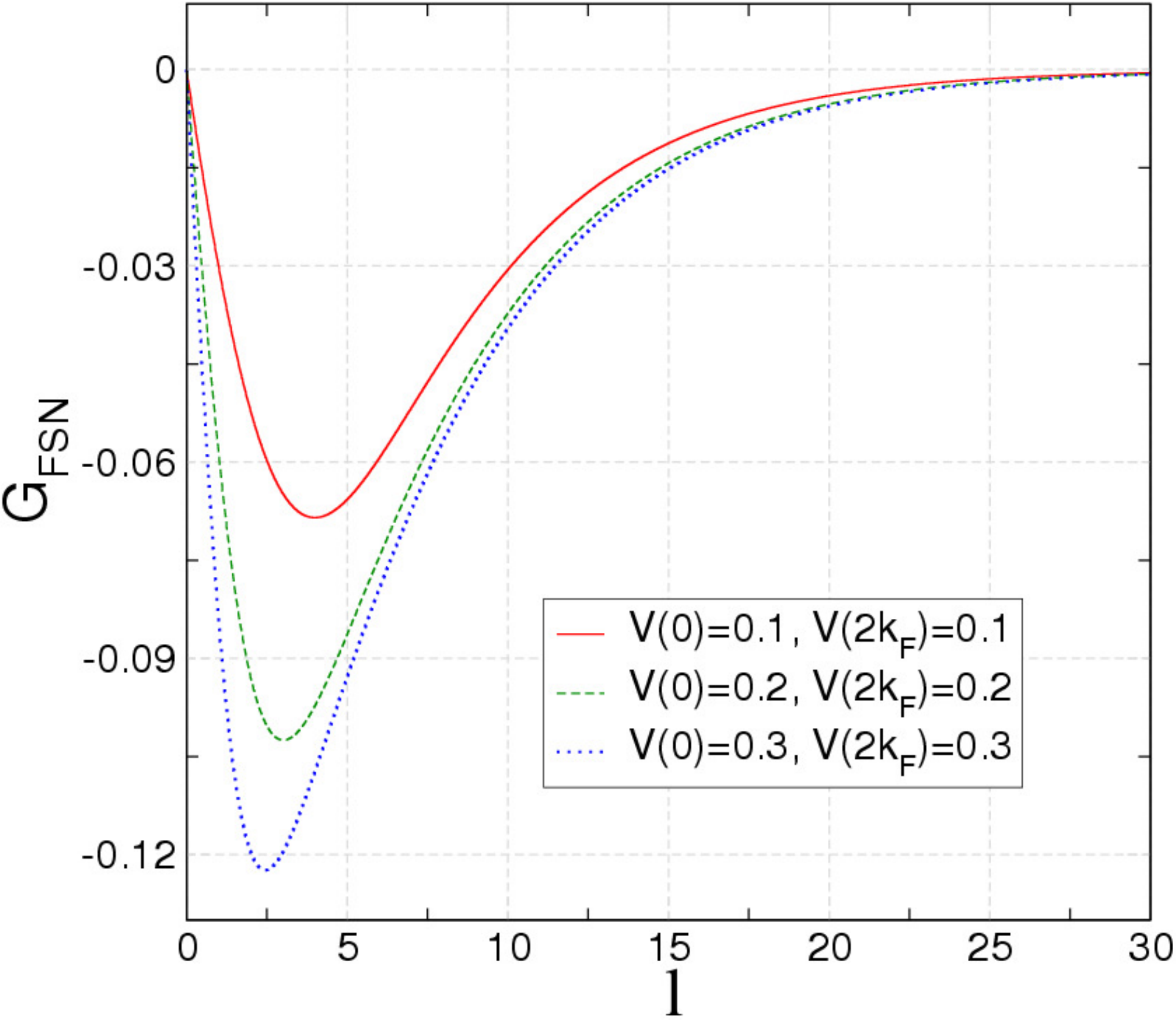}
\end{center}
\caption{(Color online) Charge conductance $G_{FSN}=|t_A|^2-|t|^2$ is plotted 
in units of $e^2/h$ for \fsn junction as a function of the dimensionless parameter 
$l$ where $l=ln(L/d)$ and $L$ is either $L_{T}=\hbar v_{F}/k_{B} T$ at zero bias or 
$L_{V}=\hbar v_{F}/eV$ at zero temperature. Here $d$ is the short distance cut-off 
for the \rg flow. The three curves correspond to three different values of $V(0)$ and $V(2k_{F})$. Figure adapted from Ref.~\protect\refcite{75}.}
\label{figthirteen}
\end{figure*}

So, the minimal \smat-matrix representing the \fsn junction is given by
\bea 
{\mathbb {S}} ~=~
\begin{bmatrix}
~r &   t & 0 &  0 & t_{A} & 0~ \\
~t^\prime &  r^\prime & 0 &  0 & r_{A} & 0~\\
~0 &  0 & r^{\prime \prime} & t_{A}^\prime & 0 & r_{A}^\prime~\\
~0 &  0 & t_{A} &  r & 0 & t~\\
~t_{A}^\prime &  r_{A}^\prime &  0 & 0 & r^{\prime \prime} & 0~\\
~0 &  0 & r_{A} &  t^\prime & 0 & r^{\prime}~\\
\end{bmatrix} \ ,
\label{smat4}
\eea
%
In Eq.~\ref{smat4}, we write down a representative \smat-matrix which satisfies 
all the constraints of the \fsn junction mentioned above as well as unitarity.
We study its \rg flow numerically by solving the nine coupled differential
equations. The modulus of the \smat-matrix elements are given by  
$|r_{\up\up}^{11}|=
|r_{\up\up}^{22}|=|r_{\dn\dn}^{22}|=|t_{A\up\up}^{12}|=
|t_{A\dn\dn}^{21}|=|r_{A\up\up}^{22}|=|r_{A\dn\dn}^{22}|
=|t_{\up\up}^{12}|=\lvert t_{\up\up}^{21}\rvert=1/\sqrt{3}$
and the corresponding phases associated with each of these amplitudes are 
$\pi/3,\pi,0,-\pi/3,0,\pi/3,0,\pi,-\pi/3$ respectively. Here also we observe 
a non-monotonic behavior of conductance, $G_{FSN}$ as a function of $l$ as 
shown in Fig.~\ref{figthirteen}.
\subsubsection{Non-ballistic \fsf junction \label{sec:3.2.7}}
Here, we consider the case where both the wires are spin polarized.
This case is similar to the ballistic case. Since here $r \ne 0$ and $r_A = 0$, 
we will have the usual Friedel oscillations and the conductance will go to zero 
following the stable fixed point as a power law. Here again we have two 
instructive cases:
%
\begin{enumerate}
\item[]{\textsl{(a)}} When the two wires connected to the superconductor have their 
spin polarization aligned, \ie~$t \ne 0$, but $t_A=0$ and
\vskip +0.3cm
\item[]{\textsl{(b)}} When the two wires have their spin polarization anti-aligned, 
\ie~$t=0$ but $t_A\ne 0$.
\end{enumerate}
Both these cases are examples of case II of Subsubsection.~\ref{sec:3.2.3}. 
Either $t$ or $t_A$ need to be zero in the two cases mentioned above. 
Hence the parameters which are zero will remain zero under \rg, while the non-zero
parameters will flow according to Eqs.~\ref{tnsn} and \ref{tansn} respectively. 
The conductances are the same as in the \nsn case except that the interaction parameters 
cannot flow now. The latter has already been emphasized in the insets in Figs.~\ref{figten}
and \ref{figeleven}. Since the electrons are now effectively spin-less, $\alpha$ and
$\alpha'$ do not flow, and we get a monotonic fall-off of the conductance in both the cases.

The results of this section are summarized in the table below for the readers.
\hskip -5.2217cm
\begin{table}[ph]   
{In this table we summarize all the relevant fixed points of the theory.}
\begin{tabular}{|c|c|c|c|c|c|c|}
\hline
$t$ & $t_{A}$ & $r_{A}$ & $r$ & Stability & Intermediate fixed point & Relevant physics\\ \hline
0 & 0 & 0 & 1 & Stable & $\times$ &  \rfpd \\ \hline
0 & 0 & 1 & 0 & Unstable & $\times$ &  \afpd \\ \hline
0 & 1 & 0 & 0 & Unstable & $\times$ &  \cafpd \\ \hline
1 & 0 & 0 & 0 & Unstable & $\times$ &  \tfpd \\ \hline
1/2 & 1/2 & -1/2 & 1/2 & Unstable & $\surd$ & \sfpd, Non-monotonic charge current\\ \hline
$e^{i \phi_{1}}\sin \theta$ & $e^{i \phi_{2}}\cos \theta$ & 0 & 0 &
Marginal & $ - $ & Pure spin current when $t=t_{A}$ \\ \hline
\end{tabular}
\end{table}
\subsubsection{Three wires$-$The Beam Splitter \label{sec:3.2.8} }
\begin{figure}[htb]
\begin{center}
\includegraphics[width=9.0cm,height=6.0cm]{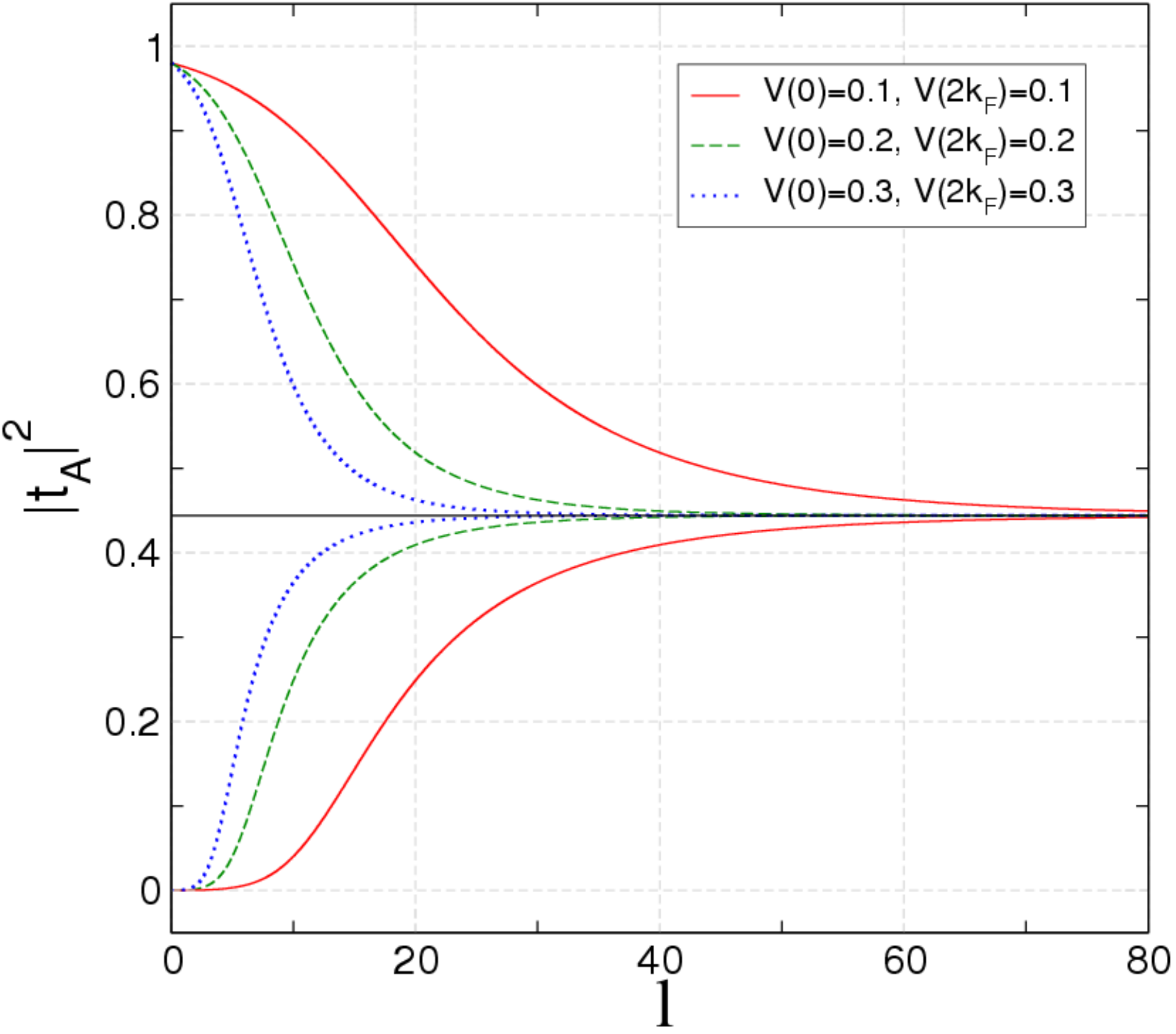}
\end{center}
\caption{$|t_A|^2$ is plotted as a function of dimensionless parameter
$l$ where $l=ln(L/d)$ and $L$ is either $L_{T}=\hbar v_{F}/k_{B} T$ at zero
bias or $L_{V}=\hbar v_{F}/eV$ at zero temperature and $d$ is the short
distance cut-off for the \rg flow. The three curves correspond to three
different values of $V(0)$ and $V(2k_{F})$.
The set of curves in the top represent the \rg flow of $|t_A|^2$ when the
starting point is in the vicinity of $r_A = 0$ fixed point while, the set
of curves in the bottom half represent the \rg flow of $|t_A|^2$ with
starting point close to $r_A = -1$. Figure adapted from Ref.~\protect\refcite{75}.}
\label{figfourteen}
\end{figure}

In this subsubsection, we consider the case of three \qws connected to
a superconductor deposited on top of them. We assume that all the wires 
are connected within the phase coherence length of the superconductor. 
Hence, \car can occur by pairing the incident electron with an electron 
from any of the other wires and emitting a hole in that wire. The 
conductance matrix can hence be extended for three wires as
\bea
\begin{bmatrix}
~I_1~ \\
~I_2~ \\
~I_3~ \\
\end{bmatrix} = \begin{bmatrix}
~G_{r\,11} & G_{t\,12} & G_{t\,13}~  \\
 ~G_{t\,12}  & G_{r\,22} &  G_{t\,23}~ \\
 ~G_{t\,13} &  G_{t\,23} &
G_{r\,33}~
\\
\end{bmatrix}
\begin{bmatrix}
~V_1~ \\
~V_2~ \\
~V_3~\\
\end{bmatrix} \ ,
\eea 
with
   %
$(G_{{CA\,ij}}+G_{{ct\,ij}})$
$G_{r\,ij} = G_{{A\, ii}} + \sum_{j}(G_{{CA\,ij}}+G_{{CT\,ij}})$
and
$G_{t\,ij} = G_{{CA\,ij}} - G_{{CT\,ij}}$. Also the generalization
to $N$ wires is obvious. Note that the conductances $G_{CA\,ij} =
G_{CA\,ji}$ and $G_{CT\,ij} = G_{CT\,ji}$. The relations of the
conductances to the reflections and transmissions is obvious, 
for e.g., $G_{A\,ii} \propto |r_{A\,ii}|^2$ as before while
$G_{CT\,ij} \propto |t_{ij}|^2$ and $G_{CA\,ij} \propto
|t_{A\,ij}|^2$. The \rg equations for the three wire case can be
written using the matrix equation as given in Eq.~\ref{smat1}
except that the \smat-matrix is now $12 \times 12$ dimensional. 
For a system with particle-hole, spin up-spin down and wire index
symmetry, the \smat-matrix is given by, 
\bea 
{\mathbb {S}}_\uparrow =
{\mathbb {S}}_\downarrow =
\begin{bmatrix}
~r & t & t^\prime & r_{A} & t_{A} & t_{A}^\prime~\\
~t &r&t & t_{A} & r_{A} & t_{A}~\\
~t^\prime & t & r & t_{A}^\prime & t_{A} & r_{A} ~\\
~ r_{A} & t_{A} & t_{A}^\prime &r & t & t^\prime~\\
~t_{A} & r_{A} & t_{A}& t & r& t~\\
~t_{A}^\prime & t_{A}& r_{A} &t^\prime &t& r~\\
\end{bmatrix} \ ,
\label{smat3} 
\eea 
where we have chosen six independent parameters, with $t_{12} =
t_{21} = t_{23} = t_{32} = t$ and $t_{13} = t_{31} = t'$
and similarly for the \car parameter $t_A$.
%
The $F$ matrix now generalizes to
\bea
F = \begin{bmatrix}
~\frac{\alpha r}{2}& 0 & 0 & \frac{-\alpha^\prime r_A} {2} &0 &0 ~\\
~0& \frac{\alpha r}{2}& 0 & 0 & \frac{-\alpha^\prime r_A}{2}&0  ~\\
~0& 0&\frac{\alpha r}{2}&0 & 0 & \frac{-\alpha^\prime r_A}{2} ~\\
~\frac{-\alpha^\prime r_A}{2} & 0 & 0& \frac{\alpha r}{2}& 0&0 ~\\
~0& \frac{-\alpha^\prime r_A}{2} & 0 & 0 & \frac{\alpha r}{2}&0 ~\\
~0&0&\frac{-\alpha^\prime r_A}{2} & 0 & 0 & \frac{\alpha r}{2} ~\\
\\
\end{bmatrix} \ ,
\label{fmat2} 
\eea
There exists possibility of emergence of many more non-trivial fixed points 
in this case. For instance, the \agfpd, as mentioned in Subsubsection.~\ref{3.1.4}. 
As discussed in Subsubsection.~\ref{3.1.4}, for the reflection-less case with
symmetry between just two wires, this complicated \smat-matrix described
by Eq.~\ref{smat3} takes a very simple form, which can be dealt analytically. 
Within the sub-space considered we found that the \agfp was a stable
fixed point. In Fig.~\ref{figfourteen}, we show the \rg flow of $|t_A|^2$ 
from two different unstable fixed points to the stable \agfpd.

The possibility of experimental detection of such a non-trivial fixed point 
with intermediate transmission and reflection is quite interesting. 
From this point of view, the \agfp is a very well-suited candidate as 
opposed to its counterpart, the Griffith's fixed point~\cite{27,65}. 
For a normal junction of three \od \qwd, the \smat-matrix corresponding 
to $r=-1/3, t=2/3$ is a fixed point (\gfpd), where $r$ and $t$ are the
reflection and the transmission for a completely symmetric three wire junction.
Even though it is an interesting fixed point, it turns out to
be a repulsive one and hence the possibility of its experimental detection 
is very low. On the contrary, the \agfpd, being an attractive fixed point, 
has a better possibility of being experimentally measured. The main
point here is that even if we begin with an asymmetric junction,
which is natural in a realistic experimental situation, the effect of interaction 
correlations inside the \qws are such that as we go down in temperature, 
the system will flow towards the symmetric junction. This feature can be inferred 
from the results shown in Fig.~\ref{figfourteen}.
\section{\label{sec:4} Stability analysis of the \rg flow for the \nsn junction of \lln wires}
In this section we study the \rg flows of the two terminal conductance of a superconducting junction 
of two \lln wires~\cite{77}. In particular we perform the stability analysis of the various fixed points 
of the theory discussed in Sec.~\ref{sec:3} and compute the power laws associated with the \rg flow around them. 
Our analysis also provides the renormalized values of the various transmission and reflection amplitudes 
around these fixed point values which can then be used to obtain the \lb conductances.
\subsection{\label{sec:4.1} Method of stability analysis}
To carry out our stability analysis we adopt the \rg method followed in Sec.~\ref{sec:2} and \ref{sec:3} 
where an \smat-matrix formulation was used to compute the linear conductance and inter-electron interactions 
inside the \qw were taken into account by allowing the \smat-matrix to flow as a function of the relevant energy 
scale (like temperature, bias voltage or system size) using an \rg procedure. This method works well when
$\ed$-$\ed$ interaction strength inside the \qw is weak so that it can be treated perturbatively.

In our stability analysis we mainly focus on three different fixed points$-$ the chiral fixed point (\cfpd) and \gfp 
of a normal junction of three \lln wires~\cite{27,99} and the \sfpd~\cite{75} 
of the \nsn junction also introduced in Sec.~\ref{sec:3}. First we discuss the stability around 
the \cfp and \gfp of a normal junction of three \lln wires (Y-junction) to benchmark our calculation 
with known results~\cite{27}. As a first step towards performing a systematic stability analysis, 
we need to obtain an \smat-matrix which results from a very small unitary deviation around the fixed point 
\smat-matrix. Given the number of independent parameters of the \smat-matrix dictated by symmetry and unitarity
constraints, the most general deviation from the fixed point \smat-matrix can be obtained by multiplying the 
fixed point ${\mathbb{S}}$-matrix by another unitary matrix which is such that it allows for a straightforward 
expansion in terms of small parameters around the identity matrix. The latter is realized as follows$-$
\bea 
{\mathbb{S}} = {\mathbb{S}}_{0} \exp\left\{i \sum_{j=1}^{9}
\epsilon_{j}\lambda_{j}\right\}~, 
\label{smatn} 
\eea
where $\mathbb{S}_0$ represents the fixed point \smat-matrix and $\lambda_{j}$'s 
(along with the identity $\lambda_0 =I$) are the eight generators of the $SU(3)$ 
group which are traceless hermitian matrices. This can be straightforwardly
generalized to the case of $N$ wires by using $SU(N)$ matrices. Perturbations around 
these fixed points are characterized in terms of the $\epsilon_{j}$'s. Of course, 
the resulting \smat-matrix obtained in this way corresponds to a small unitary 
deviation when $\epsilon_{j}$'s are small parameters. To first order in $\epsilon_{j}$'s, 
Eq.~\ref{smatn} reduces to
\bea 
{\mathbb{S}} = {\mathbb{S}}_{0}\left({\mathbb{I}} + i \sum_{j}
\epsilon_{j}\lambda_{j}\right)~, 
\label{smatd} 
\eea
where ${\mathbb{S}}_{0}$ for \cfp and \gfp fixed points are given by~\cite{27}

\bea
{\mathbb{S}}_{0}^{\cfp} = \begin{bmatrix} ~0 & 1 & 0 ~\\
~0 & 0 & 1 ~\\
~1 & 0 & 0 ~
\end{bmatrix}~; ~
{\mathbb{S}}_{0}^{\gfp} =
\begin{bmatrix} ~-1/3 & 2/3 & 2/3 ~\\
~2/3 & -1/3 & 2/3 ~\\
~2/3 & 2/3 & -1/3 ~
\end{bmatrix}~,
\label{sgfp} 
\eea 
respectively. Using Eq.~\ref{smatd}, the \rg equation (Eq.~\ref{77} given in Sec.~\ref{sec:2}) with 
$\mathbb{S}$ expanded to the linear order in $\epsilon_j$ becomes
\bea 
i \sum_{j=1}^{9} \lambda_{j} \dfrac{d\epsilon_{j}}{dl} &=&
\mathbb{S}_{0}^{\dagger} \Big[
\mathbb{I}-i\sum_{j}\epsilon_{j}\lambda_{j} \Big] ~ 
\bigg\{\mathbb{F}-\mathbb{S}_{0}
\Big[\mathbb{I}+i\sum_{j}\epsilon_{j}\lambda_{j}\Big] 
\mathbb{F}^{\dagger} \Big[\mathbb{I}+i\sum_{j}
\epsilon_{j}\lambda_{j}\Big]\bigg\}~, 
\label{eqnm} 
\eea
where $\mathbb{F}$ is the diagonal part of the following quantity
\bea 
{\mathbb{F}}=\dfrac{\alpha}{2}~
\mathbb{S}_{0}\Big[\mathbb{I}+i\sum_{j}\epsilon_{j}\lambda_{j}\Big]_{\rm {diagonal}}~. 
\label{fmatd} 
\eea
By restricting the RHS of Eq.~\ref{eqnm} to linear order in $\epsilon_j $'s,
one then obtains nine coupled linear differential equations. At the next step, 
by applying a unitary rotation, we can decouple these coupled equations
(Eq.~\ref{eqnm}) and re-cast them in terms of new variables $\epsilon_j^\prime$ 
(which are linear combinations of the original $\epsilon_j$). The equations are 
now given by 
\bea 
\dfrac{d\epsilon_{j}^{\prime}} {dl} = \mu_j \, \epsilon_{j}^{\prime} 
\eea 
where $\mu_j$ is a real number corresponding to the `power law' associated with perturbations 
turned on along each of the new nine eigen-directions $\epsilon^\prime_j$. $\mu_j<0$ indicates 
that the given direction is stable and $\mu_j>0$ indicates that it is unstable.
Here the non-diagonal $\epsilon_{j}$ are related to the diagonal $\epsilon_{j}^{\prime}$ by  
$\epsilon_{j}=\sum_{i} \mathbb{U}_{ji} \, \epsilon_{i}^{\prime}$ where $\mathbb{U}$ is the 
diagonalizing rotation matrix.

\subsection{\label{sec:4.2} Power laws around different fixed points}
Using the method discussed above we obtain all the power laws associated with the 
independent perturbations that can be switched on around a given fixed point \smat-matrix.
Now it is straightforward to show that the power laws associated with the \cfp and \gfp are 
given by [$\alpha/2$, $\alpha/2$, 0, $\alpha/2$, $\alpha/2$, $\alpha/2$, $\alpha/2$, 0, 0] and
[0, 0, 0, 0, 0, -$\alpha/3$, $2\alpha/3$, $2\alpha/3$, $\alpha$] respectively which is consistent 
with results obtained in Ref.~\refcite{27}~\footnote{There is a correction to the power laws obtained 
for the \gfp for the three-wire junction in Ref.~\refcite{27}. Lal \etal~had predicted a stable direction
with power law $-\alpha$ which should be corrected to $-\alpha/3$~\cite{100} as is obtained 
in this section. The corrected power law is also consistent with that obtained in Ref.~\refcite{99} 
using a functional \rg procedure.}. The value zero corresponds to marginal directions while 
the values with positive or negative signs correspond to stable or unstable directions respectively. 
We do not write the explicit form of the $\mathbb{U}$ matrix for the \cfp and \gfp as they are
needed only for obtaining the explicit form of the power law correction to the fixed point conductance 
which is beyond the scope of this review.

Finally, let us discuss the stability around the different \rg fixed points of the \nsn junction. 
First we focus on the \sfp of the \nsn junction which has intermediate reflections and transmissions. 
As discussed in Sec~\ref{sec:3} in presence of $\ed$-$\ed$ interaction inside the \qw the incident 
electron (hole) not only scatters from the Friedel oscillations as an electron (hole) but also 
scatters from the proximity induced pair potential inside the \qw as a hole (electron). 
Now the amplitude of both of these scattering processes are proportional to the $\ed$-$\ed$
interaction strength inside the \qwd. The interplay between these two scattering processes which 
actually arise due to the $\ed$-$\ed$ interaction strength inside the \qwd, renormalize the bare 
scattering amplitudes at the junction and give rise to this new \sfp where all the scattering amplitudes 
have intermediate non-zero values. This fact is unique about this fixed point and hence this fixed point 
is the central focus of our discussion here.

We adopt the same procedure as described above for the three-wire junction but now with $SU(4)$ generators. 
This is so because the full $8\times 8$ \smat-matrix describing the \nsn junction has a block diagonal 
form with each spin block (up and down spin sectors) being represented by a $4\times4$ matrix. Hence we 
have a unitary starting \smat-matrix deviating from the fixed point \smat-matrix (\smat$_{0}$), as given 
before by Eq.~\ref{smatd}, except that now the sum over $j$ runs from $1$ to $16$ since $\lambda_{j}$'s 
now represent the fifteen generators of the $SU(4)$ group along with the identity matrix. The \smat$_{0}$ 
which describes the \sfpd ~\cite{75} is given by $r=1/2$, $t=1/2$, $r_{A}=-1/2$ and $t_{A}=1/2$. Note that 
the \sfp is a particle-hole, left-right symmetric fixed point and hence the entire $4\times 4 $
\smat-matrix is determined completely by the above given four amplitudes for $r,t,r_A,t_A$.

We then solve Eq.~\ref{eqnm} for the present case with sixteen coupled equations up to the first order 
in the small perturbations characterized by $\epsilon_{j}$'s. We obtain sixteen eigenvalues which correspond
to the power laws around the different sixteen eigen-directions. These power laws around the various 
eigen-directions can be listed as 
[0, 0, 0, 0, 0, $-\alpha/2$, $-\alpha/2$, $(\alpha-\alpha^{\prime})/2$,
$\alpha^{\prime}/2$, $\alpha^{\prime}/2$,
$(-\alpha+\alpha^{\prime})/2$, $(-\alpha+\alpha^{\prime})/2$,
$(\alpha+\alpha^{\prime})/2$, $(\alpha+\alpha^{\prime})/2$,
$(\alpha-\alpha^{\prime}-\sqrt{9\alpha^{2}+14\alpha
\alpha^{\prime}+9\alpha^{{\prime}^{2}}})/4$,
$(\alpha-\alpha^{\prime}+\sqrt{9\alpha^{2}+14\alpha
\alpha^{\prime}+9\alpha^{{\prime}^{2}}})/4$].\\
Hence note that there are five marginal directions, two stable directions, four unstable directions 
and four other directions whose stability depends on the sign of $\alpha - \alpha^\prime$. One of
the most striking outcomes of this analysis is the fact that we obtain two power laws which are not 
just simple linear combinations of $V(0)$ and $V(2k_F)$. Instead, they appear as square roots of 
quadratic sum of these quantities. Our analysis actually leads to the first demonstration of the existence 
of such power laws in the context of quantum impurity problems in \lln theory and this is the central
result of this section.

Having obtained the power laws around \sfp, the next task is to obtain an explicit expression 
for the Landauer-Buttiker conductance corresponding to perturbations around these fixed points 
along some of the eigen-directions. Now note that the \rg equation is expressed in
terms of $\epsilon^\prime$'s whereas the \smat-matrix representing small deviations from the 
fixed point is expressed in terms of $\epsilon$'s. The two terminal linear conductance across the
junction depends explicitly on the \smat-matrix element which are expressed in terms of $\epsilon$'s 
(see Eq.~\ref{smatd}). Hence in order to obtain an expression for conductance in terms of the
temperature or the applied bias voltage dependence induced by $\ed$-$\ed$ interaction, we need to 
first assign bare values to the various perturbations parameterized by $\epsilon'$s and then express 
the $\epsilon'$'s evolved under \rg flow in terms of these bare values of $\epsilon'$'s as 
$\epsilon'(\Lambda) = (\Lambda/\Lambda_0 )^\mu \epsilon^\prime_0$ where $\Lambda$ corresponds to the 
energy scale at which we are probing the system (which can be either voltage bias at zero temperature or 
temperature at vanishing bias voltage) and $\Lambda_0$ is the high energy cut-off expressed in terms of 
voltage or temperature. Then by using the rotation matrix which diagonalizes the coupled \rg equations, 
we express $\epsilon$'s in terms of $\epsilon'$'s written explicitly as a function of temperature or
voltage. Finally plugging these renormalized values of $\epsilon$'s into the \smat-matrix given by 
Eq.~\ref{smatd}, we get all the transmission and reflection amplitudes for the system as explicit
functions of the temperature or voltage carrying the specific power laws associated with perturbations 
switched on along the eigen-directions. These amplitudes are now directly related to the linear conductances.

Now we calculate expression for conductance for a simple case where only one of the 
$\epsilon^{\prime}$($=\epsilon_{15}^{\prime}$) is turned on. For this particular case we need 
the $\mathbb{U}$ matrix which is given by

\bea 
\mathbb{U} &=& \left[\begin{array}{cccccccccccccccc} ~0 & 0
& 0 & 0 & 0 & 0 & 0 & 0 & 0 & 0 & 0 & 0 & 0 & 0 & 0 & 1~
\\
~  0  &  0  &  \sqrt{{3}/{2}}  & 0 & 0  & 0  & 0  & -1/\sqrt{2} & 0
& 0  & 0  & 0 & 0
 & 0  & 1 & 0  ~
\\
~-1 & 0 & 1 & 0 & 0 & 0 & 0 & -\sqrt{3} & 0 & 0 & 0 & 0 & 1 & 0 & 0
& 0~
\\
~ 0 & 0 & 0 & -1 & 0 & 0 & 0 & 0 & 0 & 0 & 1 & 0 & 0 & 0 & 0 & 0 ~\\
~ 0 & 0 & 2 & 0 & 0 & -1 & 0 & 0 & 1 & 0 & 0 & 0 & 0 & 0 & 0 & 0 ~\\
~ 0 & -1 & 0 & 0 & -2 & 0 & -1 & 0 & 0 & -1 & 0 & 0 & 0 & 1 & 0 & 0 ~\\
~ 0 & 0 & 0 & 0 & 1 & 0 & 1 & 0 & 0 & 1 & 0 & 1 & 0 & 0 & 0 & 0 ~\\
~ 0 & 1 & 0 & 0 & 0 & 0 & -1 & 0 & 0 & 1 & 0 & 0 & 0 & 1 & 0 & 0 ~\\
~ 0 & 0 & -\sqrt{3/2} & 0 & 0 & -\sqrt{3/2} & 0 & -1/\sqrt{2} &
\sqrt{3/2} & 0 & 0 & 0 & 0 & 0 & 1 & 0 ~\\
~ -1 & 0 & 1 & 0 & 0 & 1 & 0 & \sqrt{3} & -1 & 0 & 0 & 0 & 1 & 0 & 0
&
0~\\
~ 0 & 1 & 0 & 0 & 0 & 0 & 1 & 0 & 0 & -1 & 0 & 0 & 0 & 1 & 0 & 0 ~\\
~ 1 & 0 & 0 & 0 & 0 & -1 & 0 & 0 & -1 & 0 & 0 & 0 & 1 & 0 & 0 & 0 ~\\
~ 0 & -1 & 0 & 0 & 2 & 0 & -1 & 0 & 0 & -1 & 0 & 0 & 0 & 1 & 0 & 0 ~\\
~ 0 & 0 & 0 & 0 & 1 & 0 & -1 & 0 & 0 & -1 & 0 & 1 & 0 & 0 & 0 & 0 ~\\
~ 1 & 0 & 0 & A & 0 & 1 & 0 & 0 & 1 & 0 & A & 0 & 1 & 0 & 0 & 0 ~\\
~ 1 & 0 & 0 & B & 0 & 1 & 0 & 0 & 1 & 0 & B & 0 & 1 & 0 & 0 & 0 ~\\
\end{array} \right]\,
\label{dumatrix} 
\eea
where, $A={4(\alpha + \alpha^{\prime})}/
{(\alpha-\alpha^{\prime}-\sqrt{9\alpha^{2}+14\alpha
\alpha^{\prime}+9\alpha^{{\prime}^{2}}})/4}$ and 
$B= {4(\alpha
+ \alpha^{\prime})}/
{(\alpha-\alpha^{\prime}+\sqrt{9\alpha^{2}+14\alpha
\alpha^{\prime}+9\alpha^{{\prime}^{2}}})/4}$.

We choose this specific direction to perturb the system as this corresponds to a power law 
which is not a linear function of $V(0)$ and $V(2K_F)$ and hence interesting to study. 
The \smat-matrix to quadratic order in $\epsilon'_{15}$ is given by
\vskip 0.4cm
\bea 
\mathbb{S} = \left[
\begin{array}{cccc} \frac{[1-(1-i)\epsilon_{15}^{\prime} -
{\epsilon_{15}^{\prime}}^2]}{2} &
{\frac{[1+\epsilon_{15}^{\prime}]}{2}}-{\frac{[1-i]{\epsilon_{15}^{\prime}}
^2}{4}}
& -\frac{1}{2} &
{\frac{[1+i\epsilon_{15}^{\prime}]}{2}}-{\frac{[1+i]{\epsilon_{15}^{\prime}
}^2}{4}}\\\\
\frac{[1-(1+i)\epsilon_{15}^{\prime} -
{\epsilon_{15}^{\prime}}^2]}{2} &
{\frac{[1+\epsilon_{15}^{\prime}]}{2}}-{\frac{[1+i]{\epsilon_{15}^{\prime}}
^2}{4}}&
\frac{1}{2} &
{\frac{[1-i\epsilon_{15}^{\prime}]}{2}}+{\frac{[1-i]{\epsilon_{15}^{\prime}
}^2}{4}}\\\\
-\frac{[1+(1-i)\epsilon_{15}^{\prime} -
{\epsilon_{15}^{\prime}}^2]}{2} &
 {\frac{[1-\epsilon_{15}^{\prime}]}{2}}-{\frac{[1-i]{\epsilon_{15}^{\prime}
}^2}{4}}
& \frac{1}{2} &
{\frac{[1-i\epsilon_{15}^{\prime}]}{2}}-{\frac{[1-i]{\epsilon_{15}^{\prime}
}^2}{4}} \\\\
\frac{[1+(1+i)\epsilon_{15}^{\prime} -
{\epsilon_{15}^{\prime}}^2]}{2} &
-{\frac{[1-\epsilon_{15}^{\prime}]}{2}}+{\frac{[1+i]{\epsilon_{15}^{\prime}
}^2}{4}}
& \frac{1}{2} &
{\frac{[1+i\epsilon_{15}^{\prime}]}{2}}+{\frac{[1-i]{\epsilon_{15}^{\prime}
}^2}{4}}
\end{array} \right]\ ,
\label{sdeviation}
\eea

So, the scaling of sub-gap conductance (to ${\mathcal O}(\epsilon_{15}^{\prime 2})$) for an incident 
electron and a hole taking into account both spin-up and spin-down contributions in units of $2e^2/h$ 
is given by
\bea
G_{12}^{e} = -\dfrac{\epsilon_{15}^{\prime}} {2}~; &&
G_{21}^{e} = \dfrac{\epsilon_{15}^{\prime}}{2}\ , 
\eea
\bea G_{12}^{h} = -\dfrac{\epsilon_{15}^{\prime}}{2}~;  &&
G_{21}^{h} = \dfrac{\epsilon_{15}^{\prime}}{2}\ ,
\eea
where $\epsilon_{15}^{\prime}=\epsilon_{15,\,0}^{\prime}(\Lambda/\Lambda_0)^{(\alpha-\alpha^{\prime}
-\sqrt{9\alpha^{2}+14\alpha\alpha^{\prime}+9\alpha^{{\prime}^{2}}})/4}$. Here the superscripts $e$
and $h$ stand for electrons and holes while the subscripts $1$ and $2$ stand for first and second \qw respectively. 
Also, $G_{12}^{e}=|t^{eh}_{A,\,12}|^{2} - |t^{ee}_{12}|^{2}$ where $t^{ee}$ is the transmission amplitude for 
electrons and $t^{eh}_{A}$ represents \car amplitude for electrons. Similar expressions also hold for the holes. 
In the expressions of power laws given above, $\alpha=(g_{2}-2g_{1})/2\pi \hbar v_{F}$ and
$\alpha^{\prime}=(g_{1}+g_{2})/2\pi \hbar v_{F}$ where the bare values of $g_{1}(d)=V(2k_{F})$ and $g_{2}(d)=V(0)$. 
In our stability analysis, we have assumed $\alpha < \alpha^{\prime}$ which is consistent with experimental 
observations~\cite{4}. For the special case when $g_{2}=2 g_{1}$, $\alpha$ vanishes and only
$\alpha^{\prime}$ survives.

It is very interesting to note that even though the $\mathbb{S}$-matrix corresponding to perturbation along
$\epsilon^{\prime}_{15}$ breaks both time reversal and electron-hole symmetry, the two terminal linear conductance 
restores particle-hole symmetry. Secondly it might be of interest to note the fact that the fixed point conductance 
admits correction along $\epsilon^{\prime}_{15}$ which is linear in $\epsilon^{\prime}_{15}$ and not quadratic. 
Normally when we perform a stability analysis around a fixed point $\mathbb{S}$-matrix whose elements are
constituted out of unimodular numbers (representing disconnected or perfectly connected fixed points), 
it is always possible to identify various terms of the $\mathbb{S}$-matrix, representing small unitary 
deviations from the fixed point ${\mathbb{S}}_0$-matrix in terms of various tunneling operators which are 
perturbatively turned on around the fixed point Hamiltonian. Hence a straight forward perturbative linear
conductance calculation using the Hamiltonian along with the tunneling parts will suggest that the correction 
due to the $\mathbb{S}$-matrix representing small deviation from fixed point ${\mathbb{S}}_0$-matrix must 
introduce correction to fixed point conductance which are quadratic in terms of the deviation parameter.
Although this argument applies only to those fixed points which correspond to completely connected or disconnected 
wires and not applicable to fixed points which have intermediate values for various transmission and reflection 
amplitudes like the \sfpd. In other words, an arbitrary deviation from \sfp may not be easily representable as a
tunneling operator. This argument explains why the linear dependence of the conductance on $\epsilon^{\prime}$ 
and hence the corresponding power law dependence looks unconventional.

As a cross check, we observe that we get back the power laws associated with the symmetric fixed point~\cite{27} 
of the four-wire junction once we substitute $\alpha^{\prime}=0$ in the expression for the power laws of the \sfp 
for the \nsn junction. Although our geometry does not correspond to the real junction of four \qwsd, the presence 
of both electron and hole channel mimics the situation of a four-wire junction. More specifically, the \sfp 
of the \nsn junction turns out to be identical to the symmetric fixed point of the four-wire 
junction due to perfect particle-hole symmetry of the \sfp when $\alpha^{\prime}$ is set equal to zero.

Here we enumerate and discuss the stability of the other fixed points (\rfpd, \afpd, \tfp and \cafpd)
discussed earlier in Sec~\ref{sec:3} as well as obtained in Ref.~\refcite{75} for the \nsn junction :
\begin{enumerate}
\item[] (a) $t=t_{A}=r_A=0, r=1$  (\rfpd) :  This fixed point turns out to be stable against perturbations 
in all directions. There are ten directions for which the exponent  is -$\alpha$ while two others with the 
exponents -($\alpha$+$\alpha^{\prime}$). The remaining four directions are marginal.
\vskip +0.3cm
\item[] (b) $t=t_{A}=r=0, r_A=1$ (\afpd) : This is unstable against perturbations in twelve directions. 
There are ten directions with exponent $\alpha$ and two directions with exponent
($\alpha$+$\alpha^{\prime}$). The remaining four directions are marginal, as for \rfpd.
\vskip +0.3cm
\item[] (c)  $r_{A}=t_{A}=r=0, t=1$ (\tfpd) : This fixed point has four unstable directions with exponent $\alpha$, 
two stable directions with the exponent -$\alpha^{\prime}$ and  the remaining directions are marginal.
\vskip +0.3cm
\item[] (d) $r_{A}=t=r=0, t_A=1$ (\cafpd) : This  has four unstable directions with exponent $\alpha$ and two stable 
directions with the exponent  -$\alpha^{\prime}$ and the remaining directions are marginal.
\end{enumerate}
Note that the close similarity in stability between \cafp and \tfp fixed points can be attributed to the fact 
that both these fixed points belong to the continuous family of marginal fixed points defined by the condition 
$|t|^2 + |t_{A}|^2=1$. The entire family of fixed points is marginal because for these fixed points, the 
amplitudes for Friedel oscillation and pair potential in the \qws vanish identically.

Hence, we notice that for the \afp only the scattering amplitude from the pair potential inside the \qw 
is non-zero as only $r_A$ is non-zero. On the other hand for \rfp only the scattering amplitude from 
Friedel oscillations are non-zero as only $r$ is nonzero. Furthermore, both for \cafp and \tfpd, 
the amplitude for scattering from the Friedel oscillations as well as from the pair potential is zero 
as in these cases both $r$ and $r_A$ are zero. Hence \sfp is the only fixed point for which both the amplitude 
for scattering from the Friedel oscillations and the pair potential are finite; hence, this fixed point is 
nontrivial and elegant in the literature of \lln wires. Its very existence can be attributed to the interplay 
of these two different scattering processes arising from Friedel oscillations and the  pair potential inside the \qwd. 
The two terminal conductance at this fixed point gets contribution from both the \ct of electrons through the superconductor 
as well as through the non local \car process. Since both electron and hole channels contribute with opposite 
signs to conductance, if we apply a small perturbation around this fixed point, we get an interesting non-monotonic 
behavior of the conductance $G_{\nsn}=G_{\car}~-~G_{\ct}$ as depicted in Fig.~\ref{figtwelve}. This effect emerges 
due to the competition between the electron and the hole channels and it can be of interest from an experimental 
point of view as well as application point of view which we discuss in the next section. 

\section{\label{sec:5} Spintronics with \nsn Junction of \od quantum wires}
Two fundamental degrees of freedom associated with an electron that are of direct interest to 
condensed matter physics are its charge and spin. Until recent times, all conventional electron-based 
devices have been solely based upon the utilization and manipulation of the charge degree of 
freedom of an electron. However, the realization of the fact that devices based on the spin degree of 
freedom can be almost dissipation-less and with very fast switching times, has led to an upsurge in
research activity in this direction from the last decade~\cite{101,102,103}. The first step towards
realization of spin-based electronics (spintronics) would be to produce pure spin current (\scd).

In this section we demonstrate possible scenarios for production of pure \sc and large tunnelling 
magnetoresistance (\tmrd) ratios from \ct and \car across a superconducting junction comprising 
of normal metal-superconductor-normal metal (\nsnd), where, the normal metal is a \od interacting 
\qwd. Here we demonstrate that some of the fixed points associated with the \nsn junction of \od
\qw discussed in Sec.~\ref{sec:3} correspond to the case of pure \scd. We also analyze the influence 
of $\ed$-$\ed$ interaction and see how it stabilizes or de-stabilizes the production of pure \scd.

Formally, it is straightforward to define a charge current as a product of local charge density 
with the charge velocity, but such a definition cannot be straight-forwardly extended to the
case of \scd. This is because both spin $\vec S$ and velocity $\vec v$ are vector quantities 
and hence the product of two such vectors will be a tensor. In our study, we adopt the simple 
minded definition of \scd, which is commonly used~\cite{102,104}. It is just the product of the 
local spin polarization density associated with the electron or hole, (a scalar $s$ which is 
either positive for up-spin or negative for down-spin) and its velocity~\cite{102,104}.
The most obvious scenario in which one can generate a pure \sc in the way defined above would be 
to have {\it{(a)}} an equal and opposite flow of identically spin-polarized electrons through a channel, 
such that the net charge current through the channel is nullified leaving behind a pure \scd, or 
{\it{(b)}} alternatively, an equal flow of identically spin polarized electrons and holes in the same
direction through a channel giving rise to pure \sc with perfect cancellation of charge current. 
In this section, we explore the second possibility for generating pure \sc using a \nsn junction 
of \od \qwsd.
\begin{figure*}
\begin{center}
\includegraphics[width=12.0cm,height=5.0cm]{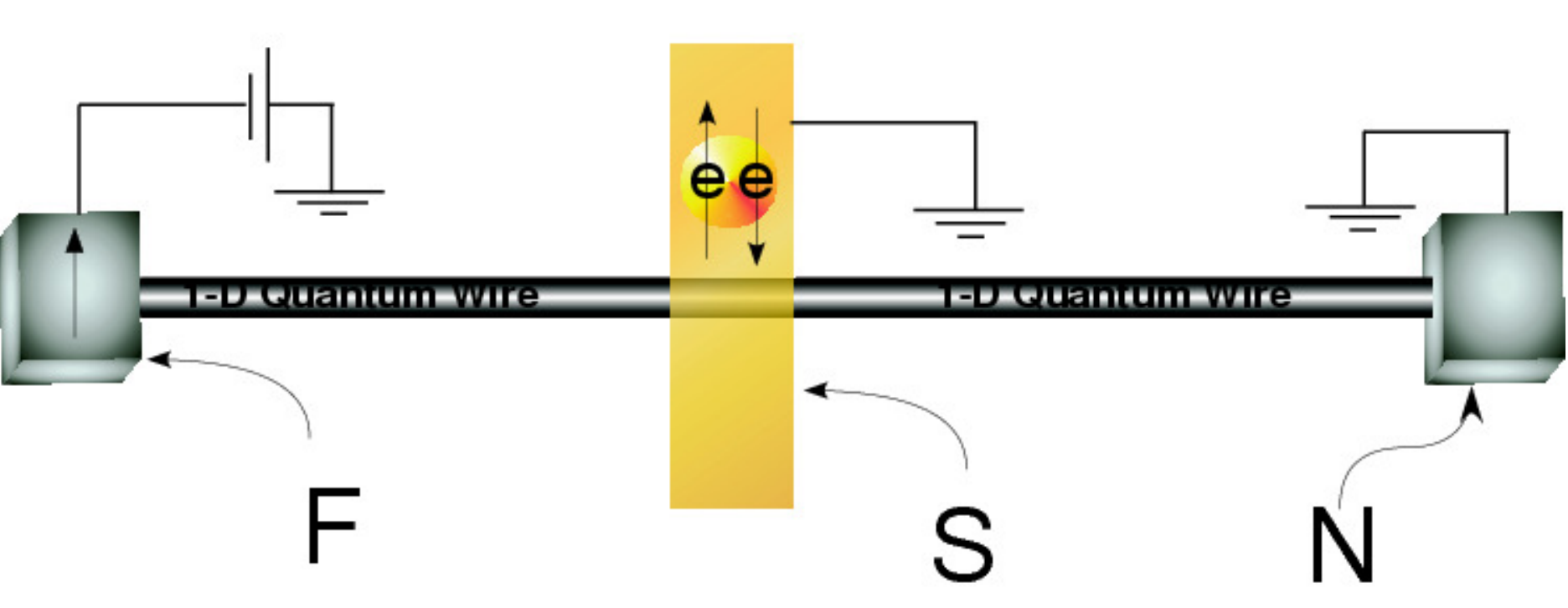}
\caption{(Color online) A \od quantum wire (carbon nanotube) connected to a ferromagnetic (F) lead on the 
left and a normal (N) lead on the right. The thin strip in the middle of the \qw depicts a \td layer 
of bulk superconducting material deposited on top of the wire. Figure adapted from Ref.~\protect\refcite{76}.}
\label{figfifteen}
\end{center}
\end{figure*}
\subsection{\label{sec:5.1} Proposed device and its theoretical modelling}
The configuration we have in mind for the production of pure \sc is shown in Fig.~\ref{figfifteen}. 
The idea is to induce a pair potential in a small region of a quantum wire (\qwd) by
depositing a superconducting strip on top of the wire (which may be, for instance, a carbon nanotube or
$GaAs$ \qwd) due to proximity effects. If the strip width on the wire is of the order of the phase coherence
length of the superconductor, then both direct \ct as well as \car to hole tunnelling can occur across the 
superconducting region~\cite{40,41}. It is worth pointing out that in the case of a singlet superconductor, 
which is the case we consider, both the tunnelling processes will conserve spin. In order to describe
the mode of operation of the device (see Fig.~\ref{figfifteen}), we first assume that the \smat-matrix representing 
the \nsn junction described above respects parity symmetry about the junction, particle-hole symmetry and 
spin-rotation symmetry. Considering all the symmetries, we can describe the superconducting junction
connecting the two half wires by an \smat-matrix with only four independent parameters namely, {\it{(i)}} 
the normal refection amplitude ($r$) for \e (\hd), {\it{(ii)}} the transmission amplitude ($t$) for \e (\hd), 
{\it{(iii)}} \ar amplitude ($r_A$) for \e (\hd), and {\it{(iv)}} the \car amplitude ($t_A$) for \e (\hd). 
If we inject spin polarized electron ($\up$ \ed) from the left \qw using a ferromagnetic contact~\cite{113} and tune 
the junction parameters such that $t$ and $t_A$ are equal to each other, it will lead to a pure \sc flowing in the 
right \qw (see Fig.~\ref{figfifteen}). This is so because, on an  average, an equal number of electrons
($\up$ \ed) (direct electron to electron tunnelling) and holes ($\up$ \hd) (crossed Andreev electron to hole tunnelling) 
are injected from the left wire to the right wire resulting in production of pure \sc in the right wire. Note that 
spin up holes ($\up$ \hd) implies a Fermi sea with an absence of spin down electron (which is what is needed for 
the incident electron $(\up \ed)$ to form a Cooper pair and enter into the singlet superconductor). Generation
of pure \sc by superconducting proximity effect has also been proposed very recently in the context of 
$\tstub$-stub geometry~\cite{108}, superconducting double barrier structure~\cite{109}, interacting quantum dots~\cite{110} 
and ferromagnetic quantum point contact~\cite{111}.

\subsection{\label{sec:5.2} Results for Spin current and Tunneling magnetoresistance}
\begin{figure}
\begin{center}
\includegraphics[width=10.0cm,height=7.5cm]{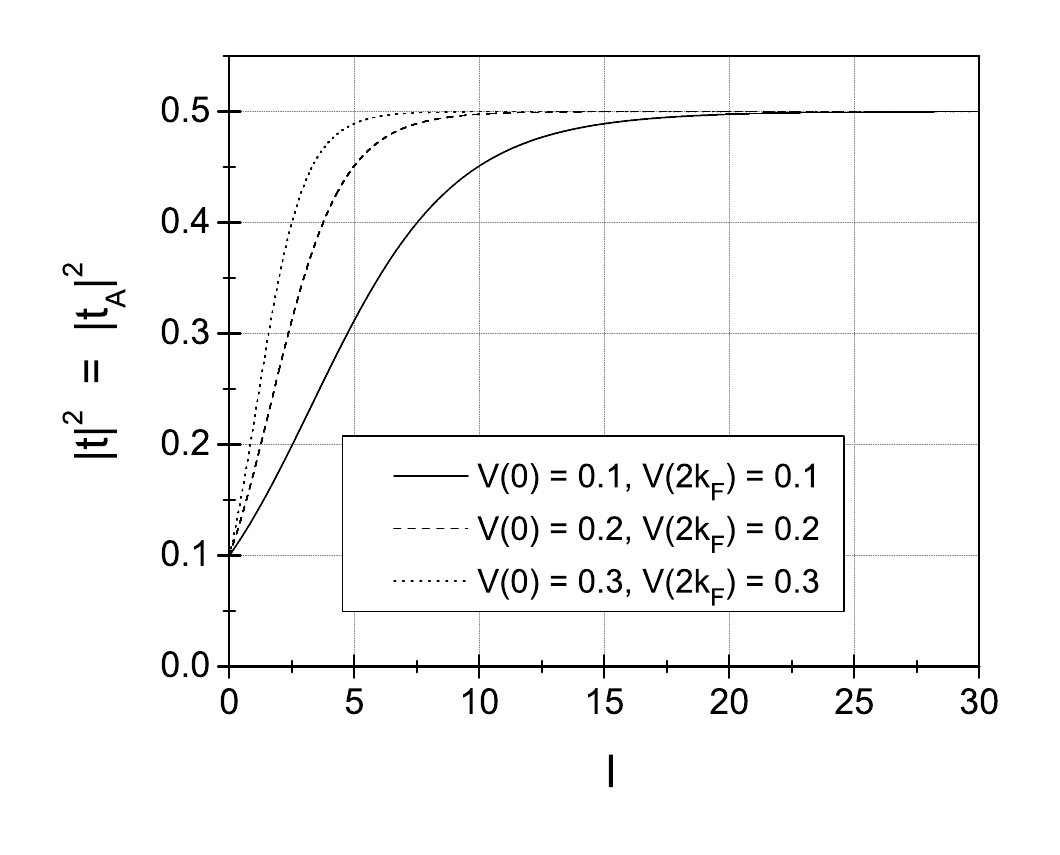}
\caption{(Color online) The variation of $|t|^2$ (=$|t_A|^2$) is plotted as a function of the dimensionless 
parameter $l$ where $l=ln(L/d)$ and $L$ is either $L_{T}=\hbar v_{F}/k_{B} T$ at zero bias or $L_{V}=\hbar v_{F}/eV$ 
at zero temperature and $d$ is the short distance cut-off for the \rg flow. The three curves correspond to the
three different values of $V(0)$ and $V(2k_{F})$ for the \nsn junction. This case corresponds to the \smat-matrix 
given by $\mathbb {S}_{1}$. Figure adapted from Ref.~\protect\refcite{76}.}
\label{figsixteen}
\end{center}
\end{figure}
To discuss pure \sc in \nsn junction we propose two possible \smat-matrices (${\mathbb {S}}_{1}$ 
and ${\mathbb {S}}_{2}$) that can be realized within our set-up which will finally lead to the production 
of pure \scd. The {\it{spin conductance}} is defined as $G^{S}_{\up}(G^{S}_{\dn}) \propto |t|^2 + |t_A|^2$ 
whereas the {\it{charge conductance}} is given by $G^{C}_{\up} (G^{C}_{\dn}) \propto -(|t|^2-|t_A|^2)$.
The $\up$ and $\dn$ arrows in the subscript represent the spin polarization of the injected electrons from 
the ferromagnetic reservoir (see Fig.~\ref{figfifteen}). The negative sign in the expression for $G^{C}_{\up} 
(G^{C}_{\dn})$ arises because it is a sum of contribution coming from two oppositely charged particles 
(electrons and holes). The first \smat-matrix, ${\mathbb {S}}_{1}$ has $r=0$ (reflection-less), $r_A \neq 0$ and $t=t_A$. 
The latter is not a fixed point as listed in Sec.~\ref{sec:3} and hence the parameters of the \smat-matrix flows
under the \rg scheme described in Subsec.~\ref{sec:3.1}. It is easy to see from Eqs.~\ref{rnsn} - \ref{tansn},
that for this case, the \rg equations for $t$ and $t_A$ are identical, and hence it is ensured that the \rg flow 
will retain the equality of the $t$ and $t_A$ leading to the preservation of pure \sc. Physically this implies that  
of a situation in which if we start the experiment with this given \smat-matrix (${\mathbb {S}}_{1}$) at the high energy
scale (at finite bias voltage and zero temperature {\it{or}} at zero bias and finite temperature), then, as we reduce 
the bias in the zero temperature case (or reduce the temperature in the zero bias case), the correlations arising due 
to inter-electron interactions in the wire are such that the amplitude of $t$ and $t_A$ will remain equal to each other. 
The quantity which increases with increasing length scale $L$ is the absolute value of the amplitude $t$ or $t_A$ leading 
to a monotonic increase of pure \sc till it saturates at the maximum value allowed by the symmetries of the \smat-matrix, 
${\mathbb {S}}_{1}$ (depicted in Fig.~\ref{figsixteen}). Here all the \smat-matrix elements are assumed to be 
energy independent and hence the bias dependence is solely due to \rg flow. Of course the bias window has to be small 
enough so that the energy dependence of $t$, $t_A$, $r$ and $r_A$ can be safely ignored.
This saturation point is actually a stable fixed point of the theory if the junction remains reflection-less ($r=0$). 
So we observe that the transmission (both $t$ and $t_A$) increases to maximum value while the \ar amplitude scales down 
to zero leading to pure \scd. Note that the interaction induced correction enhances the amplitude for pure \sc and 
also stabilizes the pure \sc operating point. This makes the operating point, ${\mathbb {S}}_{1}$ quite well-suited 
for an experimental situation. Fig.~\ref{figsixteen} shows the variation of the pure spin conductance 
($= 2 \times |t|^2$ in units of $e^2/h$) as a function of the relevant length scale, $L$ of the problem.
\begin{figure}
\begin{center}
\vskip -1.2cm
\hskip-1cm
\includegraphics[width=14.0cm,height=8.5cm]{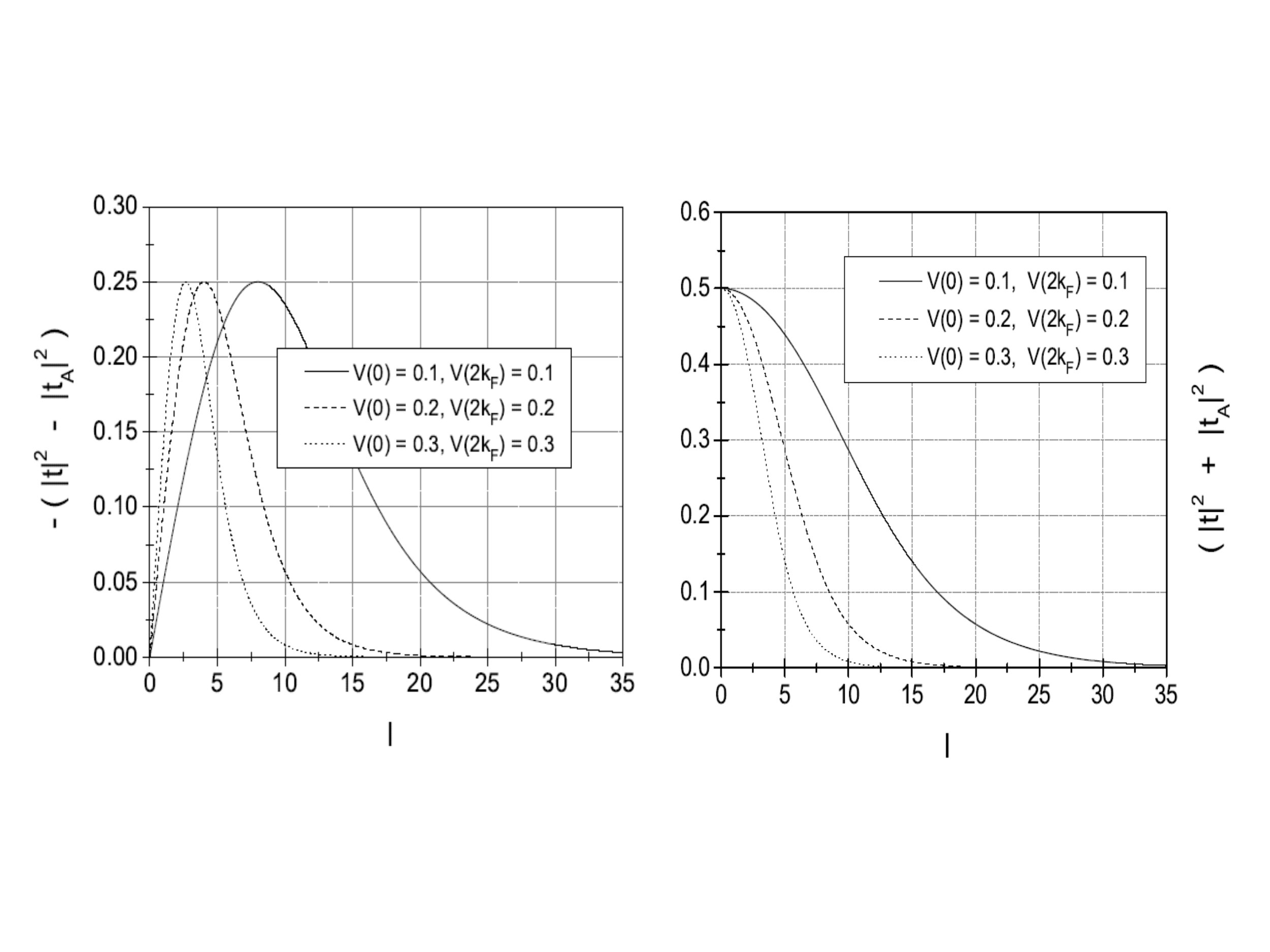}
\caption{(Color online) The variation of $-(|t|^2 - |t_A|^2)$ $\propto G^{C}_{\up}$ or $G^{C}_{\dn}$ and the variation of 
$(|t|^2 + |t_A|^2)$ $\propto G^{S}_{\up}$ or $G^{S}_{\dn}$ are plotted as a function of the dimensionless parameter $l$ 
where $l=ln(L/d)$ and $L$ is either $L_{T}=\hbar v_{F}/k_{B} T$ at zero bias or $L_{V}=\hbar v_{F}/eV$ at zero temperature 
and $d$ is the short distance cut-off for the \rg flow. The three curves in each plot correspond to three different 
values of the interaction parameter $V(0)$ and $V(2k_{F})$ for the \nsn junction. These plots correspond to the 
\smat-matrix given by $\mathbb {S}_{2}$.  Figure adapted from Ref.~\protect\refcite{76}.}
\end{center}
\label{fig17}
\end{figure}

The second case corresponds to the most symmetric \smat-matrix ($\mathbb {S}_{2}$). It is a fixed point of the 
\rg equations and is given by $r_{}=1/2, r_{A}=-1/2, t_{}=1/2, t_{A}=1/2$ discussed earlier in Sec.~\ref{sec:3}.
Here also $t$ is equal to $t_A$ as in the previous case and thus the junction will act like a perfect charge filter 
resulting in pure \sc in the right wire (if spin polarized charge current is injected in the left wire). 
However, this \smat-matrix ($\mathbb {S}_{2}$) represents an unstable fixed point. Due to the implementation of any 
small perturbation, the parameters tend to flow away from this unstable fixed point to the most stable
disconnected fixed point given by $|r|=1$ as a result of \rg flow. So this \smat-matrix ($\mathbb {S}_{2}$) is not a 
stable operating point for the production of pure \scurrentd. Although, it is interesting to note that
if we switch on a small perturbation around this fixed point, the charge conductance exhibits a non-monotonic behavior 
under the \rg flow (Fig.~\ref{figtwelve} and also the left diagram of Fig.~\textcolor{blue}{17}). As discussed earlier,
this non-monotonicity results from two competing effects {\textit {viz.}}, transport through both electron and hole 
channels and, the \rg flow of $g_1$, $g_2$. The latter essentially leads to negative differential conductance 
(\ndcd)~\cite{112}. Elaborating it further, all it means is that if we start an experiment with this given \smat-matrix 
($\mathbb {S}_{2}$) at zero temperature and at finite bias, then as we go towards zero bias, the conductance will 
show a rise with decreasing bias for a certain bias window. This feature can be seen from Fig.~\textcolor{blue}{17}. 
The following aspect of the \rg flow can be of direct relevance for manipulating electron and spin transport in some 
mesoscopic devices.
\begin{figure}
\begin{center}
\vskip -1.2cm
\hskip -1cm
\includegraphics[width=14.0cm,height=8.5cm]{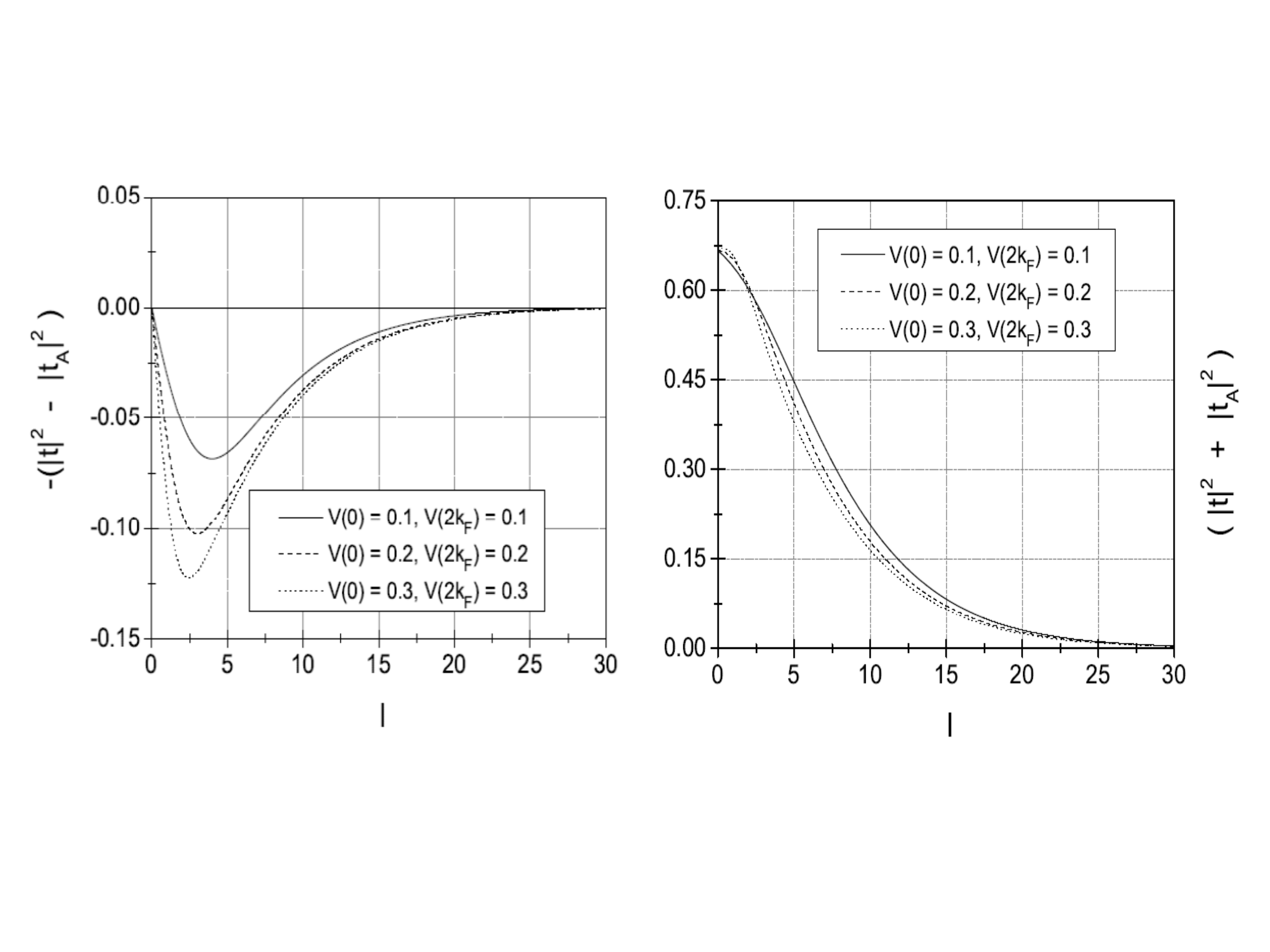}
\caption{(Color online) The variation of $-(|t|^2 - |t_A|^2)$ $\propto G^{C}_{\up}$ or $G^{C}_{\dn}$ and the variation 
of $(|t|^2 + |t_A|^2)$ $\propto G^{S}_{\up}$ or $G^{S}_{\dn}$ are plotted in left and right panel plots as a function 
of the dimensionless parameter $l$ where $l=ln(L/d)$ and $L$ is either $L_{T}=\hbar v_{F}/k_{B} T$ at zero bias
or $L_{V}=\hbar v_{F}/eV$ at zero temperature and $d$ is the short distance cut-off for the \rg flow. The three 
curves in each plot correspond to three different values of $V(0)$ and $V(2k_{F})$ for the \fsn junction. 
These plots correspond to the \smat-matrix given by $\mathbb {S}_{3}$.  Figure adapted from Ref.~\protect\refcite{76}.}
\end{center}
\label{fig18}
\end{figure}
electron and spin transport in some mesoscopic devices.

Now we switch to the case of \fsn junction which comprises of a \od ferromagnetic half metal (assuming $\up$
polarization) on one side and a normal \od metal on the other side (in a way similar to the set-up shown in 
Fig.~\ref{figfifteen}). This case is very complicated to study theoretically because the minimal number of 
independent complex-valued parameters that are required to parameterize the \smat-matrix is nine as opposed 
to the previous (symmetric) case which had only four such parameters. These are given by 
$r_{\up\up}^{11}, ~r_{\up\up}^{22},
~r_{\dn\dn}^{22}, ~t_{A\up\up}^{12}, ~t_{A\dn\dn}^{21},
~r_{A\up\up}^{22}, ~r_{A\dn\dn}^{22}, ~t_{\up\up}^{12},
~{\mathrm{and}}~ t_{\up\up}^{21}$.
Here, $1$ ($2$) is the wire index for the ferromagnetic (normal) wire while, $\up$ and $\dn$ are the respective
spin polarization indices for the electron. The large number of independent parameters in this case arise
because of the presence of ferromagnetic half-metallic wire which destroys both the spin rotation symmetry 
and the left-right symmetry. The only remaining symmetry is the particle-hole symmetry. Analogous to the \rg equations 
(given by Eqs.~\ref{rnsn}-\ref{tansn}) for the \nsn case, it is also possible to write down all the nine \rg equations 
for \fsn case and solve them numerically to obtain the results as shown in Fig.~\textcolor{blue}{18}. In this case, 
the elements of a representative \smat-matrix ($\mathbb {S}_{3}$) which correspond to the production of pure \sc are 
$ |r_{\up\up}^{11}|=|r_{\up\up}^{22}|=|r_{\dn\dn}^{22}|=|t_{A\up\up}^{12}|=
|t_{A\dn\dn}^{21}|=|r_{A\up\up}^{22}|=|r_{A\dn\dn}^{22}|
=|t_{\up\up}^{12}|=\lvert t_{\up\up}^{21}\rvert=1/\sqrt{3}$
and the corresponding phases associated with each of these amplitudes are $\pi/3,\pi,0,-\pi/3,0,\pi/3,0,\pi,-\pi/3$
respectively. By solving the nine coupled \rg equations for the above mentioned nine independent parameters, we have 
checked numerically that this is {\it{not}} a fixed point of the \rg equation and hence it will flow under \rg 
under the application of small perturbation around it and finally reach the trivial stable fixed point
given by $r_{\up\up}^{11} = r_{\up\up}^{22} = r_{\dn\dn}^{22} =1$.
Now if we impose a bias on the system from left to right, it will create a pure \sc on the right wire because
$|t_{A\up\up}^{12}|$ is exactly equal to $|t_{\up\up}^{12}|$. On the other hand, of course, this is a highly 
unstable operating point for production of pure \sc as this is not even a fixed point and hence will always flow 
under any variation of temperature or bias destroying the production of pure \scurrentd. In this case also,
the spin conductance shows a monotonic behavior while, the charge conductance is non-monotonic and hence will have \ndc 
in some parameter regime. It is worth noticing that in this case the interaction parameters $g_1$ and $g_2$ both do not 
scale on the left wire as it is completely spin polarized while $g_1$ and $g_2$ do scale on the right wire as it is not 
spin polarized. Hence even if we begin our \rg flow with symmetric interaction strengths on both left and right wires, 
they will develop an asymmetry under the \rg flow. The following features are depicted in Fig.~\textcolor{blue}{18}. 

Finally, we consider another important aspect that nicely characterizes these hybrid structures from a spintronics 
application point of view. If the \qw on the two sides of the superconductor are ferromagnetic half metals then we have 
a junction of \fsfd. We calculate the tunnelling magnetoresistance ratio (\tmrd)~\cite{102} which is defined as follows
\bea
\tmr = \left[\dfrac{{G_{\up \up} ~-~ G_{\up \dn}}}{G_{\up\dn}}\right]\ ,
\label{tmr}
\eea
Here, $G_{\up \up}$ corresponds to the conductance across the junction when both left and right wires are in parallel
spin-polarized configurations. $G_{\up \dn}$ corresponds to the case when the left and right wires are in anti-parallel
spin-polarized configurations. Thus, \tmr is the maximum relative change in resistance in going from the parallel to the
anti-parallel configuration. For the parallel case, the \car amplitude ($t_A$) is zero and the only process which 
contributes to the conductance is the direct tunnelling process. This is because the \car process involves non-local 
pairing of $\up\e$ in the left wire with $\dn\e$ in the right wire to form a Cooper pair. However for $\dn\e$, the density 
of states is zero in the right wire which makes this process completely forbidden. Hence, $G_{\up \up} \propto |t|^2$. 
On the other hand, for the anti-parallel case, $G_{\up \dn} \propto -|t_A|^2$ as there is no density of states for the 
$\up\e$ in the right lead and so no direct tunnelling of $\up\e$ across the junction is allowed; hence \car is the only 
allowed process. Note that the negative sign in $G_{\up\dn}$ leads to a very large enhancement of \tmr (as opposed
to the case of standard ferromagnet$-$normal metal$-$ferromagnet (\fnfd) junction) since the two contributions will add up. 
A related set-up has been studied in~\cite{105} where also a large \tmr has been obtained.

One can then do the \rg analysis for both parallel and anti-parallel \fsf cases. It turns out that the equations for 
$|t|$ and $|t_A|$ are identical leading to identical temperature (bias) dependance. The \rg equation for $|t_A|$ is
\bea 
\frac{dt_A}{dl} = - \beta\, t_A \, \left[1-|t_A|^2\right]\ ,
\eea
Here, $\beta=(g_2 - g_1)/ 2 \pi \hbar v_F$. Also $|t|$ satisfies the same equation. So, in a situation where the 
reflection amplitudes at the junction for the two cases are taken to be equal then it follows from Eq.~\ref{tmr} 
that the \tmr will be pinned to its maximum value \ie~magnitude of $\tmr=2$ and the temperature dependence will be 
flat even in the presence of $\ed$-$\ed$ interactions inside the \qwd. The latter feature can be very useful in
fabricating spintronic devices based on \fsf junctions of \od \qwsd.  

\section{\label{sec:6} Summary and Outlook}
To summarize, in this review we have presented the effects of $\ed$-$\ed$ interactions on transport properties 
of mesoscopic hybrid structures which include superconducting junctions of multiple \od \qwsd. We start with 
an elementary introduction to the technique called weak interaction renormalization group (\wirgd) used 
to tackle $\ed$-$\ed$ interactions inside the \qw in this review. Within this approach one can obtain the
\rg equations for the effective \smat-matrix and calculate the various fixed point \smat-matrices representing 
the junction. Then we have generalized the \wirg approach to study transport through a superconducting junction 
of multiple \od interacting \qwsd. Applying the \wirg approach we have derived the \rg equations
for the effective \smat-matrix and obtained the various fixed point \smat-matrices representing the 
superconducting junction. Our analysis led to the finding of a novel fixed point with intermediate \ar 
and \car named \sfp which turns out to be unstable and hence experimentally hard to access. We also compute the 
length scale (or temperature) dependance of the Landauer$-$Buttiker conductance taking into account $\ed$-$\ed$ 
interaction induced forward and back-scattering processes. The conductance shows a non-monotonic behaviour around 
the \sfp even for the case of two-wire superconducting junction (\nsnd) in contrast to the \ns junction where the 
dependence is purely monotonic. When more than two wires are attached to the superconductor, we found even more exotic 
fixed points like the \agfp which turns out to be a stable fixed point for a reflection-less ($r=0$) symmetric junction. 
Thus it would be an interesting experimental challenge to look for signature of the \agfp at intermediate temperatures.

As the next step, we have laid down a scheme to perform a systematic stability analysis which works well for both 
normal and superconducting junctions of multiple \lln \qwd. We compute the power laws associated with the renormalization
group flow around the various fixed points of this system using the generators of the $SU(3)$ and $SU(4)$ group to 
generate the appropriate parameterization of a \smat-matrix representing small deviations from a given fixed point 
\smat-matrix. Using our procedure, we reproduce the known power laws for the \gfp of a three-wire junction. Then we apply 
the same precedure to a superconducting junction of two \lln wires and obtain the non-trivial power laws around the 
\sfp which are non-linear functions of $V(0)$ and $V(2 k_F)$. Finally, we calculate the Landauer-Buttiker conductance 
associated with the perturbations switched on around these fixed points and found the explicit voltage or temperature 
power law dependence.

From the application point of view we demonstrate possible scenarios for production of pure \sc and
large \tmr ratios from \ct and \car processes across a superconducting junction
comprising of \od \qwsd. In particular we have studied both spin and charge transport in \nsnd, \fsnd, and \fsf 
structures in the context of \od \qwd. We show that there are fixed points in the theory which correspond to the 
production of pure \scd. We describe the effect of $\ed$-$\ed$ interaction and note the stability of the production 
of pure \sc against temperature and voltage variations. Finally, we also show that the presence of the \car process 
heavily enhances the \tmr in such geometries and also calculate its power law temperature dependence.

Before we conclude, it is worth mentioning that the geometry discussed in this review is also of direct interest 
for the production of non-local entangled electron pairs propagating in two different wires. These electron pairs
are produced by Cooper pair splitting via \car processes when the superconductor is biased with respect to the wires 
comprising the junction. One can then ask if $\ed$-$\ed$ interaction in the \qws actually leads to enhancement of 
entangled electron pair production via the \car processes. For example, we have observed that for the \nsn junction 
with $r=0, t=0, r_{A} \neq 0, t_{A} \neq0 $, interaction can lead to enhancement of the \car amplitude ($t_{A}$) 
under the \rg flow. The latter scenario implies that for the case when the superconductor is biased with respect to 
the wires, inter-electron interactions enhance the production of non-local entangled pairs, for which the amplitude 
$t_A$ is relevant, as compared to the local entangled pairs for which the amplitude $r_A$ would be relevant. 
This is consistent with the results of Recher and Loss~\cite{106,107} who also argued that it is energetically 
more favourable for the two entangled electrons of the Cooper pair to go into different wires, rather than the same wire. 
Finally the \rg flow leads to a fixed point with $t_{A}=1$ where the system becomes a perfect entangler. 
A more general case would be when $r \neq 0, t \neq 0, r_{A} \neq 0, t_{A} \neq 0$. To study this case, one can start 
from the two wires \sfp \smat-matrix and study the \rg flow of $t_{A}$ for an \smat-matrix which is in the close 
vicinity of this fixed point. The result of this study is shown in Fig.~\ref{figtwelve}. We show that starting from 
the short-distance cut-off $d$ the \rg flow initially leads to enhancement of $t_{A}$ which will lead to an enhancement
in the production of non-local entangled pairs. Hence, these studies establish the fact $\ed$-$\ed$ interactions inside
the \qws can actually lead to an enhancement of non-local entangled electron pair production. The above mentioned
scenario of production of entangled electron pairs in Cooper pair beam splitter geometry has been put forwarded
very recently via the differential conductance and shot noise charge as well as spin cross-correlation measurements~\cite{53,54,57,114}.

\section*{Acknowledgements}
We would like to thank our collaborators Sourin Das and Sumathi Rao. 
We especially thank Sumathi Rao for careful reading of the manuscript and valuable comments. 
This work is supported by the Swiss NSF, NCCR Nanoscience, and NCCR QSIT.

\appendix{\rg equations for \fsn junction}
Here we present the \rg equations for nine independent parameters in case of a \fsn junction.
\bea
\dfrac{dr}{dl} &=& \Big[\dfrac{\beta}{2} ~r
\left(1~-~|r|^2\right)-\dfrac{\alpha}{2}~\left(t r^{\prime \star}
t^\prime ~+~ r^{\prime\prime\star} t_A t_{A}^\prime \right)
~+~ \dfrac{\alpha \prime}{2}~ \left(tr_{A}^{\star}t_{A}^\prime ~+~
t_{A} r_{A}^{\prime\star} t^\prime \right)\Big]\ ,
\label{a1} \\
\dfrac{dt}{dl} &=& -\, \Big[ \dfrac{\beta}{2}~ |r|^2 t ~+~
\dfrac{\alpha}{2}~ \left(|r'|^{2}t~+~t_{A}r''^{
\star}r_{A}'\right)
 ~-~
\dfrac{\alpha'}{2}~\left(r_{A}^{\star}r_{A}'t~+~r'r_{A}'^{\star}t_{A}\right)\Big]\ ,
\label{a2}\\
\dfrac{dt_{A}}{dl} &=& -\,\Big[\dfrac{\beta}{2} |r|^2 t_{A}
~+~\dfrac{\alpha}{2}~ \left(|r''|^{2} t_{A} ~+~ t r^{\prime \star}
r_{A}\right)
~-~ \dfrac{\alpha'}{2}~ \left( r_{A}^{\star} r'' t ~+~ r_{A}
r_{A}^{\prime \star} t_{A} \right) \Big]\ ,
\label{a3}\\
\dfrac{dr'}{dl} &=& -\,\Big[\dfrac{\beta}{2}~r^{\star}tt' + \dfrac
{\alpha}{2}~ [r''^{\star}r_{A}r_{A}'-r'(1-|r'|^2)]
~-~ \dfrac
{\alpha'}{2}~r'(r_{A}r_{A}'^{\star}+r_{A}^{\star}r_{A}')\Big]\ ,
\label{a4}\\
\dfrac{dr_{A}}{dl} &=& -\,\Big[\dfrac{\beta}{2}~r^{\star}tt_{A} +
\dfrac{\alpha}{2}~r_{A}(|r'|^{2}~+~|r''|^{2})
~+~
\dfrac{\alpha'}{2}~(r_{A}-r_{A}^2r_{A}'^{\star}-r_{A}^{\star}r'r'')\Big]\ ,
\label{a5}\\
\dfrac{dt'}{dl} &=& -\,\Big[\dfrac{\beta}{2}~|r|^2t' + \frac
{\alpha}{2}~ (|r'|^2t'+r_{A}r''^{\star}t_{A}')
 ~-~\dfrac
{\alpha'}{2}~(r_{A}r_{A}'^{\star}t'~+~r'r_{A}^{\star}t_{A}')\Big]\ ,
\label{a6}\\
\dfrac{dr''}{dl} &=& -\,\Big[\dfrac{\beta}{2}~r^{\star}t_{A}t_{A}'
~+~ \dfrac{\alpha}{2}~ [r'^{\star}r_{A}r_{A}'-r''(1-|r''|^2)]
~-~ \dfrac
{\alpha'}{2}~r''(r_{A}r_{A}'^{\star}~+~r_{A}^{\star}r_{A}')\Big]\ ,
\label{a7}\\
\dfrac{dr_{A}'}{dl} &=&
-\,\Big[\dfrac{\beta}{2}~r^{\star}t_{A}'t~+~
\dfrac{\alpha}{2}~r_{A}'\left(|r''|^2~+~|r'|^2\right)
~+~ \dfrac
{\alpha'}{2}~\left(r_{A}'-r_{A}^{\star}r_{A}'^{2}~-~r_{A}'^{\star}r'r''\right)\Big]\ ,
\label{a8}
\\
\dfrac{dt_{A}'}{dl} &=& -\,\Big[\dfrac {\beta}{2}~|r|^2t_{A}' ~+~
\dfrac {\alpha}{2}~\left(|r''|^2t_{A}'~+~r'^{\star}t'r_{A}'\right)
~-~ \dfrac
{\alpha'}{2}\left(r_{A}'^{\star}r''t'~+~r_{A}'r_{A}^{\star}t_{A}'\right)\Big]\ ,
\label{a9}
\eea
\newpage

\appendix{\rg equations for 3 wire \nsn junction}
Here we give the \rg equations for six independent parameters in case of a symmetric 3 wire \nsn junction.
\bea
\dfrac{dr}{dl} &=& -\,\Big[\dfrac {\alpha}{2} [ r^{\star} \left(
r_{A}^{2}~+~t^{2}+t'^{2}~+~t_{A}^{2}~+~t_{A}'^{2}\right)
~-~ r\left(1~-~|r|^{2}\right)] ~-~ \alpha'[
r|r_{A}|^{2}~+~r_{A}^{\star}\left(tt_{A}~+~t't_{A}'\right)]\Big]\ ,
\label{b1}
\\
\frac{dt}{dl} &=& -\,\Big[\alpha
\left[|r|^{2}t~+~r^{\star}\left(r_{A}t_{A}+t'^{2}~+~t_{A}'^{2}\right)\right]
~-~\alpha' \left[|r_{A}|^{ 2}
t~+~r_{A}^{\star}\left(rt_{A}~+~t't_{A}'\right)\right]\Big],
\label{b2}
\\
\dfrac{dt'}{dl} &=& -\,\Big[\dfrac {\alpha}{2}\left[2|r|^{2}t'~+~
r^{\star}\left(tt'+t_{A}t_{A}'~+~2r_{A}t_{A}'\right)\right]
~-~ \dfrac {\alpha'}{2}\{ 2|r_{A}|^{2}t'~+~r_{A}^{\star}[r
\left(t'~+~t_{A}'\right)
 ~+~\left(t't_{A}~+~tt_{A}'\right) ] \} \Big]\ , \label{b3}
\\
\frac{dr_{A}}{dl} &=&-\,\Big[\frac
{\alpha}{2}\{
2|r|^{2}r_{A}~+~r^{\star}\left[2tt_{A}~+~t_{A}'(t+t')\right] \}
~+~\dfrac
{\alpha'}{2}[r_{A}-r_{A}^{\star}(r^{2}+r_{A}^{2}~+~t_{A}^{2}
~+~ t_{A}'^{2}~+~2tt')]\Big]\ , \label{b4}
\\
\dfrac{dt_{A}}{dl}&=&- \,\Big[\dfrac
{\alpha}{2}\left[2(|r|^{2}t_{A}~+~r^{\star}r_{A}t)~+~r^{\star}t't_{A}'+r_{A}^
{\star}t'^{2}\right]
~-~\dfrac {\alpha'}{2}\left[2(|r_{A}|^{2}t_{A}~+~r t r_{A}^
{\star})~+~r_{A}^
{\star}\left(t'^{2}~+~t_{A}'^{2}\right)\right]\Big]\ , \label{b5}
\\
\dfrac{dt_{A}'}{dl} &=& -\,\Big[\frac
{\alpha}{2}\left[2\left(|r|^{2}t_{A}'~+~
r^{\star}r_{A}t')~+~r^{\star}(t't_{A}~+~tt_{A}'\right)\right]
~-~\dfrac{\alpha'}{2} [ 2\left(|r_{A}|^{2}t_{A}'~+~ r
t'r_{A}^{\star} \right)
 +r_{A}^{\star} \left(tt'~+~t_{A}t_{A}'\right)]\Big]\ ,
\label{b6}
\eea

\section*{References}









\end{document}